\newif\ifllncsloaded\llncsloadedfalse
\newif\ifieetranloaded\ieetranloadedfalse
\newif\ifacmartloaded\acmartloadedfalse

\documentclass[american,usenames]{llncs}
\llncsloadedtrue

\usepackage[utf8]{inputenc}
\usepackage[T1]{fontenc}
\usepackage{microtype}
\usepackage{xspace}
\usepackage{relsize}
\usepackage[normalem]{ulem}
\usepackage{ragged2e}           
\usepackage{url}
\usepackage[font=small,labelfont=bf]{caption}

\usepackage[group-separator={,}]{siunitx}
\usepackage{xifthen}            
\ifacmartloaded                 
\usepackage[MnSymbols,rawdefs]{u8chars}
\else                           
\usepackage[scaled=0.88]{helvet}
\usepackage{mathtools}
\ifllncsloaded\else
\usepackage{amsmath}
\usepackage{amsthm}
\fi
\usepackage{amssymb}
\usepackage[MnSymbols,rawdefs,noscr]{u8chars}
\fi                             
\usepackage{halloweenmath}
\usepackage[caption=false, labelformat=simple]{subfig}

\usepackage{wrapfig}
\usepackage[inline,shortlabels]{enumitem}
\usepackage[ruled,vlined]{algorithm2e}
\usepackage{listings}
\usepackage{comment}            
\usepackage[usenames]{xcolor}
\usepackage{xargs}
\usepackage{graphicx}
\usepackage{tikz}
\usepackage[textsize=footnotesize,textwidth=1.5cm]{todonotes}
\usepackage{booktabs}
\usepackage{tabularx}
\usepackage[flushleft]{threeparttable}
\usepackage{multirow}
\usepackage{array}
\ifieetranloaded
\usepackage[%
backend=bibtex,                 
natbib=true,                    
citestyle=numeric-comp,
sorting=none,
maxcitenames=2,                 
url=false,                      
doi=false,                      
isbn=false,                     
hyperref=true,
]{biblatex}
\fi
\ifllncsloaded
\usepackage[%
  square,        
  comma,         
  numbers,       
  sort,          
  sort&compress, 
]{natbib}

\fi
\usepackage[%
enums,%
scfg-noassertions,%
fsm-noassertions,%
]{symbolic-notations}
\usepackage{galois}
\ifacmartloaded\else
\usepackage[
colorlinks,
linkcolor={red!50!black},
citecolor={blue!50!black},
urlcolor={blue!80!black},
]{hyperref}
\fi

\ifllncsloaded			
\else
\fi

\newcommand{\itodo}[2][]{\todo[inline,caption={2do}, #1]{%
    \begin{minipage}{\textwidth-4pt}#2\end{minipage}}}
\newcommand\nberth[1]{{\color{green!40!black}{#1}}}%
\newcommand\rmrk[2][]{\todo[color=green!20,caption={rmrk}, #1]{%
    \begin{minipage}{\textwidth-4pt}\smaller#2\end{minipage}}}%

\newcommand\Symmaries{\textsf{TOOL}\xspace}
\newcommand\Guardies{\textsf{Guardies}\xspace}

\usepackage{fancyhdr}
\usepackage[24hr]{datetime}
\AtEndDocument{\label{page:last}}
\ifacmartloaded
\fi

\newcolumntype{R}{>{\raggedleft\arraybackslash}X}
\newcolumntype{D}{>{$}c<{$}}
\newcolumntype{Q}{>{$}l<{$}}

\ifacmartloaded\else
\ifllncsloaded\else
\newtheorem{example}{Example}
\newtheorem{definition}{Definition}
\newtheorem{theorem}{Theorem}
\newtheorem{proposition}{Proposition}
\newtheorem{lemma}{Lemma}

\fi
\fi

\ifacmartloaded
\theoremstyle{acmplain}
\fi
\ifllncsloaded\else

\theoremstyle{remark}

\fi

\newcommand{\adhoc}		{{\itshape	 ad hoc\/}\xspace}

\newcommand{\vs}		{{\itshape	       vs}\xspace}

\newcommand{\eg}		{{e.g.,}\xspace}

\newcommand{\ie}		{{i.e.,}\xspace}

\newcommand{\wrt}		{{w.r.t.}\xspace}
\newcommand{\st}		{{\itshape	    s.t\/}\xspace}

\newcommand\A{\ensuremath{\mathcal A}\xspace}%
\newcommand\B{\ensuremath{\mathcal B}\xspace}%
\providecommand\G{\ensuremath{\mathcal G}\xspace}%
\newcommand\I{\ensuremath{\mathcal I}\xspace}%
\newcommand\Q{\ensuremath{\mathcal Q}\xspace}%
\newcommand\R{\ensuremath{\mathcal R}\xspace}%
\renewcommand\S{\ensuremath{\mathcal S}\xspace}%
\newcommand\T{\ensuremath{\mathcal T}\xspace}%
\newcommand\V{\ensuremath{\mathcal V}\xspace}%
\newcommand\X{\ensuremath{\mathcal X}\xspace}%
\newcommand\BB{\ensuremath{\mathbb B}\xspace}%
\newcommand\HH{\ensuremath{\mathbb H}\xspace}%
\newcommand\LL{\ensuremath{\mathbb L}\xspace}%
\newcommand{\ReaX}{{\sffamily ReaX}\xspace}
\newcommand{\Java}{{\sffamily Java}\xspace}
\newcommand{\Jimple}{{\sffamily Jimple}\xspace}
\newcommand{\JVM}{{\sffamily  JVM}\xspace}
\newcommand{\soot}{{\sffamily soot}\xspace}
\newcommand{\KeY}{{\scshape   KeY}\xspace}
\newcommand{\Joana}{{\scshape Joana}\xspace}
\newcommand{\PIDGIN}{{\scshape Joana}\xspace}

\newcommand{\Cassandra}{{\scshape Cassandra}\xspace}
\newcommand{\IFSPEC}{\textsc{IFSpec}\xspace}
\newcommand\IFSpec\IFSPEC

\newcommandx\adjustedbg[3][1=0pt,2=white]{%
  \setlength{\fboxsep}{#1}%
  \colorbox{#2}{#3}%
}

\newcommand\lambdaone[2][]{\ensuremath{\ifthenelse{\isempty{#2}}{}{#1（#2）}}}
\newcommand\lambdatwo[3][]{\ensuremath{\ifthenelse{\isempty{#1#2}}{}{#1（#2, #3）}}}
\newcommand\llambdaone[1]{\ensuremath{\ifthenelse{\isempty{#1}}{}{(#1)}}}
\newcommand\llambdatwo[2]{\ensuremath{\ifthenelse{\isempty{#1#2}}{}{(#1, #2)}}}
\newcommand\subst[3]{\ensuremath{#1\left[#2 ⟼ #3\right]}}%

\newlength\myrulespace%
\newcommand\rname[1]{\textsc{#1}\xspace}%

\newif\ifflushrulesleft
\flushrulesleftfalse

\newcommand{\tuple}[1]{\ensuremath{\left(#1\right)}\xspace}%
\newcommand{\semloc}[1]{\tuple{#1}}%
\newcommand{\myfreerule}[3]{\ensuremath{\rname{#1}\dfrac{#2}{#3}}}
\newenvironment{freeruleset}{
}{}
\ifflushrulesleft
  \newenvironment{ruleset}{
    \align}{\endalign}
  \newcommand{\myrule}[3]{\rname{#1}&\dfrac{#2}{#3\hfill}\nonumber}
\else
  \newenvironment{ruleset}{
    \centering}{\\}
  \newcommand{\myrule}[3]{\myfreerule{#1}{#2}{#3}}
\fi

\newcommand{\trans}[1]{%
  \raisebox{-.5ex}{\ensuremath{\xrightarrow{#1}}}
}

\newenvironment{centermath}{\begin{array}{@{}c@{}}}{\end{array}}

\newcommand{\myoverline}[2][3]{%
  {}\mkern#1mu\overline{\mkern-#1mu#2}}
\newcommand{\myoverrightarrow}[2][1]{
  \mkern#1mu%
  \overscriptrightarrow{
    \mkern-#1mu#2
  }}

\newcommandx\SecuritySemantics[2][1=\hd]{\ensuremath{\mbox{\textsc{Encode}}^{#1}\lambdaone{#2}}\xspace}
\newcommandx\Coreach[1]{\ensuremath{\mbox{\textsc{Coreach}}\lambdaone{#1}}\xspace}
\newcommandx\ComputeGuard[2]{\ensuremath{\mbox{\textsc{SynthesizeGuard}}\lambdatwo{#1}{#2}}\xspace}

\newcommand\void{\ensuremath{\mathsf{void}}\xspace}%
\newcommand\PrimTypes{\ensuremath{\mathsf{PrimTyps}}\xspace}%
\newcommand\Classes{\ensuremath{\mathsf{Classes}}\xspace}%
\newcommand\PrimFields[1]{\ensuremath{\mathsf{PFields}}\xspace}%
\newcommand\RefFields{\ensuremath{\mathsf{RFields}}\xspace}%
\newcommand\MethBody[1]{\ensuremath{\stms_{#1}}\xspace}%
\newcommand\stms{\ensuremath{\mathbb{S}}\xspace}%
\newcommand\lbl[1]{\ensuremath{\mathit{#1}}\xspace}%
\newcommand\Labels[1]{\ensuremath{\mathit{Labels}_{#1}}\xspace}%
\newcommand\TargetName[1]{\ensuremath{\mathsf{target}_{#1}}\xspace}%
\newcommand\Target[2]{\ensuremath{\TargetName{#1}(#2)}\xspace}%
\newcommand\MethSign[1]{\ensuremath{\mathit{Sign}_{#1}}\xspace}%
\newcommand\MethClass[1]{\ensuremath{\mathit{Cls}_{#1}}\xspace}%
\newcommand\MethRet[1]{\ensuremath{\mathit{Ret}_{#1}}\xspace}%
\newcommand\MethFormalArgs[1]{\ensuremath{\mathit{Args}_{#1}}\xspace}%
\newcommand\lits{\red{\ensuremath{w}}\xspace}%
\newcommand\elts{\red{\ensuremath{y}}\xspace}%
 \newcommand\s[1]{\ensuremath{⟪#1⟫}\xspace}%
\newcommand\stm{\ensuremath{\mathit{a}}\xspace}%
\newcommand\cgoto{\ensuremath{\mathtt{goto}}\xspace}%
\newcommand\cif{\ensuremath{\mathtt{if}}\xspace}%
\newcommand\coutput[1]{\ensuremath{\mathtt{output}_{#1}}\xspace}%
\newcommand\cnew{\New}%
\newcommand\sep{~|~}

\newcommand\SumGuard[1]{\ensuremath{\mathit{Grd}_{#1}}\xspace}%
\newcommand\SumEffect[1]{\ensuremath{\mathit{Effect}_{#1}}\xspace}%
\newcommand\HDom[1]{\ensuremath{\mathbb{H}
  }\xspace}%
\newcommand\HFoot[1]{\ensuremath{\mathbb{H}_{\mathit{ret}}%
  }\xspace}%
\newcommand\hfvar[1]{\ensuremath{\hvar_{\mathit{ret}}}\xspace}%

\definecolor{bleudefrance}{rgb}{0.19, 0.55, 0.91}
\definecolor{burgundy}{rgb}{0.5, 0.0, 0.13}
\definecolor{darkcandyapplered}{rgb}{0.64, 0.0, 0.0}
\colorlet{hdval}{bleudefrance!80!black}
\colorlet{hvar}{darkcandyapplered}

\newcommand\hdval[1]{\ensuremath{\textcolor{hdval}{#1}}\xspace}

\newcommand\hd{\hdval{\mathsf{hd}}}
\newcommand\hb{\hdval{\mathsf{*}}}
\newcommand\hdeep{\hdval{\mathsf{deep}}}
\newcommand\hshal{\hdval{\mathsf{shal}}}
\newcommand\hdumb{\hdval{\mathsf{dumb}}}
\newcommand\hsyma{\hdval{\mathsf{syma}}}
\newcommand\hshre{\hdval{\mathsf{shre}}}
\newcommand\hashr{\hdval{\mathsf{ashr}}}

\newcommandx\SymbolicAbstractHeapDom[2][1=\hd]{\ensuremath{\mathsf{HeapDom}_{#1}}\xspace}
\newcommandx\Rels{\ensuremath{\mathcal{R}}\xspace}%
\newcommandx\HeapDomRelations[1][1=\hd]{\ensuremath{\Rels_{\mathsf{Sen}}}\xspace}%
\newcommandx\HeapDomConstRelations[1][1=\hd]{\ensuremath{\Rels_{\mathsf{Insen}}}\xspace}%
\newcommandx\ConcreteHeapDom[1]{\ensuremath{\SymbolicAbstractHeapDom[\concrete]{#1}}\xspace}

\newcommand\HeapDom[1]{\ensuremath{\mathbb{H}_{#1}}\xspace}%
\newcommand\Transformers[1]{\ensuremath{\mathbb{T}_{#1}}\xspace}%

\newcommandx\TypeAnalysis[3][1=\hd]{\ensuremath{\mathsf{CanRelate}\lambdaone{#3}}\xspace}%
\newcommand\TAtauto{\ensuremath{\mathsf{yes}}\xspace}
\newcommand\TAunsat{\ensuremath{\mathsf{no}}\xspace}
\newcommand\TAmaybe{\ensuremath{\mathsf{maybe}}\xspace}

\newcommand\TAbase[2]{\TypeAnalysis[
\hb
]{#1}{#2}}
\newcommand\TAsyma[2]{\TypeAnalysis[\{\ConnectRelSymbol\}]{#1}{#2}}

\newcommandx\DeepHeapAbstractDom[1]{\SymbolicAbstractHeapDom[\hdeep]{#1}}
\newcommandx\DeepHeapDom[1]{\HeapDom[\hdeep]{#1}}
\newcommandx\DeepTransformers[1]{\Transformers[\hdeep]{#1}}

\newcommand\hv[1]{\ensuremath{\textcolor{hvar}{#1}}\xspace}
\newcommand\hvar{\ensuremath{\hv{\mathfrak h}}\xspace}%
\newcommand\fvar{\ensuremath{\hv{\mathfrak f}}\xspace}%

\newcommand\hlvlH[2][\mathclap{\phantom{\hvar}}]{\ensuremath{\ifthenelse{\isempty{#1}}{%
      \vec{#2}%
    }{%
      \myoverrightarrow{#2}\mkern-2mu%
      ^{\mathchoice%
	{\raisebox{-2pt}{$\displaystyle#1$}}%
	{\raisebox{-1pt}{$#1$}}%
	{\raisebox{-0.5pt}{$\scriptstyle#1$}}%
	{\raisebox{-0.2pt}{$\scriptscriptstyle#1$}}}%
    }}\xspace}

\newcommand\hlvl[2][\mathclap{\phantom{\hvar}}]{\ensuremath{\ifthenelse{\isempty{#1}}{%
      \vec{#2}%
    }{%
      \myoverrightarrow{#2}\mkern-2mu%
    }}\xspace}
\newcommand\flvl[3][]{\hlvl[#1]{#2}}
\newcommand\chlvl[2][\mathclap{\phantom{\hvar}}]{\hlvl[#1]{\cv{#2}}}%
\newcommand\HeapPredicate[1][\sHeap]{\ensuremath{#1}\xspace}%
\newcommand\GenericHLvlVars[1][\sHeap]{\ensuremath{\Varset
_{\hlvl L}}\xspace}%
\newcommand\GenericLvlVars[1][\sHeap]{\ensuremath{\Varset_{\plvl L}}\xspace}%

\newcommand\GenericRelSymb{\ensuremath{\smallfrown}\xspace}%
\newcommand\GenericRel[2]{\ensuremath{#1 \GenericRelSymb #2}\xspace}%
\newcommand\GenericRelVar[3][\sHeap]{\ensuremath{#2\!\GenericRelSymb^{\!\hv{#1}}\!\ifthenelse{\isempty{#1}}{\,}{}#3}\xspace}%
\let\GenericRelConst\GenericRelVar 
\newcommand\PrimVars{\ensuremath{P}\xspace}%

\newcommand\GeneralAliasVars[2][\sHeap]{\ensuremath{\Varset
_{#2}}\xspace}%
\newcommand\Varset{\ensuremath{\mathbb V}\xspace}%
\newcommand\GenericRelVars[1][\sHeap]{\GeneralAliasVars[#1]\Rels}
\newcommand\GenericRelTauto[1][\sHeap]{\ensuremath{\Varset
_{\tt}}\xspace}%
\newcommand\GenericRelUnsat[1][\sHeap]{\ensuremath{\Varset
_{\ff}}\xspace}%

\newcommand\genericRel[3][]{\GenericRelVar[#1]{#2}{#3}} 
\let\HLvl\hlvl

\newcommandx\GenericUpdateRels[3][1=\hd]{\ensuremath{\mathsf{UpdHpRel}(#2)}\xspace}%
\newcommandx\NoHeapEffect[3][1=\hd]{\ensuremath{⦇#2⦈\xspace}}%
\newcommandx\OneHeapEffect[4][1=\hd]{\ensuremath{⦇#2, \uparrow#3⦈}\xspace}%
\newcommandx\HeapUpgrade[4][1=\hd]{\ensuremath{⦇#2 ⬿\,#3⦈_{#4}^{\hdval{#1}}}\xspace}%
\newcommandx\GenericHeapInit[3][1=\hd,2=\sHeap]{\NoHeapEffect[#1]{\mathsf{null}~#3}{#2}}%
\newcommandx\HeapInit[2][1=\sHeap]{\GenericHeapInit[\hd][#1]{#2}}
\newcommandx\DeepInit[2][1=\sHeap]{\GenericHeapInit[\hdeep][#1]{#2}}
\newcommandx\ShalInit[2][1=\sHeap]{\GenericHeapInit[\hShal][#1]{#2}}

\newcommand\Null{\ensuremath{\mathtt{null}}\xspace}%
\newcommand\New{\ensuremath{\mathtt{new}}\xspace}%

\newcommandx\NullRef[3][1=\hd]{\NoHeapEffect[#1]{#2 = \Null}{#3}}%
\newcommandx\CopyRef[4][1=\hd]{\NoHeapEffect[#1]{#2 = #3}{#4}}%
\newcommand\LoadRef[4]{\NoHeapEffect{#1 = #2.#3}{#4}}%
\newcommand\CopyRefOp[2]{\ensuremath{#1 = #2}}%
\newcommand\LoadRefOp[3]{\ensuremath{#1 = #2.#3}} %
\newcommand\NewRefOp[2]{\ensuremath{#1 = \New}\xspace}%
\newcommand\NewRef[4]{\OneHeapEffect{\NewRefOp{#1}{#2}}{#3}{#4}}%
\newcommand\StorePrimOp[3]{\ensuremath{#1.#2 ⬿}\xspace}%
\newcommand\StoreRefOp[3]{\ensuremath{#1.#2 = #3}\xspace}%
\newcommand\StorePrim[5]{\OneHeapEffect{\StorePrimOp{#1}{#2}{#3}}{#4}{#5}}%
\newcommand\StoreRef[5]{\OneHeapEffect{\StoreRefOp{#1}{#2}{#3}}{#4}{#5}}%
\newcommand\AliasRelSymbol{\ensuremath{∼}\xspace}%
\ifacmartloaded
\newcommand\AliasRel[3][]{\ensuremath{{#2\!\AliasRelSymbol^{#1}\!\ifthenelse{\isempty{#1}}{}{\!}#3}}\xspace}%
\else
\newcommand\AliasRel[3][]{\ensuremath{{#2\!\!\AliasRelSymbol^{#1}\!\!#3}}\xspace}%
\fi%
\newcommand\cAliasRel[3][]{\AliasRel[#1]{\cv{#2}}{\cv{#3}}}
\newcommand\FieldAliasXRelSymbol{\ensuremath{\stackrel{.*}{\hookrightarrow}}\xspace}%
\newcommand\FieldAliasXRel[3][]{\ensuremath{#2{\FieldAliasXRelSymbol}^{#1}#3}\xspace}%
\newcommand\FieldAliasXVars[1][\hvar]{\ensuremath{{#1}_{\FieldAliasXRelSymbol}}\xspace}%
\newcommand\cFieldAliasXRel[3][]{\FieldAliasXRel[#1]{\cv{#2}}{\cv{#3}}}%
\newcommand\FieldAliasRelSymbol{\ensuremath{\stackrel{.f}{\hookrightarrow}}\xspace}%
\newcommand\FieldAliasVars[1][\sHeap]{\ensuremath{\Varset
_{\FieldAliasRelSymbol}}\xspace}%

\newcommand\ShareRelSymbol{\ensuremath{→\mkern-15mu←}\xspace}%

\newcommand\ShareRel[3][]{\ensuremath{#2{\ShareRelSymbol}^{#1}#3}\xspace}%
\newcommand\FieldShareXRelSymbol{\ensuremath{\stackrel{.*}\ShareRelSymbol}\xspace}%

\newcommand\FieldShareXRel[3][]{\ensuremath{#2{\FieldShareXRelSymbol}^{#1}#3}\xspace}%
\newcommand\FieldShareXVars[1][\hvar]{\ensuremath{{#1}_{\FieldShareXRelSymbol}}\xspace}%

\newcommand\ConnectRelSymbol{\ensuremath{\stackrel{*}{\backsim}}\xspace}%
\newcommand\ConnectRel[3][]{\ensuremath{{#2{\ConnectRelSymbol}^{#1}\ifthenelse{\isempty{#1}}{}{\!\!}#3}}\xspace}%
\newcommand\ConnectVars[1][\hvar]{\ensuremath{{#1}_{\ConnectRelSymbol}}\xspace}%

\newcommand\jsone{\ensuremath{\mathsf{njb}}\xspace}%
\newcommand\jstwo{\ensuremath{\mathsf{nb}}\xspace}%
\newcommand\newloc[1]{\semloc{#1, \jsone}}%
\newcommand\guard{\ensuremath{\mathit{guard}}\xspace}%
\newcommand\cv[1]{\ensuremath{\mathtt{#1}}\xspace}%
\newcommand\plvl[1]{\ensuremath{\mathop{\myoverline[1]{#1}}}\xspace}%

\newcommand\cplvl[1]{\plvl{\cv{#1}}}%
\newcommand\pc{\ensuremath{\mathit{pc}}\xspace}%
\newcommand\Chkpt[1]{\guard}%
\newcommand\uVar{Υ\xspace}%
\newcommand\assignv{\assign_{\uVar}}%
\newcommand\nomlvl[1]{\ensuremath{\lfloor#1\rfloor_{\uVar}}\xspace}%
\newcommand\rVar{\ensuremath{\mathit{hr}}\xspace}%
\newcommand\rMode[1]{\ensuremath{\mathsf{P}_{#1}}\xspace}
\newcommand\rNom{\rMode{\vphantom{ρ}⊥}}%
\newcommand\TBrch[1]{\ensuremath{\mathrm{Brch}\lambdaone{#1}}\xspace}%
\newcommand\Startua{\ensuremath{\mathrm{Start\mbox-ua}}\xspace}%
\newcommand\Endua[1]{\ensuremath{\mathrm{End\mbox-ua}
  }\xspace}%
\newcommand\Regions[1]{\ensuremath{\mathsf{CDRs}(#1)}\xspace}%
\newcommand\regionName[1]{\ensuremath{\mathsf{CDR}_{#1}}\xspace}%
\newcommand\region[2]{\ensuremath{\regionName{#1}(#2)}\xspace}%
\newcommand\juncName[1]{\ensuremath{\mathsf{junc}_{#1}}\xspace}%
\newcommand\junc[2]{\ensuremath{\juncName{#1}(#2)}\xspace}%
\newcommand\juncRName[1]{\ensuremath{\mathsf{junc}^{-1}_{#1}}\xspace}%
\newcommand\juncR[2]{\ensuremath{\juncRName{#1}(#2)}\xspace}%
\newcommand\inducingName[1]{\ensuremath{\mathsf{inducing}_{#1}}\xspace}%
\newcommand\inducing[2]{\ensuremath{\inducingName{#1}(#2)}\xspace}%
\newcommand\Junc[2]{\ensuremath{\mathrm{Junc}（#1,#2）}\xspace}%
\newcommand\JuncL[1]{\ensuremath{\mathrm{Junc}（#1）}\xspace}%
\newcommand\NonJuncL{\ensuremath{¬\mathrm{Junc}（ℓ）}\xspace}%
\newcommand\SaveHeap[1]{\ensuremath{{#1}' \assign #1}\xspace}%
\newcommand\SwapHeap[1]{\ensuremath{{#1}' \assign #1, #1 \assign {#1}'}\xspace}%
\newcommand\GotoRule{\rname{Goto}}%
\newcommand\OutputRule{\rname{Sink}}
\newcommand\AssignRule{\rname{Assign}}%
\newcommand\JunctionRule{\rname{Junc}}%
\newcommand\BranchRule{\rname{Branch}}%
\newcommand{\mode}{\uVar}

\newcommand{\low}{\ensuremath{\bot}\xspace }
\newcommand{\high}{\ensuremath{\top}\xspace }

\newcommand{\qstate}{q}

\newcommand\evalStm[1]{\ensuremath{
   \mathsf{Apply}_{#1}
  }\xspace}

\newcommand{\mStates}{\ensuremath{\Gamma}}

\newcommand\concrete{\ensuremath{\mathsf{crt}}\xspace}

\newcommand{\mHeap}{\ensuremath{\hslash}\xspace}

\newcommand{\sStates}{\ensuremath{\overline{\mStates}\xspace}}

\newcommand{\sHeap}{\hvar}
\newcommand{\isState}[1]{\ensuremath{\sStates_{0}}\xspace}

\newcommand{\smTrans}{\ensuremath{\to}}
\newcommand{\smStates}{\ensuremath{\mathcal{Q}}\xspace}

\newcommand{\mAssignRule}{\textsc{m-Assign}\xspace}

\newcommand{\mStmRule}{\textsc{m-Stm}\xspace}

\newcommand{\mBranchRule}{\textsc{m-Brch}\xspace}

\newcommand{\mUpgrade}{\textsc{m-Upgrade}\xspace}
\newcommand{\mJunctionRule}{\textsc{m-Junc}\xspace}

\newcommand\mframe[4][m]{\ensuremath{\tuple{#2, #3,#4}_{#1}}\xspace}%
\newcommand\mframeX[5]{\ensuremath{〈#2,#3,#1,#4,#5〉}\xspace}%

\ifllncsloaded\else
\DeclareRobustCommand{\qed}{%
  \ifmmode           
  \else \leavevmode\unskip\penalty9999 \hbox{}\nobreak\hfill
  \fi
  \quad\hbox{\qedsymbol}}
\fi


\newcommand{\ttrans}[2]{\raisebox{-.7ex}{\ensuremath{\xRightarrow{\mbox{\smaller\keymathbox{#1}\,\keymathbox{#2}}}}}\xspace}%

\newcommand\lowbisim{\ensuremath{\sim_{\textit{low}}}\xspace}

\newcommand\Nil{\ensuremath{\mathsf{nil}}\xspace}%
\newcommand\GeneralH  {\ensuremath{\mathsf{h}}\xspace}%

\newcommand{\RefGraph}[2]{\ensuremath{{G_{#2}^{#1}}}\xspace}
\newcommand{\Nodes}[1]{\ensuremath{N_{#1}}\xspace}
\newcommand{\Edges}[1]{\ensuremath{E_{#1}}\xspace}

\newcommand{\EquivSymbol}{∼}

\newcommand\FieldAliasRelConcreteName[1]{\ensuremath{\stackrel{.#1}{\hookrightarrow}}\xspace}%
\newcommand\FieldAliasRelConcrete[3]{\ensuremath{#1\!\FieldAliasRelConcreteName{#3}\!#2}\xspace}%

\newcommandx\GenericUpgradeObjLevel[4][1=\hd,2=\sHeap]{\ensuremath{\mathsf{UpdHpLev}\llambdatwo{#3}{#4}}\xspace}%
\newcommandx\HeapUpgradeObjLevel[3][1=\sHeap]{\GenericUpgradeObjLevel[\hd][#1]{#2}{#3}}%
\newcommandx\DeepUpgradeObjLevel[3][1=\sHeap]{\GenericUpgradeObjLevel[\hdeep][#1]{#2}{#3}}%
\newcommandx\ShalUpgradeObjLevel[3][1=\sHeap]{\GenericUpgradeObjLevel[\hshal][#1]{#2}{#3}}%
\newcommandx\SymaUpgradeObjLevel[3][1=\sHeap]{\GenericUpgradeObjLevel[\hsyma][#1]{#2}{#3}}%

\newcommandx\GenericBulkUpgradeFrom[3][1=\hd,2=\sHeap]{\ensuremath{\mathsf{BulkUpgr}_{#2←#3}}\xspace}%
\newcommandx\HeapBulkUpgradeFrom[2][1=\sHeap]{\GenericBulkUpgradeFrom[\hd][#1]{#2}}
\newcommandx\DeepBulkUpgradeFrom[2][1=\sHeap]{\GenericBulkUpgradeFrom[\hdeep][#1]{#2}}
\newcommandx\ShalBulkUpgradeFrom[2][1=\sHeap]{\GenericBulkUpgradeFrom[\hshal][#1]{#2}}

\newcommandx\GenericRestoreObjLevels[3][1=\hd,2=\sHeap]{\ensuremath{\mathsf{RstrLev}_{#2←#3}}\xspace}%
\newcommandx\HeapRestoreObjLevels[2][1=\sHeap]{\GenericRestoreObjLevels[\hd][#1]{#2}}
\newcommandx\DeepRestoreObjLevels[2][1=\sHeap]{\GenericRestoreObjLevels[\hdeep][#1]{#2}}
\newcommandx\ShalRestoreObjLevels[2][1=\sHeap]{\GenericRestoreObjLevels[\hshal][#1]{#2}}
\newcommandx\SymaRestoreObjLevels[2][1=\sHeap]{\GenericRestoreObjLevels[\hsyma][#1]{#2}}

\newcommandx\GenericCopyAliases[3][1=\hd,2=\sHeap]{\ensuremath{\mathsf{CopyRels}%
_{#2←#3}}\xspace}%
\newcommandx\HeapCopyAliases[2][1=\sHeap]{\GenericCopyAliases[\hd][#1]{#2}}
\newcommandx\DeepCopyAliases[2][1=\sHeap]{\GenericCopyAliases[\hdeep][#1]{#2}}
\newcommandx\ShalCopyAliases[2][1=\sHeap]{\GenericCopyAliases[\hshal][#1]{#2}}

\newcommandx\GenericNullRefs[3][1=,2=\sHeap]{\ensuremath{\mathsf{NullRefs}%
_{#2}\lambdaone{#3}}\xspace}%
\newcommandx\HeapNullRefs[2][1=\sHeap]{\GenericNullRefs[\hd][#1]{#2}}
\newcommandx\DeepNullRefs[2][1=\sHeap]{\GenericNullRefs[\hdeep][#1]{#2}}
\newcommandx\ShalNullRefs[2][1=\sHeap]{\GenericNullRefs[\hshal][#1]{#2}}

\newcommandx\GenericHeapResetRef[3][1=\hd,2=\sHeap]{\ensuremath{\GenericUpdateRels[#1]{#3 = \_}{#2}}\xspace}%
\newcommandx\HeapResetRef[2][1=\sHeap]{\GenericHeapResetRef[\hd][#1]{#2}}
\newcommandx\DeepResetRef[2][1=\sHeap]{\GenericHeapResetRef[\hdeep][#1]{#2}}
\newcommandx\ShalResetRef[2][1=\sHeap]{\GenericHeapResetRef[\hshal][#1]{#2}}
\newcommandx\SymaResetRef[2][1=\sHeap]{\GenericHeapResetRef[\hsyma][#1]{#2}}

\newcommandx\GenericHeapCopyRef[4][1=\hd,2=\sHeap]{\ensuremath{\GenericUpdateRels[#1]{\CopyRefOp{#3}{#4}}{#2}}\xspace}%
\newcommandx\HeapCopyRef[3][1=\sHeap]{\GenericHeapCopyRef[\hd][#1]{#2}{#3}}
\newcommandx\DeepCopyRef[3][1=\sHeap]{\GenericHeapCopyRef[\hdeep][#1]{#2}{#3}}
\newcommandx\ShalCopyRef[3][1=\sHeap]{\GenericHeapCopyRef[\hshal][#1]{#2}{#3}}
\newcommandx\SymaCopyRef[3][1=\sHeap]{\GenericHeapCopyRef[\hsyma][#1]{#2}{#3}}

\newcommandx\GenericHeapLoadRef[5][1=\hd,2=\sHeap]{\ensuremath{\GenericUpdateRels[#1]{\LoadRefOp{#3}{#4}{#5}}{#2}}\xspace}%
\newcommandx\HeapLoadRef[4][1=\sHeap]{\GenericHeapLoadRef[\hd][#1]{#2}{#3}{#4}}
\newcommandx\DeepLoadRef[4][1=\sHeap]{\GenericHeapLoadRef[\hdeep][#1]{#2}{#3}{#4}}
\newcommandx\ShalLoadRef[4][1=\sHeap]{\GenericHeapLoadRef[\hshal][#1]{#2}{#3}{#4}}
\newcommandx\SymaLoadRef[4][1=\sHeap]{\GenericHeapLoadRef[\hsyma][#1]{#2}{#3}{#4}}

\newcommandx\GenericHeapStoreRef[5][1=\hd,2=\sHeap]{\ensuremath{\GenericUpdateRels[#1]{\StoreRefOp{#3}{#4}{#5}}{#2}}\xspace}%
\newcommandx\HeapStoreRef[4][1=\sHeap]{\GenericHeapStoreRef[\hd][#1]{#2}{#3}{#4}}
\newcommandx\DeepStoreRef[4][1=\sHeap]{\GenericHeapStoreRef[\hdeep][#1]{#2}{#3}{#4}}
\newcommandx\ShalStoreRef[4][1=\sHeap]{\GenericHeapStoreRef[\hshal][#1]{#2}{#3}{#4}}
\newcommandx\SymaStoreRef[4][1=\sHeap]{\GenericHeapStoreRef[\hsyma][#1]{#2}{#3}{#4}}

\newcommandx\Connect[2]{\ensuremath{\mathsf{Connect}\lambdatwo{#1}{#2}}\xspace}%

\newcommandx\Share[2]{\ensuremath{\mathsf{Share}\lambdatwo{#1}{#2}}\xspace}%

\newcommand{\InRelation}{\ensuremath{\beta}\xspace}

\newcommand\AliasSubCmd[1]{\csname #1AliasSubCmd\endcsname}
\newcommand\DefAliasSubCmd[1]{%
	\expandafter\newcommand\csname #1AliasSubCmd\endcsname[1]{%
		\ensuremath{\expandafter\csname
			#1\endcsname{##1}^{\AliasRelName}}}}

\newcommand{\MemHeapSem}[2]{\ensuremath{\lfloor#1 \rfloor_{#2}}\xspace}
\newcommand{\RefContent}[2]{\ensuremath{#1.#2}\xspace}
\newcommand{\LowRefs}[1]{\ensuremath{R_{\bot}^{#1}}\xspace}

\newcommand{\ConcreteHeapTransformer}[2]{\ensuremath{\lfloor #1 \rfloor_{#2}
}\xspace}
\newcommand{\ConcreteHeapOp}[2]{\ensuremath{\GenericUpdateRels[\concrete]{#1}{#2}}\xspace}%

\newcommand\IUpdatePrimFields[2]{\ensuremath{\mathsf{UpdPFields}(#2)\xspace}}%
\newcommand\IResetRef[2]{\ensuremath{\ConcreteHeapOp{#1}{#2}}\xspace}%
\newcommand\ICopyRef[3]{\ensuremath{\ConcreteHeapOp{#1 =#2}{#3}}\xspace}%
\newcommand\IStoreRef[4]{\ensuremath{\ConcreteHeapOp{#1.#2=#3}{#4}}\xspace}%
\newcommand\ILoadRef[4]{\ensuremath{\ConcreteHeapOp{#1 =#2.#3}{#4}}\xspace}%

\newcommandx\ConcCopyRef[3][1=\sHeap]{\red{\HeapCopyRef[\hd][#1]{#2}{#3}}}
\newcommandx\ConcLoadRef[4][1=\sHeap]{\red{\HeapLoadRef[#1]{#2}{#3}{#4}}}
\newcommandx\ConcStoreRef[4][1=\sHeap]{\red{\HeapStoreRef[\hd][#1]{#2}{#3}{#4}}}
\newcommand\UndefVal {\ensuremath{\mathsf{und}}\xspace}
\newcommand\DefVal {\ensuremath{\mathsf{default}}\xspace}

\newcommand{\sHeapValLevs}[2]{\ensuremath{{#1}_{\hlvl{L}}}\xspace}%
\newcommand{\mHeapVal}{\ensuremath{\mathbf{h}}\xspace}
\newcommand{\VarsTypes}{\ensuremath{\Omega}\xspace}

\setlist{noitemsep}
\setlist[1]{labelindent=\parindent} 
\newlist{compactitem}{itemize}{4}
\setlist[compactitem,1]{nolistsep,label=\textbullet}
\setlist[description]{font=\mdseries\itshape}
\newlist{mathdesc}{description}{4}
\newlist{mathdesc*}{description*}{4}
\newlist{mathpardesc}{description}{4}
\newlist{mathpardesc*}{description}{4}
\newlist{mathpardesc**}{description*}{4}
\newcommand*{\keymathbox}[1]{%
  \mdseries%
  \upshape%
  \setlength{\fboxsep}{.4pt}%
  \fcolorbox{gray}{white}{%
    \strut\ensuremath{#1}%
  }%
}%
\setlist[mathdesc]{format=\ensuremath}
\setlist[mathdesc*]{format=\ensuremath,mode=unboxed}
\setlist[mathpardesc]{format=\keymathbox,
  leftmargin=\parindent,
  labelindent=0pt,
}
\setlist[mathpardesc*]{format=\keymathbox,
  nosep,
  leftmargin=0pt,
  labelindent=\parindent,
}
\setlist[mathpardesc**]{format=\keymathbox,mode=unboxed}

\newlist{bolddescr}{description}{4}
\newcommand*{\mybolddescritem}[1]{\bfseries\upshape{#1}:}%
\setlist[bolddescr]{style=sameline, nosep, format=\mybolddescritem,
  leftmargin=0pt, labelindent=0pt,
}

\usepackage{etoolbox}

\SetCommentSty{AlgoCommandFont}
\SetFillComment
\SetKwComment{Comment}{\{\,}{\,\}}
\DontPrintSemicolon

\LinesNumbered
\makeatletter
\patchcmd{\@algocf@start}
{-1.5em}
{-.2em}
{}{}
\makeatother

\makeatletter
\newcommand{\removelatexerror}{%
  \ifieetranloaded%
  \let\@latex@error\@gobble%
  \fi%
}
\makeatother

\usetikzlibrary{fit}
\usetikzlibrary{arrows, arrows.meta}
\usetikzlibrary{decorations,decorations.pathmorphing}
\usetikzlibrary{calc}

\usetikzlibrary{external}
\AtBeginDocument{\tikzexternaldisable} 
\newif\ifusetikzexternal
\usetikzexternalfalse

\edef\defaultpgflinewidth{\the\pgflinewidth}
\tikzset{
  every path/.style = {
    line width=1pt,
    cap=round,
    join=round,
  },
  rho-highlight/.style = {
    rectangle,
    rounded corners=2pt,
    draw,
    thin,
    inner sep=0pt,
    outer sep=0pt,
  },
  rho-annot-arrow/.style = {
    thin,
    -stealth,
  },
}
\newcommand{\lstkwstyle}{\color{blue}\bfseries}
\newcommand{\lstbasicstyle}{\fontfamily{lmvtt}\selectfont%
\upshape%
}
\newcommand{\lstbasicsize}{
\footnotesize\linespread{0.96}%
}
\newcommand{\lstinlinesize}{
}
\lstdefinestyle{numbered}{%
  numbers=left,
  numberstyle=\tiny,
  numbersep=2pt,
  firstnumber=1,
  numberfirstline=true,
  xrightmargin=0pt,%
  framesep=0pt,%
}

\makeatletter
\gdef\lst@numberfirstlinefalse{\global\let\lst@ifnumberfirstline\iffalse}
\lst@AddToHook{Init}{\global\let\lst@ifnumberfirstline\iftrue}
\makeatother

\lstdefinestyle{nonumbers}{%
  numbers=none,
  xleftmargin=0pt,
}

\lstdefinestyle{inlined}{%
  basicstyle=\lstinlinesize\lstbasicstyle,%
  breakatwhitespace,%
}

\definecolor{darkcyan}{rgb}{0.0, 0.55, 0.55}
\newcommand\lstatenable[1][]{%
  \protect\lstMakeShortInline[style=inlined,mathescape,#1]@%
}

\catcode`~=11 
\newcommand{\mytilde}{
    \texttt{~
  }}
\catcode`~=13 

\lstdefinelanguage{meth}[]{Java}{%
  inputencoding=utf8,
  literate=%
  {¬}{{\(\mybld\neg\)}}1%
  {∨}{{\(\mybldbin\vee\)}}1%
  {∧}{{\(\mybldbin\wedge\)}}1%
  {⊔}{{\(\mybldbin\sqcup\)}}1%
  {⊓}{{\(\mybldbin\sqcap\)}}1%
  {⊑}{{\(\mybldbin\sqsubseteq\)}}1%
  {⊥}{{\(\mybld\bot\)}}1%
  {⊤}{{\(\mybld\top\)}}1%
  {~}{{\mytilde}}1%
  ,
  classoffset=1,
  morekeywords={output,checkpoint},keywordstyle=\bfseries\color{red!60!black},
  classoffset=0,
  emph={this},emphstyle={},
  tabsize=4
}

\lstdefinelanguage{secsum}[]{meth}{%
  inputencoding=utf8,
  tabsize=4
}

\newcommand\inlinemeth[1]{\lstinline[style=inlined,language=meth,mathescape]!#1!}%

\colorlet{rho3color}{green!46!black}%
\colorlet{rho5color}{red!74!black}%

\ifacmartloaded

\else                           

\fi

\ifllncsloaded
\AtBeginDocument{%
}
\fi

\newcommand{\green}[1]{{\leavevmode\color{green!50!black}#1}\xspace}%
\newenvironment{redenv}{
}{}

\newcommand\removable[1]{{
    {#1}}}%
\newcommand\nakh[1]{{\color{purple}#1}}
\newcommand{\red}[1]{{\color{red}#1}}

\newcommand{\nbrem}[2][]{%
  \ifthenelse{\equal{#1}{yep}}\relax{{\leavevmode
      #2}}}%

\ifieetranloaded
\IEEEoverridecommandlockouts
\IEEEpubid{\makebox[\columnwidth]{000-0-0000-0000-0/00/\$00.00~\copyright~YYYY
    IEEE \hfill}%
  \hspace{\columnsep}\makebox[\columnwidth]{ }}
\fi

\ifacmartloaded
\setcopyright{none}
\acmYear{2021}

\acmConference[Woodstock '18]{Woodstock '18: ACM Symposium on Neural
  Gaze Detection}{June 03--05, 2018}{Woodstock, NY}
\acmBooktitle{Woodstock '18: ACM Symposium on Neural Gaze Detection,
  June 03--05, 2018, Woodstock, NY}
\acmPrice{XX.XX}
\acmISBN{978-1-4503-XXXX-X/18/06}

\fi

\ifieetranloaded
\bibliography{IEEEabrv,ifc,sa,mc,des,rv,other}
\fi

\excludecomment{leaveout}
\excludecomment{forappendix}

\renewcommand\red[1]{#1}
\renewcommand\green[1]{#1}
\renewcommand\nakh[1]{#1}
\renewcommand\nberth[1]{#1}
\renewcommand\itodo[1]{}
\renewcommand\todo[1]{{\color{red}#1}}
\renewcommand\note[1]{}
\renewcommand\rmrk[1]{}

\usetikzexternalfalse

\ifieetranloaded\else
\begin{document}
\fi

\title{%
  Symbolic Abstract Heaps for Polymorphic Information-flow Guard Inference\newline(Extended Version)%
}%

\ifacmartloaded                 
\author{Nicolas Berthier}
\affiliation{%
  \institution{University of Liverpool}
  \city{Liverpool}
  \country{UK}
}
\orcid{0000-0002-0933-8193}
\email{nicolas.berthier@ocamlpro.com}
\author{Narges Khakpour}
\affiliation{%
  \institution{Newcastle University\\
  Linn\ae us University}
  \city{Newcstle upon Tyne}
  \country{UK}
}
\email{narges.khakpour@ncl.ac.uk}
\fi

\author{%
  \ifieetranloaded
  \IEEEauthorblockN{\strut}
  \IEEEauthorblockA{\strut\\\strut}
  \else
  Nicolas Berthier\inst{1} \and
  Narges Khakpour\inst{2}
  \fi
  %
}
\institute{
  OCamlPro (France), University of Liverpool (UK)\\
  \and
  Newcastle University (UK), Linn\ae us University (Sweden)
}

\lstatenable[language=meth]

\ifllncsloaded
\maketitle
\fi

\ifieetranloaded
\begin{document}
\maketitle
\IEEEpubidadjcol
\fi

\begin{abstract}
  In the realm of sound object-oriented program analyses for information-flow control, very few approaches adopt flow-sensitive abstractions of the heap that enable a precise modeling  of implicit flows.
  To tackle this challenge, we advance a new symbolic abstraction approach for modeling the heap in \Java-like programs.
  We use a store-less representation that is parameterized with a family of relations among references to offer various levels of precision based on user preferences.
  This enables us to automatically infer polymorphic information-flow guards for methods via a co-reachability analysis of a symbolic finite-state system.
We instantiate the heap abstraction with three different families of relations.
We prove the soundness of our approach and compare the precision and scalability obtained with each instantiated heap domain by using the \IFSpec benchmarks~and~real-life~applications.
\end{abstract}

\ifieetranloaded
\fi

\ifacmartloaded

\ccsdesc{Security and privacy}  

\keywords{{information flow control, language-based security, modular static analysis, discrete controller synthesis, low-level code, heap abstraction, symbolic summaries}}

\maketitle
\fi

\section{Introduction}
\label{sec:introduction}
\emph{Information Flow Control} (IFC) mechanisms offer an effective approach to 
prevent unwanted disclosure of confidential information, or illegal tampering of data.
Their task is to ensure confidentiality and/or integrity, which are usually formalized as noninterference baseline properties~\citep{GuoguenMeseguer1982SecPolsNSecMods}.
Confidentiality demands that \emph{high-sensitive} (secret) inputs do not influence \emph{low-sensitive} (public) outputs.
\nbrem{This means that any change in the value of a secret input must not induce a change in any public output.}
In other words, there must be no \emph{information flow} from any secret 
to any public output.
In software programs, information may flow \emph{explicitly} via direct assignments, \eg from \cv y to \cv x in @x = y;@, or \emph{implicitly} when the execution of statements is guarded by a condition, \eg from \cv c to \cv x in @if (c > 0) x = 42;@.

Many static analysis approaches to ensure noninterference have been advanced, that rely on type-systems~\citep{pottier2003information,sabelfeld2003language,Barthe:2007:CLN:1762174.1762189,LiuMilanova2010StaticIFC4JavaWithImplicitFlows,Hedin:2012:ISC:2354412.2355236}, self-composition~\citep{Barthe2011SecureInformationFlowBySelfComposition,Terauchi:2005:SIF:2156802.2156828,Barthe2011RelationalVerificationUsingProductPrograms}, theorem-proving~\citep{Darvas2005TheoremProvingW4SecureIF-KeY}, and abstract interpretation~\citep{MizunoSchmidt1992SecurityFlowControlAlgo,Mirko2002SecurityTypingsbyAbstractInterpretation,GiacobazziMastroeni2004AbstractNonInterference,Zanioli2012SAILS
}.
%
\emph{Flow-insensitive} static analyses (with ``flow'' as in ``control-flow'') deal with a single set of \red{facts} that is valid for all possible executions of the whole program
, whereas \emph{flow-sensitive} analyses 
provide one set of facts for each statement.
In general, 
flow-sensitivity increases precision, yet
comes with an additional computational cost.
Heap abstractions as computed by alias or points-to analyses obey the same principle
~\citep{Kanvar:2016:HAS:2966278.2931098}: a flow-insensitive heap analysis provides a single, finite representation of the conceptually infinite set of memory locations manipulated by all entire executions of the program, while a flow-sensitive variant gives an abstraction of the heap at each statement.
%
Note that a flow-sensitive analysis for object-oriented programs may rely on a flow-insensitive heap abstraction; this means that the analysis must remain imprecise when dealing with heap-allocated structures.
We consider a \Java-style low-level object-oriented language whose syntax is close to \Jimple's~\citep{vallee2010soot},
and design a static IFC analysis that \emph{automatically} decides whether a program \(P\) implemented in this language \emph{is secure}, \ie \(P\) satisfies a desired noninterference property.
Several approaches have been proposed to verify such properties for \Java-style languages~\citep{Barthe:2007:CLN:1762174.1762189,Hammer:2009:FCO:1667545.1667547,LiuMilanova2010StaticIFC4JavaWithImplicitFlows,GordonKPGNR15DroidSafe,Artz2014FlowDroid,Johnson2015PIDGIN,Lortz2014Cassandra,Darvas2005TheoremProvingW4SecureIF-KeY,Zanioli2012SAILS}.
To the best of our knowledge, however, none of the sound and scalable solutions rely on a flow-sensitive heap abstraction.
Our analysis is therefore \emph{the first of its kind}, as it is \emph{sound}, shows potential for \emph{scalability} since it supports \emph{modularity}, 
and both: 
\begin{enumerate*}[(i)]
\item \emph{captures implicit flows}; and
\item relies on a \emph{flow-sensitive heap abstraction}.
\end{enumerate*}
Achieving these goals in combination is challenging because 
the analysis must
track the information flows that result from
manipulations of object fields and references performed in the program branches that are taken \emph{as well as} in any program branch that is \emph{not} taken\red{: it must therefore reflect about the states of the heap in both taken and non-taken branches  simultaneously}.
\nbrem[yep]{The only work we have found that does consider implicit flows with a flow-sensitive heap abstraction was developed by \citet{Zanioli2012SAILS}
; it however only operates intra-procedurally and on high-level structured programs.}%

%
To do so,
we use a \emph{security typing environment} that associates each memory location manipulated by the program \(P\) %
with a \emph{security level}.
In the case of confidentiality, such a level indicates whether the memory location may hold high-sensitive (secret) data, and \(P\) is secure if
no value from a high-sensitive \emph{source}
flows to a \emph{sink} statement.
%
To deal with all information flows in the heap, we introduce the notion of \emph{symbolic abstract heap domains}, that combine a flow-sensitive security typing environment for every object reachable from a given set of reference variables \(R\), with a flow-sensitive representation of a set of \emph{heap-related relations} pertaining to \(R\) (\eg aliasing).
The domains are parametric in a \emph{family of heap-related relations}, which defines the relations that are captured flow-sensitively by the domain.
This 
allows us to define multiple heap abstractions, each one with its level of precision.
Abstract heaps in such a domain are \emph{predicates} in a propositional logic, that provide a \emph{store-less} model of the heap where irrelevant details related to the behaviors of the memory allocator and garbage collection are safely abstracted away.
The semantics of reference and object mutations are specified using \emph{predicate transformers}.
These can be used to encode the security semantics of any method \(m\) of the program by means of a symbolic transition system \(S_m\), 
where the desired noninterference property is reduced to a safety property \(φ_m\)~\citep{Boudol2009SecureInfoFlowAsSafetyProperty}
.
\nbrem[yep]{We construct our symbolic abstraction based on a \emph{store-less} model of the heap that provides us with canonical representations for semantically equivalent \emph{heap graphs} (up to isomorphism\todo{this probably needs reformulation}), and where irrelevant details related to the behaviors of the memory allocator and garbage collection are safely abstracted away.
  \nbrem[yep]{Our construction relies on a simplified variant of the store-less structures developed by~\citet{BozgaIosifLaknech2003StorelessSemanticsAndAliasLogic}, where we distinguish every \emph{set of reachable objects} as equivalence classes defined using heap-related relations among reference variables, instead of \eg more complex symbolic access paths.}}

Artifacts that we can infer for a method \(m\) include an \emph{information-flow guard}, that is a predicate expressed on propositional ``facts'' about security levels and heap-related relations pertaining to \(m\)'s formal arguments.
This guard describes \red{(sufficient)} circumstances upon which \(m\) is secure, and it is \emph{polymorphic} since it is valid in \emph{any calling context}.
More elaborate artifacts that additionally describe the \emph{effects} of \(m\) on the heap enable \emph{sound} inter-procedural analyses.
\red{In the present paper, however, we focus on our approach for abstracting the heap 
and concentrate our exposition on the inference of 
guards; we leave the inference of polymorphic effects for a future publication.}
%
We compute the guard for a method \(m\)\ via a co-reachability analysis of the system \(S_m\) \wrt its safety property \(φ_m\). \red{A co-reachability analysis finds all states from which a given set of states may be reached, and is typically solved using a fixed-point~\citep{Ramadge89,PnueliRosner}.}
We have implemented the guard inference algorithm in a prototype tool called \Guardies, available at \url{http://nberth.space/symmaries}, that is equipped with several instantiations of symbolic abstract heap domains using various families of heap-related relations.



\subsubsection*{Summary of Contributions}
\begin{itemize}[nosep,left=0pt]
\item We introduce a novel notion of symbolic abstract heap domains that
  uses a set of relations to represent the heap, and is the first flow-sensitive heap model used for information-flow control analysis.
  We define three different instances of this domain (\hdeep, \hshal, \hdumb), each with a different set of relations (Section~\ref{sec::heap.domains})\nberth{, and show that \hdeep constitutes a secure heap abstraction (Section~\ref{sec:secure.heap.abstraction.def})};
\item We symbolically specify the security semantics of our input language to capture explicit and \emph{implicit} flows via the heap, and infer polymorphic information-flow guards via a co-reachability analysis (Section~\ref{sec:secur-guard-synth}).
\nakh{We prove that our analysis under a secure heap abstraction guarantees termination-insensitive noninterference~\cite{sabelfeld2003language};}
\item  We 
  empirically study the respective impacts of our three heap domains, in terms of precision on \IFSpec benchmarks~\citep{HamannHMM0T18}, and in terms of scalability with 60 real-life ABM applications~\citep{do2016toward} 
  (Section~\ref{sec:experimental-results}).
\nakh{Our experiments show that our approach offers the best precision, and the heap model precision has an inverse relationship with scalability. The heap domains \hdumb and \hdeep, that are resp. the least- and most-precise heap model, improve the state-of-the-art precision by 2.4\% and 4.2\% respectively}.

\item \nakh{While the existing approaches use an \adhoc (flow-insensitive) heap model, \Guardies  offers six different heap models, thereby allowing the user to choose a suitable heap model based on her preferences for precision and scalability.}

\end{itemize}




\section{Preliminaries}
\label{sec:preliminaries}

\newcommand\Sm{\ensuremath{S_{\mathtt m}}\xspace}%

\subsubsection*{Input Programs}
\begin{forappendix}
	The \emph{signature} of a method consists in a tuple \(\MethSign{} ≝ 〈\MethClass{}, \MethFormalArgs{}, \MethRet{}〉\) where \(\MethClass{} ∈ \Classes\) is the name of its enclosing class, \(\MethFormalArgs{}
	\) is a mapping from each formal argument to its respective type, and \(\MethRet{} ∈ \PrimTypes ∪ \Classes ∪ ｛\void｝\) is its return type.
\end{forappendix}
We consider a \Java-style low-level language where the code of a
\begin{wrapfigure}{r}{.44\linewidth}
  \begingroup%
  \smaller
  \vspace*{-\intextsep}%
  \setlength\abovedisplayskip{0pt}%
  \setlength\belowdisplayskip{0pt}%
  \[
    \begin{array}{r@{\ }r@{\ }l}
      \stms &⩴& [\lbl{lbl}:]~\stm;\stms \sep \surd \\
      \stm &⩴& v = e \sep v = r.f_p \sep r.f_p = e \\
	    &|& r = r \sep r = r.f_r \sep r.f_r = r \\
	    &|& r = \cnew~c \sep r = \Null \sep r.m(\lits)\\
	    &|& \cgoto~\lbl{lbl} \sep \cif~(e)~\cgoto~\lbl{lbl} \\
	    &|& \coutput l (v) \sep \coutput l (r) \\
      e &⩴& p \sep v \sep ⊖ e \sep e ⊕ e \sep r == r \\
    \end{array}
  \]
  \endgroup%
  \vspace*{-\intextsep}%
\end{wrapfigure}
method is a non-empty finite semicolon-separated sequence of statements built according to \stms in the grammar on the right, where square brackets denote optional constructs.
\(\surd\) is an empty sequence of statements, and \(\lbl{lbl} ∈ \Labels{}\) is a label that uniquely identifies a statement.
\(c 
\) is a class name, \(f_p\) (resp. \(f_r\)) is a primitive (resp. reference) field name, \(p\) is a scalar constant, and \(e\) is an expression.
\(v\) (resp. \(r\)) depicts any local primitive (resp. reference) variable or formal argument used of the method in \MethBody {}.
\nbrem[yep]{; we use \(x\) to denote any such variable}%
  \(\coutput{l}(v)\) sends data $v$ over a channel with the security label \(l\). The label $l$ belongs to a \emph{two-level security domain} which is 
  formalized as a 
  lattice \(〈\LL, ⊑, ⊔
  〉\), where $\LL=\{⊥, ⊤\}$ is the set of security levels, $\low$ is the low-sensitive label, $\high$ is the high-sensitive label, \(⊑\) is a partial order defined over \LL with \(⊥ ⊑ ⊤\), and ⊔ gives the least-upper-bound%
  \footnote{%
    As is traditional, we will present our work by focusing on a standard two-level lattice \(\LL ≝ ｛⊥,⊤｝\); minor adaptations would be necessary to support more complex lattices.}.
Information may become public via \emph{sink} statements, that we denote  $\coutput{⊥}(x)$.
Therefore, \(\coutput{⊥}(v)\) is a sink for the value of \(v\), and \(\coutput{⊥}(r)\) is a sink for \emph{every object that is reachable in the heap} via the reference \(r\).

\subsubsection*{Symbolic Control-flow Graphs --- SCFGs}
The transition systems that we use to encode the security semantics are 
traditional labeled transition systems augmented with sets of \emph{state} and \emph{input} variables, respectively denoted \(X\) and \(I\).
The values for input variables can be seen as coming from the environment of the system.
\begin{definition}[Symbolic Control-flow Graph]
  A \emph{symbolic control-flow graph} is a tuple \(S = 〈Λ, X, I, Δ, ℓ_0, X_0〉\) where:
  \begin{mathdesc*}
  \item[Λ] is a finite non-empty set of locations;
  \item[X] and \(I\) respectively denote state and input variables;
  \item[Δ] is a set of transitions labeled with a \emph{guard} that is a predicate on state or input variables, and a possibly empty set of \emph{assignments} to state variables, noted \([v_0 \assign e_0, …, v_n \assign e_n]\), or ∅ if empty;
  \item[ℓ_0 ∈ Λ] is the initial location; and
  \item[X_0] is a predicate that describes the entire set of possible initial valuations for all the state variables.
  \end{mathdesc*}
\end{definition}
Predicates and right-hand-side expressions in assignments are built using traditional logical connectives (\ie ¬, ∨, ∧ and ⇒), along with a ternary conditional construct ``\ite{⋅}{⋅}{⋅}'' with an obvious meaning. 
The symbolic variables we make use of typically take their values in the security domain \LL, or the set of Booleans \(\BB ≝ ｛\ff, \tt｝\)
.
We use a \emph{merge operation} \SQMergeiu to merge variable assignments
.
This operation is obtained as a union where multiple expressions assigned to a variable \(v\) are combined using some 
connective \(⊔_v\).
The latter depends on the semantics of each variable:
as we only use variables to hold over-approximations in our encoding, we use the dis\-junction ∨ for Booleans, and the least-upper-bound ⊔ for security levels.
For instance, \(\{a \assign \tt, b \assign \ff\} \SQMergeiu \{b\assign \tt\} = \{a \assign \tt, b \assign \ff ∨ \tt\}\).
An \emph{invariant} \(φ 
\) 
for the SCFG \(S\) is a mapping from locations to predicates on state variables.
\(S\) \emph{satisfies} \(φ\) iff every state \(
 q
\) with location ℓ that is reachable by S is such that~\(
q
⊨ φ
(ℓ)
\).

\nakh{An SCFG \(S\) induces a model \FSM S that is a finite-state automaton whose \emph{state-space} \(\Q_S\) is the Cartesian product of the set of locations Λ and the set of all possible valuations for the state variables, \ie \(\Q_S = Λ × \Val X\), where \Val X is the set of \emph{valuations} for all variables in \(X\).
\FSM S takes one transition whenever it receives a valuation for \emph{all} the input variables, \ie an element in \Val I (an empty set for \S).
In any location, there is always exactly one transition whose guard is satisfied by the valuations for all the variables.
When this transition is taken,
its assignments are applied to update the state variables.}
In this work, we only construct SCFGs that are both \emph{deterministic} and \emph{reactive}: \ie given any location \(ℓ ∈ Λ\) and valuations for input and state variables
, there always exists a unique transition in \(Δ(ℓ)\) whose guard is satisfied.
The only source of non-determinism that we make use of in the models lies in the sets of initial values for the state variables
.
Note there is also no notion of accepting state; combined with the properties above, this means that every infinite sequence of elements in \Val I leads to a valid run%
.
%


\section{Symbolic Abstract Heap Domains}
\label{sec::heap.domains}

We first detail our design of heap abstractions that are suitable for the symbolic encoding of security semantics.
In this approach, one predicate is used to model \emph{a set of symbolic heaps}.
Each symbolic heap represents a \emph{parameterizable} set of \emph{heap-related relations} between the portions of the heap that are reachable via a given set of references \(R\), along with a \emph{security typing environment} for every reachable portion of heap.
Such predicates provide \emph{store\-less representations} since object locations are not explicitly represented.
Predicate transformers describe the \emph{effects of heap and reference variable mutations} on sets of symbolic heaps.

\subsection{Families of Heap-related Relations}
Our definition of symbolic abstract heap domains is parameterized by
a \emph{family of heap-related relations}.
A typical example of a heap-related relation is the aliasing relation, that we denote with the symbol \AliasRelSymbol, and which is defined as an equivalence relation where \(\AliasRel r s\) holds iff \(r\) and \(s\) point to the same object.
We define a family of heap-related relations 
as a pair \(\hd ≝ 〈\HeapDomRelations, \HeapDomConstRelations
〉\)
where
\HeapDomRelations is a set of \emph{flow-sensitive relations}, and \HeapDomConstRelations are \emph{constant flow-insensitive relations} (or facts).
A relation 
is formally specified as a set of Boolean variables that each indicates whether two  references taken from \(R\) are in the relation or not \red{(\ie we use predicate abstraction where a Boolean variable specifies whether a relation between two  references holds).
For instance, we need four propositions (therefore, as many Boolean variables) to represent the relation \GenericRel {} {} defined over $R=\{a,b\}$, \ie $\GenericRelVar[] a a$, $\GenericRelVar[] a b$, $\GenericRelVar[] b a$ and $\GenericRelVar[] b b$. The proposition $\GenericRel x y$ evaluates to true if $x$ is in the relation \GenericRel{}{} with $y$. Further, a relation $\GenericRel{}{} =\{(a,a),(b,a)\}$ is formalized as $\GenericRelVar[] a a ∧ ¬ {\GenericRelVar[] a b} ∧ \GenericRelVar[] b a ∧ ¬ \GenericRelVar[] b b $.}

The propositions about flow-sensitive relations may be updated by the program statements,
while the propositions about constant flow-insensitive relations are straightforwardly substituted with \tt or \ff,
and serve the sole purpose of improving the precision of the \red{predicate transformers that manipulate symbolic abstract heaps}.
For instance, in a heap domain that does not handle the aliasing relation flow-sensitively,
two references of incompatible types can never alias each other.
We formalize this with the pre-analysis function \TypeAnalysis R {}%
, that returns three-valued \emph{sound} 
facts about the relations in \Rels = \HeapDomConstRelations $\cup$ \HeapDomRelations
~\wrt the set of reference variables \(R\). 
Given any relation \(\GenericRelSymb ∈ \Rels\) and a pair of references \((r,s) ∈ R^2\), \TypeAnalysis
R {\GenericRel r s} returns \TAtauto if \GenericRel r s always holds, \TAunsat if it cannot hold, or \TAmaybe otherwise.
In its most trivial form, this pre-analysis function operates on a purely lexical level, \eg by returning \TAtauto if queried for \GenericRel r r with \GenericRelSymb a reflexive relation, \TAmaybe otherwise.
It may additionally involve an analysis of the class hierarchy and take the declared type of the reference variables into account to give more precise facts.
\red{Note that \TypeAnalysis {} {} helps us simplify the heap formulae by reducing the number of propositions used to represent heaps
. For instance, if $\GenericRel{}{}$ is a reflexive relation in our previous example, we don't need to consider the propositions \GenericRel a a and \GenericRel b b, and use the constant \tt instead.}
We leave further specifications of the pre-analysis open for the sake of modularity.

\red{We need to differentiate between ``flow-sensitive heaps'' and ``flow-sensitive heap relations''. The first case means that the heap changes during the execution while the latter states that the relation used to specify the heap structure changes during the execution (\ie \HeapDomRelations). Therefore, a flow-sensitive symbolic heap abstraction models at least one heap relation flow-sensitively.}

\subsection{Symbolic Abstract Heap Domain}
Formally, a \emph{symbolic abstract heap domain} for the family of heap-related relations \hd
is defined as a pair $\SymbolicAbstractHeapDom R ≝ \langle
\HeapDom \hd , \Transformers \hd \rangle$ %
where
\HeapDom \hd is the set of symbolic abstract heaps,
and \Transformers \hd is a set of predicate transformers to manipulate the abstract heaps.
A \emph{symbolic abstract heap} from this domain 
is a predicate \(\HeapPredicate[\sHeap] ∈ \HeapDom \hd\) 
defined on two sets of state variables
\GenericHLvlVars and \GenericRelVars. 
The set \GenericHLvlVars associates a \emph{security level variable} \hlvlH[\sHeap] r with each reference \(r ∈ R\), that represents an \emph{upper bound} on the security levels of any object that is reachable via \(r\)\nberth{: these variables constitute the \emph{security typing environment} for the abstract heap \sHeap}.
In turn, the set \GenericRelVars consists of Boolean variables that describe \emph{over-approximations of \red{flow-sensitive} heap-related relations} between the references in \(R\),
\ie a variable \(\GenericRelVar r s ∈ \GenericRelVars[\sHeap]\)
holds whenever \((r, s)\) \emph{may} be in the heap-related relation \GenericRelSymb:
%
{
\[\red{\GenericRelVars[\sHeap] ≝
｛\GenericRelVar r s ｜ (r,s)∈R^2, \GenericRelSymb ∈ \HeapDomRelations, \TypeAnalysis R {\GenericRel r s} = \TAmaybe ｝\mbox.}
\]%
}%
Further, \GenericRelUnsat and \GenericRelTauto are sets of constants that capture all relations that never hold and always hold according to the function \TypeAnalysis R {}, respectively.
(For the sake of readability, we will omit the exponent \sHeap of security-level variables when a single abstract heap \sHeap is involved, \ie \hlvlH[\sHeap] r will be denoted \hlvl[\sHeap] r.)
%
%
{
  \begin{align*}
    \GenericRelUnsat ≝
    &\ ｛\GenericRelConst r s 
      ｜ \begin{array}{@{}l@{}}
	   (r,s)∈R^2,
	   \GenericRelSymb ∈ \HeapDomRelations ∪ \HeapDomConstRelations, \TypeAnalysis R {\GenericRel r s} = \TAunsat
	 \end{array}%
    ｝
    \\
    \GenericRelTauto ≝
    &\ ｛\GenericRelConst r s 
      ｜ \begin{array}{@{}l@{}}
	   (r,s)∈R^2,
	   \begin{array}{@{}l@{}}
	     （\GenericRelSymb ∈ \HeapDomRelations, \TypeAnalysis R {\GenericRel r s} = \TAtauto） ∨ \\
	     （\GenericRelSymb ∈ \HeapDomConstRelations, \TypeAnalysis R {\GenericRel r s} ≠ \TAunsat）
	   \end{array}%
	 \end{array}%
    ｝\mbox.\\
  \end{align*}}%

%
\nbrem[yep]{\HeapPredicate can be written in \red{full disjunctive normal form} using \(N\) conjuncts as
{{\footnotesize
  \setlength{\abovedisplayskip}{1pt}%
  \setlength{\belowdisplayskip}{1pt}%
  \begin{equation*}
  \label{eq:symbolic-abstract-heap-value-predicate}
  \HeapPredicate =%
  ⋁_{i∈｛1,…,N｝}{（⋀_{\HLvl[\sHeap]r ∈ \GenericHLvlVars}{\HLvl[\sHeap]r = l_r}\ ∧
    ⋀_{\GenericRelVar r s \in \GenericRelVars}\boxed{\GenericRelVar r s}）}\mbox,
\end{equation*}%
}}\todo{Maybe use inline text description for conciseness.}%
\noindent where \red{\(l_r ∈ \LL\)} is a security level for \(r\), and \(\boxed{e}\) can be either $e$ or its negation $\neg e$.
This form highlights that an abstract heap represents \red{a \emph{set} of symbolic heaps}, each of which features a \emph{typing environment} encoded with variables \GenericHLvlVars in the left-hand side of the conjunction.}%
\begin{leaveout}
\begin{redenv}
We give the definitions of \GenericRelVars, \GenericRelTauto and \GenericRelUnsat in \figurename~\ref{fig:heap-abstr-val-alias-vars-n-consts}.
The former set encodes unknown or non-constant elements of the relations in \HeapDomRelations using variables,
whereas symbols in the latter are conservatively defined according to the pre-analysis \TypeAnalysis R {}. 
Note that some symbols in \GenericRelTauto may represent relations from \HeapDomConstRelations that cannot be safely ruled out by the pre-analysis.
\end{redenv}
\end{leaveout}
\begin{table}[t!]
  \centering%
  \smaller%
  \setlength\aboverulesep{.1em}%
  \setlength\belowrulesep{.1em}%
  \begin{threeparttable}
    \caption{Denotations for symbolic abstract heaps}
    \label{tab:symbolic-heaps}
    \begin{tabularx}{\linewidth}{@{}l@{}R@{~}|@{~}l@{}R@{}}%
      \toprule
      \multicolumn{3}{@{}l@{}}{Two symbolic abstract heaps:}
      & \(\mathllap{\hvar, \hvar' \in  \HeapDom \hd}\) \\
      \multicolumn{3}{@{}l@{}}{Set of variables encoding a heap-related relation \(\GenericRelSymb ∈ \HeapDomRelations\):}
      & \GenericRelVars\\
      \multicolumn{3}{@{}l@{}}{Set of constants encoding \red{non-membership} facts, for any relation \(\GenericRelSymb ∈ \HeapDomRelations\cup \HeapDomConstRelations \):}
      & \GenericRelUnsat
      \\
      \multicolumn{3}{@{}l@{}}{Set of constants encoding \red{membership} facts, for any relation
      \(\GenericRelSymb ∈ \HeapDomRelations \cup \HeapDomConstRelations \):}
      & \GenericRelTauto
      \\
      \multicolumn{3}{@{}l@{}}{Set of variables encoding the security levels:}
      & \GenericHLvlVars[\sHeap]\\
      \midrule
      \multicolumn{2}{@{}c@{~}|@{~}}{Variables \& Predicate}
      & \multicolumn{2}{@{}c}{Transformers (\(\Transformers R\))} \\
      \midrule
      Security level variable
      :
      & \(\hlvlH[\hvar]r\) \red{(or \hlvl r)}
      & Reference assignment:
      & \(\NoHeapEffect{\mathit{as}}{\hvar}\)
      \\
      Relation variable:
      &  \genericRel[\hvar] r s
      & Mutation and allocation:
      & \(\mathllap{\OneHeapEffect{\mathit{mu}}{l}{\hvar}}\)
      \\ 
      Initialization
      :
      & \(\HeapInit{R'}\)
      & Bulk upgrade:
      & \(\mathllap{\HeapBulkUpgradeFrom[\hvar]{\hvar'}}\)
      \\
      \bottomrule
    \end{tabularx}
    \RaggedRight%
    with \(\mathit{as} ∈ \{r = s, r = s.f_r, r = \Null\}\),
    \(\mathit{mu} ∈ \{\red{r.f_p ⬿,} r.f_r =s, r =\New\}\),
    \((s, r) ∈ R^2\), \(R' ⊆ R\),
     and any security level expression
     \(l\). 
     \par
  \end{threeparttable}
\end{table}

\begin{figure}[t]
  \centering
  \begin{minipage}[t]{.46\linewidth}%
    \raggedleft
    \subfloat[Class definitions and method \protect\inlinemeth{m}.]{%
      \input{fig/m}%
      \label{lst:m-example-code}%
    }%
    \\
  \end{minipage}%
  \hfill%
  \begin{minipage}[t][][b]{.46\linewidth}%
    \subfloat[\red{Heap Transformers and Variables}]{%
     \begin{minipage}{\linewidth}%
	\centering\input{fig/m-example-setup}%
    \smaller \begin{tabular}{c|c}
    \hline
    Statement Loc& Heap Transformer \\\hline
      3   & \( \NewRef\bl{\mathtt{B}}\pc\sHeap\)\\
      4   & \StorePrim{\cv a}{\fint}{\cv i}{\cplvl i ⊔ \pc}\sHeap\\
      5   & \StoreRef\bl\fa{\cv a}{\cplvl a ⊔ \chlvl[\sHeap] a ⊔ \pc}\sHeap\\
      6   & ∅\\\hline
    \end{tabular}
    \(\begin{array}{@{}c@{}}
        R =  ｛\cv a, \cv b, \cv r｝,\Rels = \{\cAliasRel {} {}, \cFieldAliasXRel {} {}\}\\
      \GenericHLvlVars= ｛\chlvl a, \chlvl b,  \chlvl r｝ ,
      \GenericRelVars = \{\cAliasRel b r, \cFieldAliasXRel b a, \cFieldAliasXRel r a\}
    \end{array}\)%
    \label{table:example-heap-transformers}
  \end{minipage}%
    }%
\end{minipage}%
  \caption{Method that manipulates references, with a representation of heap transformers, heap-related relationships and variables.
  }%
  \label{fig:m-example}%
\end{figure}
%
\begin{figure}[t]
  \setlength{\abovedisplayskip}{0pt}%
  \setlength{\belowdisplayskip}{0pt}%
  \footnotesize%
  \smaller
  \begin{alignat*}{4}
    \NullRef r \sHeap ≝
    &\ \HeapResetRef r
    &\ \SQMergei u& ［\hlvl[\sHeap]r \assign \red{⊥}］ \\
    \CopyRef r s \sHeap ≝
    &\ \HeapCopyRef r s
    &\ \SQMergei u& ［\hlvl[\sHeap]r \assign \hlvl[\sHeap]s］ \\
    \LoadRef r s {f_r} \sHeap ≝
    &\ \HeapLoadRef[\sHeap] r s {f_r}
    &\ \SQMergei u& ［\hlvl[\sHeap]r \assign \hlvl[\sHeap]s］ \\
    \NewRef r c l \sHeap ≝
    &\ \HeapResetRef r
    &\ \SQMergei u& ［\hlvl[\sHeap]r \assign l］ \\
    \StorePrim r {f_p} e l \sHeap ≝
    &&&\ \HeapUpgradeObjLevel[\sHeap] r l \\
    \StoreRef r {f_r} s l \sHeap ≝
    &\ \HeapStoreRef[\sHeap] r {f_r} s
    &\ \SQMergei u&\ \HeapUpgradeObjLevel[\sHeap] r l \\
    \HeapInit[\sHeap]{R'} ≝
    &\ \HeapNullRefs[\sHeap]{R'}
    &\ ∧&\ ⋀\nolimits_{r∈R'}{\hlvl[\sHeap] r = ⊥}\\
    \HeapBulkUpgradeFrom[\sHeap]{\sHeap'} ≝
    &\ \HeapCopyAliases[\sHeap]{\sHeap'}
    &\ \SQMergei u&\ \HeapRestoreObjLevels[\sHeap]{\sHeap'}
  \end{alignat*}\vspace*{-1em}
  \setlength{\abovedisplayskip}{0pt}%
  \setlength{\belowdisplayskip}{0pt}%
  \[\mbox{with~}
    \HeapCopyAliases[\sHeap]{\sHeap'} ≝
    \SQMergeu_{\mathclap
      {\GenericRelVar[\sHeap']r s ∈ \GenericRelVars[\sHeap']}}%
    ［\GenericRelVar[\sHeap] r s \assign \GenericRelVar[\sHeap']r s］
    \mbox{~and~}
    \HeapNullRefs[\sHeap]{R'} ≝
    ⋀_{\mathclap{\GenericRelVar[\sHeap] r s ∈ \GenericRelVars[\sHeap], ｛r,s｝∩R' ≠ ∅}}{%
      （\GenericRelVar[\sHeap] r s = \ff）}\mbox.
  \]\vspace*{-1em}
  \caption{Definitions of generic transformers \Transformers R 
    for any symbolic abstract heap domain \SymbolicAbstractHeapDom R
    . 
    Note \(\hvar \) and \(\hvar'\) belong to the same domain
    , \ie \((\hvar, \hvar') ∈ {\HeapDom R}^2\), and \(R' ⊆ R\).
    \nakh{See \figurename~\ref{fig:heapdom-specialized-update-helpers} for an example definition of ${\GenericUpdateRels[\hd]{⋅}{\hvar}}$. }
  }
  \label{fig::Symmaries.Abstract.Heap.Operations}
\end{figure}
%
%
 
%
\vspace{-0.25in}
We summarize in \tablename~\ref{tab:symbolic-heaps} the main denotations that we use to represent and manipulate symbolic abstract heaps.
The leftmost column shows the variables that represent security levels and heap-related relations, along with the operator \(\HeapInit[\sHeap]{R'}\).
The right-hand side column
lists the set of \emph{predicate transformers} that can be applied on a symbolic abstract heap to alter its representation.
The two first transformers in the column operate in accordance with a given reference assignment (\(\mathit{as}\)) or mutation (\(\mathit{mu}\)).
The expression \(l\) given to the latter gives the security level of the information that flows to mutated objects.
We give in \figurename~\ref{fig::Symmaries.Abstract.Heap.Operations} the definitions of all transformers
.
These definitions make use of 
functions specialized for each family of heap-related relations, detailed below.
\(\HeapInit[\sHeap]{R'}\) builds a predicate that constrains variables in \GenericHLvlVars and \GenericRelVars to account for the fact that 
a given set of references \(R'⊆R\) 
is \Null
---this notably entails that every object reachable via \(R'\) is low-sensitive.
\green{
  The \emph{bulk upgrade} is a transformer used to capture implicit flows through the heap by joining two \nberth{distinct} heap abstractions \sHeap and \(\sHeap'\) that belong to the same domain.
  More specifically, this transformer \red{assumes that \(\hlvlH[\sHeap']r ⊑ \hlvlH[\sHeap]r\) for all \(r ∈ R\)}, and:
  \begin{enumerate*}[(i)]
  \item copies the heap-related relations from \(\sHeap'\) to \(\sHeap\); and
  \item upgrades the typing environment of \sHeap via a pairwise join with the corresponding levels in \(\sHeap'\). 
  \end{enumerate*}}

\begingroup%
\def\bl{\cv r}%
\def\fint{\mathtt{fi}}%
\def\fa{\mathtt{fa}}%
 
%

\begin{example}
	 \label{expl:heapdom-instantiations}%
  Consider method @m@ given in \figurename~\ref{lst:m-example-code}.
  \figurename~\ref{table:example-heap-transformers} shows its references $R$, heap-related relations \Rels, references security levels \GenericHLvlVars and variables \GenericRelVars to specify the heap structure using relations \HeapDomRelations. Further, the table in this Figure shows the heap transformers associated with each statement to update the heap relations and reference security levels.
  As an example transformer, consider
\NewRef{\cv r}{\cv B}{\pc}\sHeap
that corresponds to the statement @r = new B;@.
With a heap domain that captures the aliasing relation flow-sensitively, the resulting set of assignments includes (at least) \(［\cAliasRel[\sHeap] b r \assign \ff, \chlvl[\sHeap] r \assign l］\) where \(l\) is a security level expression \st \(l ⊒ \pc\).
\end{example}
\begin{leaveout}
Further assuming that \(\sHeap ∧ \pc = ⊤\) holds on location \(ℓ_0\), then \(\chlvl[\sHeap] r = ⊤ ∧ ¬\cAliasRel[\sHeap]b r\) must hold on location \(ℓ_1\), which notably means that some object reachable via \cv r might be high-sensitive after the object allocation (since 
the allocation happens in high context).

\endgroup%
%
Assuming the same domain as above for \Sm, 
\(X_0\) implies a predicate 
\(\sHeap_0 = (\chlvl[\sHeap] r = ⊥ ∧ ¬\cAliasRel[\sHeap] b r)\), which represents the set of all heaps where every object reachable via \cv r is low-sensitive, and \cv b does not alias \cv r.
Neither \chlvl[\sHeap] a nor \chlvl[\sHeap] b are bound in \(\sHeap_0\), which means that \(\sHeap_0\) represents an abstract heap 
where no fact is known about the security levels of the sets of objects reachable via \cv a or \cv b.
Note, however, that if \cAliasRel[\sHeap]b r were to hold, then we would need to ensure that \(\chlvl[\sHeap] b = \chlvl[\sHeap] r\) also holds.  More generally, we also want transitive (resp. symmetric, etc) relations to be represented as such by the relation variables.  Such \emph{well-formedness} constraints 
are invariants that must be maintained by the transformers
.
\end{leaveout}

\begin{leaveout}
\begin{figure}[t]
  \centering
  \begin{minipage}[t]{.46\linewidth}%
    \raggedleft
    \subfloat[Class definitions and method \protect\inlinemeth{m}.]{%
      \input{fig/m}%
      \label{lst:m-example-code}%
    }%
    \\
    \subfloat[Invariant \(φ_{\cv m}\).]{%
      \smaller
      \begin{minipage}{\linewidth}%
	\centering
	\(φ_{\cv m}(ℓ_3) ≝ \pc ⊔ \cplvl b ⊔ \chlvl[\sHeap] b = ⊥
	\)%
      \end{minipage}%
      \label{subfig:m-example-invariant}%
    }%
  \end{minipage}%
  \hfill%
  \begin{minipage}[t][][b]{.39\linewidth}%
    \subfloat[SCFG \Sm.]{%
      \begin{minipage}{\linewidth}%
	\raggedright%
	\input{fig/m-example-setup}%
	\input{fig/m-example-scfg.tikz}%
      \end{minipage}%
      \label{fig:m-example-scfg}%
    }%
  \end{minipage}%
  \caption{Method that manipulates references, with a representation of the corresponding SCFG and invariant.  Unneeded variables involved in modeling branching behaviors and capturing implicit flows have been omitted from \(X\), \(I\), and \(X_0\) for clarity
    .%
  }%
  \label{fig:m-example}%
\end{figure}
\end{leaveout}

\subsection{Instances of Symbolic Abstract Heap Domains}

\begin{leaveout}
\begin{table}[t]
  \removable{%
  \caption{Example families of heap-related relations
    .}
  \label{tbl:heapdom-relation-parameters}
  \centering
  \begin{tabular}{@{}D|D D D D@{}}
    \hd
    & \hdeep
    & \hshal
    & \hdumb
    & \hsyma
    \\ \hline
    \HeapDomRelations[\hd]
    & \big\{\AliasRelSymbol, \FieldAliasXRelSymbol\big\}
    & \big\{\AliasRelSymbol\big\}
    & ∅
    & \big\{\ConnectRelSymbol\big\}
    \\
    \HeapDomConstRelations[\hd]
    & ∅
    & \big\{\FieldAliasXRelSymbol\big\}
    & \big\{\AliasRelSymbol, \FieldAliasXRelSymbol\big\}
    & ∅
  \end{tabular}}
\end{table}
\end{leaveout}

We present three instances of the 
domain introduced above.
We first define the \emph{transitive} ``\emph{field-aliasing}'' relation, denoted with the symbol \FieldAliasXRelSymbol,
which states for any given pair of references \(r\) and \(s\), \FieldAliasXRel r s holds whenever a reference field of an object reachable via \(r\) is an alias of \(s\).
This relation allows heap domains to capture some useful facts about the structure of the graph of objects when it comes to maintaining object types such as security levels. 

\nbrem[yep]{We also combine aliasing and field-aliasing to form 
the ``\emph{connecting}'' equivalence relation, denoted with the symbol \ConnectRelSymbol.
This relation is such that \ConnectRel r s holds whenever the object pointed to by \(s\) is also reachable via \(r\), or the converse: \ie
\(\ConnectRel r s ≝ \AliasRel r s ∨ \FieldAliasXRel r s ∨ \FieldAliasXRel s r\mbox.\)
This latter relation is inspired by the pointer interference analysis of \citet{Salagnac2007RegionBasedRTJava}, whose goal is to find groups of references that point to the same overall data-structure
.
A domain that captures this relation basically associates the same security level with every object of such data-structures.}%


\begin{table}[b]
\centering
    \begin{threeparttable}
    \caption{Variables and constants involved in representing \sHeap
      when analyzing \inlinemeth{m}, for each domain.
      \sHeap exponents have been omitted for readability.%
    }%
    \smaller
    \label{tbl:m-example-heapdom-vars-n-symbols}%
    \begin{tabular}{@{}c*{3}{|D}@{}}%
      & \hdeep
      & \hshal
      & \hdumb
      \\ \hline
      \GenericHLvlVars
      & \multicolumn 3 D {\{\chlvl a, \chlvl b, \chlvl r\}}
      \\ \hline
      \GenericRelVars
      & \{\cAliasRel b r, \cFieldAliasXRel b a, \cFieldAliasXRel r a\}
      & \{\cAliasRel b r\}
      & ∅
      \\ \hline
      \GenericRelUnsat
      & \multicolumn 3 {D} {\{\cAliasRel a b, \cAliasRel a r, \cFieldAliasXRel a a, \cFieldAliasXRel a b, \cFieldAliasXRel a r, \cFieldAliasXRel b b, \cFieldAliasXRel b r, \cFieldAliasXRel r b, \cFieldAliasXRel r r\}}
      \\ \hline
      \GenericRelTauto
      & \{\cAliasRel a a, \cAliasRel b b, \cAliasRel r r\!\!\}
      & \{\cAliasRel a a, \cAliasRel b b, \cAliasRel r r,\cFieldAliasXRel b a, \cFieldAliasXRel r a\}
      & \{\cAliasRel a a, \cAliasRel b b, \cAliasRel r r,\cAliasRel b r, \cFieldAliasXRel b a, \cFieldAliasXRel r a\}
    \end{tabular}%
  \end{threeparttable}
\end{table}

We now assume a sound pre-analysis function \TAbase{}{} over \(\{\AliasRelSymbol, \FieldAliasXRelSymbol\}\), and use the above relations to define the three families of heap-related relations based on which we shall instantiate our symbolic abstract heap domains:
%
\begin{itemize}[nosep,left=0pt]
\item
  \SymbolicAbstractHeapDom[\hdeep] R, with
\(\hdeep ≝ 〈\{\AliasRelSymbol, \FieldAliasXRelSymbol\}, ∅
〉\), uses symbolic variables to represent over-approximations of
aliasing and field-aliasing relations
;
\item 
  \SymbolicAbstractHeapDom[\hshal] R, with
  \(\hshal ≝ 〈\{\AliasRelSymbol\}, \{\FieldAliasXRelSymbol\}
  〉\), 
  only maintains a flow-sensitive over-approximation of the aliasing relation,
  yet makes use of 
  field-aliasing facts to improve the precision of transformers;
\item
  \SymbolicAbstractHeapDom[\hdumb] R, with \(\hdumb ≝ 〈∅, \{\AliasRelSymbol, \FieldAliasXRelSymbol\}
  〉\),
  does not represent any flow-sensitive heap-related relation,
  yet makes use of flow-insensitive (field-)aliasing
  relations.
%
%
%
\nbrem[yep]{\item The \emph{symmetric deep-aliaing abstract heap domain} is defined solely using the connecting relation with \(\hsyma ≝ 〈\{\ConnectRelSymbol\}, ∅, \TAsyma{}{}〉\), where
  \TAsyma{}{} is straightforwardly defined based on \TAbase{}{} as
  {\footnotesize
    \setlength{\abovedisplayskip}{1pt}%
    \setlength{\belowdisplayskip}{3pt}%
    \begin{equation*}
      \TAsyma R {\ConnectRel r s} ≝
      \begin{cases}
	\TAtauto &\!\!\text{if }⋁_{q∈\mathrm{rl}(r,s)}{\TAbase R {q} = \TAtauto} \\
	\TAunsat &\!\!\text{if } ⋀_{q∈\mathrm{rl}(r,s)}{\TAbase R {q} = \TAunsat} \\
	\TAmaybe \text{~~otherwise}\span
      \end{cases}
    \end{equation*}}
  where \(\mathrm{rl}(r,s)≝\{\AliasRel {r\,} {\,s},\FieldAliasXRel r s, \FieldAliasXRel s r\}\).}%
\end{itemize}

\begin{example}
  \label{expl:heapdom-instantiations}%
  \sloppy%
  Consider method @m@ given in \figurename~\ref{lst:m-example-code},
  and assume a class hierarchy pre-analysis.
  We instantiate the three symbolic abstract heap domains as \SymbolicAbstractHeapDom R and define an abstract heap 
  \(\sHeap ∈ \HeapDom \hd \), for each \(\hd ∈ ｛\hdeep, \hshal, \hdumb
  ｝\).
  \tablename~\ref{tbl:m-example-heapdom-vars-n-symbols} shows the sets of variables used by each one of these domains to represent \sHeap (\GenericHLvlVars and \GenericRelVars), along with the symbols that denote constants 
  involved in 
  capturing relations flow-insensitively (\GenericRelUnsat and \GenericRelTauto).
\end{example}

\begin{leaveout}
\begin{example}[\red{``Instantiating''} \DeepHeapAbstractDom{}]%
  \label{expl:deep-alias-instantiation}%
  \def\fvar{\hvar}%
  Assume a set of reference variables \(R = \{\cv r, \cv s\}\), and that any object pointed to by \(\cv s\) may only contain primitive fields---\ie no field of \cv s may point to any object, thus we can always assume \(\cFieldAliasXRel{\cv s}{\cv r} = \cFieldAliasXRel{\cv s}{\cv s} = \ff\).
  We construct the abstract domain with \red{\(\DeepHeapAbstractDom R = 〈R, \HeapDomRelations, \DeepHeapDom R, \DeepTransformers R〉\)}.
  Let \fvar ∈ \DeepHeapDom R be an abstract heap 
  from this domain. 
  Then, \fvar~\red{is a predicate built using} the variables \chlvl[\fvar] r, \chlvl[\fvar] s, \cAliasRel[\fvar] r s, \cFieldAliasXRel[\fvar] r r, and \cFieldAliasXRel[\fvar] r s.
\end{example}
\end{leaveout}


\begin{figure}[t]
\smaller
  \setlength{\abovedisplayskip}{0pt}%
  \setlength{\belowdisplayskip}{0pt}%
  \def\AliasVars{\GenericRelVars}%
  \def\FieldAliasXVars{\GenericRelVars}%
  \def\ConnectVars{\GenericRelVars}%
  \footnotesize
  \begin{align*}
    \GenericHeapResetRef r ≝
    &~
      \SQMergeu_{d∈R}%
      ［
      \AliasRel d r \assign \ff ,
      \FieldAliasXRel r d \assign \ff,
      \FieldAliasXRel d r \assign  \ff
      ］
    \\
    \GenericHeapCopyRef r s ≝
    &~
      \SQMergeu_{d∈R}%
      ［
      \AliasRel d r \assign  \AliasRel d s ,
      \FieldAliasXRel r  d \assign  \FieldAliasXRel s d ,
      \FieldAliasXRel d r \assign  \FieldAliasXRel d s
      ］
    \\
    \GenericHeapLoadRef r s {f_r} ≝
    &~
      \SQMergeu_{d∈R}%
      ［
      \AliasRel d r \assign  \FieldAliasXRel s d  ,
      \FieldAliasXRel r d \assign  \FieldAliasXRel s d  ,
      \FieldAliasXRel d r \assign , \AliasRel d s ∨ \FieldAliasXRel d s
      ］
    \\
    \GenericHeapStoreRef r {f_r} s ≝
    &~
      \SQMergeu\nolimits_{
      {
      {\FieldAliasXRel a b}} ∈ \FieldAliasXVars}
      ［
      \begin{array}{@{}r@{\,}l@{}}
        \FieldAliasXRel a b \assign
        & \FieldAliasXRel a b ∨ （
	  (\AliasRel a r ∨ \FieldAliasXRel a r) ∧ 
	  (\AliasRel b s ∨ \FieldAliasXRel s b)
	）
      \end{array}］%
    \\
    \GenericUpgradeObjLevel r l ≝
    &~
      \SQMergeu\nolimits_{s∈R}\!［%
      \hlvl s \assign \hlvl s\,⊔\ \ite{\AliasRel s r ∨ \FieldAliasXRel s r}{l}{⊥}
      ］
    \\
    \GenericRestoreObjLevels{\sHeap'} ≝
    &~
      \SQMergeu\nolimits_{
      {(r,s)}∈R^2}\!［%
      \hlvlH[\sHeap]s \assign \hlvlH[\sHeap]s\,⊔\,\ite{\AliasRel[\sHeap'] s r
      ∨ \FieldAliasXRel[\sHeap'] s r}{\hlvlH[\sHeap']r}{⊥}
      \right]
  \end{align*}
  \hrule width 0pt              
 \caption{
   Specialized functions for updating security level and relation variables for each domain defined with \(\hd ∈ ｛\hdeep, \hshal, \hdumb
   ｝\); \hvar exponents have been omitted when a single abstract heap is involved.}
  \label{fig:heapdom-specialized-update-helpers}
\end{figure}
\begin{leaveout}
\begin{figure}[t]
\begin{redenv}
  \setlength{\abovedisplayskip}{0pt}%
  \setlength{\belowdisplayskip}{0pt}%
  \def\AliasVars{\GenericRelVars}%
  \def\FieldShareXVars{\GenericRelVars}%
  \def\ShareVars{\GenericRelVars}%
  \footnotesize
  \def\hd{d}%
  \begin{align*}
    \GenericHeapResetRef[\hshre] r ≝
    &~
      \SQMergeu\nolimits_{d∈R}%
      ［\ShareRel d r \assign \ff\ (\ShareRel d r ∈ \ShareVars)］
    \\
    \GenericUpdateRels[\hshre]{\mathit{op}}\sHeap ≝
    &~
	\SQMergeu\nolimits_{d∈R}%
	［\ShareRel d r \assign \ShareRel d s\ (\ShareRel d r ∈ \ShareVars)］
	\qquad\smash{（\mathit{op}∈｛\begin{centermath}
	    \CopyRefOp r s, \\
	    \LoadRefOp r s {f_r}
	  \end{centermath}｝）}
    \\
    \GenericUpdateRels[\hshre]{\StoreRefOp r {f_r} s}\sHeap ≝
    &~
      \SQMergeu\nolimits_{d∈R}%
      ［\ShareRel d r \assign \ShareRel d r ∨ \ShareRel d s\ (\ShareRel d r ∈ \ShareVars)］
    \\
    \GenericUpgradeObjLevel[\hshre] r l ≝
    &~
      \SQMergeu\nolimits_{s∈R}\!［%
      \hlvl s \assign \hlvl s \,⊔\ \ite{\ShareRel s r}{l}{⊥}
      ］
    \\
    \GenericRestoreObjLevels[\hshre][\sHeap]{\sHeap'} ≝
    &~
      \SQMergeu\nolimits_{
      {(r,s)}∈R^2}\!［%
      \hlvl[\sHeap]s \assign \hlvl[\sHeap]s \,⊔\,\ite{\ShareRel[\sHeap'] s r}{\hlvl[\sHeap']r}{⊥}
      ］
    \\\hline
    \GenericHeapResetRef[\hashr] r ≝
    &~
      \SQMergeu_{d∈R}%
      ［\begin{array}{@{}r@{\ }l@{\ }r@{}}
          \AliasRel d r \assign & \ff & (\AliasRel d r ∈ \AliasVars) \\
          \FieldShareXRel r d \assign & \ff & (\FieldShareXRel r d ∈ \FieldShareXVars) \\
          \FieldShareXRel d r \assign & \ff & (\FieldShareXRel d r ∈ \FieldShareXVars) \\
        \end{array}］
    \\
    \GenericHeapCopyRef[\hashr] r s ≝
    &~
      \SQMergeu_{d∈R}%
      ［\begin{array}{@{}r@{\ }l@{\ }r@{}}
          \AliasRel d r \assign & \AliasRel d s & (\AliasRel d r ∈ \AliasVars) \\
          \FieldShareXRel r  d \assign & \FieldShareXRel s d & (\FieldShareXRel r d ∈ \FieldShareXVars) \\
          \FieldShareXRel d r \assign & \FieldShareXRel d s & (\FieldShareXRel d r ∈ \FieldShareXVars) \\
        \end{array}］
    \\
    \GenericHeapLoadRef[\hashr] r s {f_r} ≝
    &~
      \SQMergeu_{d∈R}%
      ［
      \begin{array}{@{}r@{\ }l@{\ }r@{}}
        \AliasRel d r \assign & \FieldShareXRel s d  & (\AliasRel r d ∈ \AliasVars) \\
        \FieldShareXRel r d \assign & \FieldShareXRel s d  & (\FieldShareXRel r d  ∈ \FieldShareXVars) \\
        \FieldShareXRel d r \assign & \AliasRel d s ∨ \FieldShareXRel d s & (\FieldShareXRel d r ∈ \FieldShareXVars) \\
      \end{array}］
    \\
    \GenericHeapStoreRef[\hashr] r {f_r} s ≝
    &~
      \SQMergeu\nolimits_{
      {
      {\FieldShareXRel a b}} ∈ \FieldShareXVars}
      ［
      \begin{array}{@{}r@{\,}l@{}}
        \FieldShareXRel a b \assign
        & \FieldShareXRel a b ∨ （
	  \begin{array}{@{}l@{}}
	    (\AliasRel a r ∨ \FieldShareXRel a r) ∧ \\
	    (\AliasRel b s ∨ \FieldShareXRel s b ∨ \FieldShareXRel b s)
	  \end{array}）
      \end{array}］%
    \\
    \GenericUpgradeObjLevel[\hashr] r l ≝
    &~
      \SQMergeu\nolimits_{s∈R}\!［%
      \hlvl s \assign \hlvl s\,⊔\ \ite{\AliasRel s r ∨ \FieldShareXRel s r}{l}{⊥}
      ］
    \\
    \GenericRestoreObjLevels[\hashr]{\sHeap'} ≝
    &~
      \SQMergeu\nolimits_{
      {(r,s)}∈R^2}\!［%
	  \hlvl[\sHeap]s \assign \hlvl[\sHeap]s\,⊔\,\ite{\AliasRel[\sHeap'] s r
	  ∨ \FieldShareXRel[\sHeap'] s r}{\hlvl[\sHeap']r}{⊥}
	  ］
 \end{align*}
 \caption{
   Specialized functions for updating security level and relation variables for each domain defined with \(\hd ∈ ｛\hshre, \hashr｝\); \hvar exponents have been omitted when a single abstract heap is involved.}
\end{redenv}
  \label{fig:heapdom-specialized-update-helpers-sharing}
\end{figure}
\end{leaveout}
Regarding transformers, we give in \figurename~\ref{fig:heapdom-specialized-update-helpers} the specialized functions used by their definitions in \figurename~\ref{fig::Symmaries.Abstract.Heap.Operations}
.
\begingroup%
The domains defined with \(\hd ∈ ｛\hdeep, \hshal, \hdumb｝\) share these definitions.
\GenericHeapResetRef r updates relation variables in \sHeap to reflect the erasing of a given reference \(r\) to either \Nil or a fresh reference by clearing variables from \GenericRelVars[\sHeap]
.
In turn, \GenericHeapCopyRef r s updates the variables in \GenericRelVars to encode the copy of a reference \(s\) to \(r\).
%
\(\GenericHeapLoadRef r s {f_r}\) makes any reference \(d\) that may be an alias of one of \(s\)'s field a potential alias of \(r\) (when a corresponding variable \AliasRel[\sHeap] d r belongs to \GenericRelVars[\sHeap]), and updates any variable that 
represents the \FieldAliasXRelSymbol relation to reflect that \(s\) becomes a field-alias of \(r\).
\(\GenericHeapStoreRef r {f_r} s\) makes \(s\) a field-alias of \(r\) while maintaining transitivity of \FieldAliasXRelSymbol.
Storing a reference may only add elements in relation \FieldAliasXRelSymbol, hence the disjunction in every assignment defined by this operation.
Observe that, as can be seen in the definition of \GenericHeapLoadRef r s {f_r} for \(\hd = \hshal\), 
instead of simply setting \(\AliasRel d r \assign \tt\) for every potential alias \(d\) of \(r\) (\ie blindly assuming that no information is known about the potential aliasing relation), a constant \FieldAliasXRel s d \(∈ \GenericRelTauto ∪ \GenericRelUnsat\) is used instead to further restrict the new potential aliases to the cases that have not been ruled out by 
the pre-analysis.
The \hdumb domain involves some pre-established facts via a similar mechanism.
%
\(\GenericUpgradeObjLevel r l\) takes a reference \(r\) and a security level expression \(l\), and upgrades the security level associated with the objects reachable via \(r\) as well as that of every reference \(s\) that may transitively field-alias \(r\) (\ie \(\FieldAliasXRel s r\)).
\(\GenericRestoreObjLevels{\sHeap'}\) 
upgrades the typing environment of \sHeap according to that of \(\sHeap'\).
\endgroup
\nbrem[yep]{The functions specialized for the \hsyma family of heap-related relations operate similarly.}

\begin{leaveout}
\begin{example}[Continuing Example~\ref{expl:deep-alias-instantiation}]
  \def\fvar{\hvar}%
  The set of variables given in the previous example are used by the operations in \tablename~\ref{tab:symbolic-heaps} to encode queries and predicates on \hvar.
  We can for instance constrain \hvar so that it represents the set of all concrete heaps where \cv r is known to be \Null using \(\HeapInit[\fvar]{\{\cv r\}} = (\chlvl[\fvar] r = ⊥ ∧ ¬\cAliasRel[\fvar] r s ∧ ¬\cFieldAliasXRel[\fvar] r r ∧ ¬\cFieldAliasXRel[\fvar] r s)\mbox,%
  \) or build the abstract transformer \(\NoHeapEffect{\cv r ← \cv s}{\fvar} = [\chlvl[\fvar]r\assign\chlvl[\fvar]s, \cAliasRel[\fvar]r s \assign \tt, \cFieldAliasXRel[\fvar] r r \assign \ff, \cFieldAliasXRel[\fvar] r s \assign \ff]\) that encodes the semantics of statement @r = s@.
\end{example}
\end{leaveout}

\section{Secure Heap Abstraction}\label{sec:secure.heap.abstraction.def}

To specify the semantics of heap operations performed by a program, we define a \emph{concrete   heap domain} that maintains the value of \emph{primitive fields} in addition to \emph{the precise heap-related relations}.
A concrete heap domain is defined similarly to that of abstract heap domains introduced in Section~\ref{sec::heap.domains}, \emph{with the difference that the heap maintains the primitive fields instead of security levels}.
The concrete heap domain is defined as {$\ConcreteHeapDom {} = \langle\HeapDom \concrete , \Transformers \concrete  \rangle$}
\nakh{where \(\concrete ≝ 〈\{\AliasRelSymbol, \FieldAliasRelSymbol\}, ∅ 〉\)},
the relation \AliasRelSymbol is an ordinary reference aliasing relation,
 and \FieldAliasRelSymbol is a field-aliasing relation, \ie \(\FieldAliasRelConcrete r s f\) holds iff
the field $f$ of the object referenced by \(r\) {is an alias of
	\(s\)}. 
\figurename~\ref{fig::concrete.alias-variable-update-helpers}
presents the operations on the concrete heap domain.
The notation \ConcreteHeapTransformer {\mathit{op}} \mHeap shows the predicate transformer that corresponds to the operation $\mathit{op}$ on a concrete heap $\mHeap \in \HeapDom \concrete$. 
The functions \ConcreteHeapOp{op}{\mHeap} and \IUpdatePrimFields{\mHeap}{op} respectively specify the updates to the \red{concrete} heap-related relations and the primitive fields of the heap  \mHeap by performing the heap operation $\mathit{op}$.
\RefFields, \DefVal and \UndefVal  show  the set of reference fields, the default value and the undefined value, respectively.
\begin{figure}[t!]
  \setlength{\abovedisplayskip}{0pt}%
  \setlength{\belowdisplayskip}{0pt}%
		\footnotesize%
		\begin{align*}
		\ConcreteHeapTransformer {op} \mHeap
		&≝\left\{\!%
		\begin{array}{@{}l@{\,}l@{\,}l@{~\text{if~}}l}
		\IStoreRef r {f_r} s {\mHeap} & & &  \mathit{op}= \s{r.f_r = s} \\
		  \ILoadRef{r} s {f_r}{\mHeap}  &&
		  & 		\mathit{op}= \s{r = s.f_r}\ \\
		\ICopyRef {r}{s}{\mHeap}  &\SQMergei u &
		\IUpdatePrimFields {\mHeap} {op} %
		& \mathit{op}= \s{r = s}  \\
		\IResetRef {r = \_}{\mHeap} & \SQMergei u &
		\IUpdatePrimFields {\mHeap} {op}
		 &\mathit{op}=\s{r = \cnew} \\
		\IResetRef {r = \_} {\mHeap} &	\SQMergei  u &
		\IUpdatePrimFields {\mHeap} {op}
		& \mathit{op}= \s{r = \Null}  \\
		&& [v:= \RefContent {s}{f_p}  ]
		& 		\mathit{op}= \s{v =  {s.f_p}} \\
		&&		\IUpdatePrimFields {\mHeap} {op}
		&  \mathit{op}= \s{r.f_p = v} \\
		\end{array}
		  \right.
		\end{align*}
		{where}
		\begin{align*}
		\IResetRef{r = \_}{\mHeap} &≝
		\SQMergeu_{\mathllap{ r'∈R, f \in} \RefFields}%
		［\begin{array}{@{}r@{\ }l@{\ }r@{}}
		\AliasRel r {r'} \assign  \ff ,
		\FieldAliasRelConcrete r {r'} f \assign  \ff ,
		\FieldAliasRelConcrete {r'} r f \assign  \ff
		\end{array}］
		\\
		\ICopyRef{r}{s}{\mHeap} &≝
		\SQMergeu_{\mathllap{r'∈R, f \in} \RefFields}%
		［\begin{array}{@{}r@{\ }l@{\ }r@{}}
		\AliasRel   r {r'} \assign \AliasRel s  {r'} ,
		\FieldAliasRelConcrete r {r'} f  \assign  \FieldAliasRelConcrete s {r'} f  ,
		\FieldAliasRelConcrete {r'}  r  f \assign  \FieldAliasRelConcrete {r'}  s f
		\end{array}］
		\\
		\ILoadRef{r} s {f_r}{\mHeap}  &≝
		\SQMergeu_{\mathllap{r'∈R, f' \in} \RefFields}%
		［\begin{array}{@{}r@{\ }l@{\ }r@{}}
 		\AliasRel r {r'}\assign  \FieldAliasRelConcrete  s {r'} {f_r} ,
 		    \FieldAliasRelConcrete r {r'} {f'} \assign \FieldAliasRelConcrete s {s'} {f_r} \wedge \FieldAliasRelConcrete {s'} {r'} {f'} ,
		\FieldAliasRelConcrete {r'} r {f} \assign  \AliasRel {r'} s
		\end{array}］%
		\\
		\IStoreRef r {f_r} s {\mHeap} &≝
		\SQMergeu_{
			\mathllap{\FieldAliasRelConcrete {r'} {s'} {f_r} ∈} \GenericRelVars}%
		［\begin{array}{@{}r@{\ }l@{}}
		\FieldAliasRelConcrete {r'} {s'} {f_r} \assign & \AliasRel {r'} r \wedge \AliasRel {s'} s 
		\end{array}］%
		\end{align*}
		and
		\begin{align*}
		\IUpdatePrimFields {\mHeap} {op} &≝
		\SQMergeu_{\mathllap{f_p \in} \PrimFields {s}}%
		\left\{\!\!%
		\begin{array}{@{}r@{\ }l@{\ }l@{}}
		& [\RefContent {s}{f_p} :=\RefContent {s}{f} ]   &\quad \text{if} \quad \mathit{op}= \s{r = s} \\
		  & [\RefContent {s}{f_p} :=\DefVal ]   &\quad \text{if} \quad\mathit{op}= \s{r = \cnew} \\
			 & [\RefContent {r'}{f_p} :=v ] &\quad \text{if}  \quad\mathit{op}= \s{r.f_p = v}, \mHeap \models \AliasRel r {r'} \\
			 & [\RefContent {s}{f_p} :=\UndefVal ]   &\quad \text{if} \quad\mathit{op}= \s{r = \Null}
		\end{array}%
		 \right.%
		\end{align*}%
	\caption{
	  Update of alias variables in a heap \mHeap from the concrete heap domain.}
	\label{fig::concrete.alias-variable-update-helpers}
\end{figure}


Since we use an abstract heap domain to model and analyze information flow via heap, we should ensure that the analysis under abstract heap domains guarantees noninterference.
\red{To this end, we first define the notion of \emph{indistinguishable heaps} and then prove that two indistinguishable heaps from the concrete domain remain indistinguishable after applying a heap transformer.}
Let \red{$\GeneralH$}  be a heap from an arbitrary heap domain and \(R' \subseteq R\) be a set of references.
The reference graph over \(R'\) induced by \GeneralH is a labeled digraph \(\RefGraph{\GeneralH}{R'} ≝ \tuple{\Nodes \GeneralH, \Edges\GeneralH}\) where $\Nodes\GeneralH = R'$ is the set of nodes, and the edges \Edges\GeneralH show the heap-related relations between them, \ie
 \(\Edges\GeneralH =\{ (r,\GenericRelSymb,r') ~|~ \GeneralH \models r \GenericRelSymb r', r \in R' \vee r' \in R'\} \).
Let \sHeap be an abstract heap and \RefGraph{\GeneralH}{{\low {}}} be a sub-graph of \RefGraph{\GeneralH}{R} containing the low-sensitive references \(\LowRefs \sHeap= \{r ~|~\sHeap \models  (\red{\hlvl[\mHeap]r}=\low)\}\).
We say two concrete heaps are indistinguishable, if heap-related relations and  primitive fields of their low-sensitive references are identical, \ie (i) the reference graphs corresponding to their low-sensitive portions of the heaps are \emph{isomorphic}, and (ii) the valuation of primitive fields of their low-sensitive references are~identical.


\begin{definition}[{
    Indistinguishable Heaps}]\label{def::heap-low-equivalence-relation}
	We say 
	two concrete heaps 
	\mHeap and $\mHeap'$ from \HeapDom \concrete,
	are indistinguishable \wrt an abstract heap \sHeap, noted by \({\mHeap =_{\sHeap} \mHeap'}\),
	iff
	(i)  \RefGraph{\mHeap}{\low{}} and \RefGraph{\mHeap'}{\low{}} are isomorphic, denoted by \(\RefGraph{\mHeap}{\low{}} \cong \RefGraph{\mHeap'}{\low{}}\), and
	(ii)
	\(
 \forall x.~ \mHeap \models	\RefContent{r}{f_p}= x \Leftrightarrow  \mHeap' \models	\RefContent{r} {f_p}=x \), for all $r \in \LowRefs{\sHeap}$ where $f_p$ is a primitive field.
\end{definition}

We define the concept of \emph{secure heap abstraction}, which states that two indistinguishable heaps should remain indistinguishable after applying a heap operation and its corresponding transformer at the abstract heap domain level:

\begin{definition}[Secure Heap Abstraction]\label{def::secure.heap.abstraction}
\sloppy  The concrete heap domain \ConcreteHeapDom {} is \emph{secure} \wrt an abstract heap domain \SymbolicAbstractHeapDom R, if and only if it preserves the heap indistinguishability relation, \ie given any concrete heaps 
  \((\mHeap_1, \mHeap_2) ∈ \HeapDom \concrete \times \HeapDom \concrete \)%
  , and an abstract heap 
  \(\sHeap ∈ \HeapDom \hd \) \st \({\mHeap_1 =_{\sHeap} \mHeap_2}\), it holds that:
  \begin{enumerate}[nosep,leftmargin=1.8em]
  \item[(a)] for all pair \((\mathit{as}, \mathit{as}'
    ) \) of reference assignment statements and their corresponding operations on abstract heaps where %
    \(\mathit{as} \in \{\s{r = s}, \s{r = s.f_r}, \s{r = \Null} \} \), \({\mHeap'_1 =_{\sHeap'} \mHeap'_2}\) holds where \( \mHeap'_i= {\ConcreteHeapTransformer {\mathit{as}}{ \mHeap_i}}\), $i \in \{1,2\}$, and \(\sHeap' = \NoHeapEffect{\mathit{as}'}{\sHeap}\);
  \item[(m)] for all pair \((\mathit{mu}, \mathit{mu}' ) \) of mutation statements and their corresponding operations on heaps where %
    \(\mathit{mu} \in \{%
    \s{r.f_p = e}, \s{r.f_r = s}, \s{r = \New\,c}\}\), {for all \(l ∈ \LL\)} where \({\hlvl[ {\sHeap}]s } \sqsubseteq l\) if \(\mathit{mu} \) is \(\s{r.f_r = s}\), and \({\plvl e} \sqsubseteq l\) if \(\mathit{mu}\) is \(\s{r.f_p = e}\), it holds that \({\mHeap'_1 =_{{\sHeap'}} \mHeap'_2}\) where \( \mHeap'_i= {\ConcreteHeapTransformer {\mathit{mu}}{ \mHeap_i}}\), $i \in \{1,2\}$, and \(\sHeap' = \OneHeapEffect{\mathit{mu}'}{l}{\sHeap}\);
  \end{enumerate}
\end{definition}


\begin{theorem}\label{thm::secure.heap.model}
	The concrete heap domain \ConcreteHeapDom R  is secure \wrt
	the deep abstract heap domain \DeepHeapAbstractDom {R} according to Definition~\ref{def::secure.heap.abstraction}.
\end{theorem}
\begin{proof}
	See Appendix~\ref{sec::secure.heap.asbtraction-proof}.
\end{proof}

\begin{leaveout}
\begin{theorem}\label{thm::secure.heap.model1}
\red{On the relations of two heap domains: a secure heap domain w.r.t. the concrete one will be }
\end{theorem}

%
\nb{Partial order below is only used for computing summaries. I also don't think initialisation operator is strictly required; but that latter one may still be helpful for comprehension.}
\red{Further, \(\hvar ⊑_{\mathsf H} \hvar'\) denotes a partial order that holds whenever the abstract heap \hvar, restricted to the set of references \(R' ⊆ R\) pertaining to \(\hvar'\), over-approximates a set of heap configurations, \red{that is greater of equal than the set of heap configurations represented by \(\hvar'\)}: this amounts to a pairwise comparison of security levels, \ie one must have \(\hvar ⊑_{\mathsf H} \hvar' ⇒ (∀r∈R', \hlvl[\hvar]r ⊑ \hlvl[\hvar']r)\) (and similarly for any over-approximated fact about the heap, like a may-alias relation).}
%

\end{leaveout}


\section{Inferring Polymorphic Information-flow Guards}
\label{sec:secur-guard-synth}



Let us now put the heap abstraction aside, and focus on our approach for computing guards and capturing implicit flows.
We start by considering the method @f@ given in \figurename~\ref{lst:f-example-code}, that does not involve any reference variable.
@f@ implements a canonical pattern used to illustrate implicit flows: if its argument \cv v is high-sensitive, then the information output on line~\ref{f-line:output-l} is also high-sensitive via an implicit flow induced by the assignment guarded by the condition on line~\ref{f-line:v-cond}.
For this example, our requirement demands that executing the sink statement on line~\ref{f-line:output-l} does not leak confidential information.
Therefore, the guard that we want to compute for @f@ is a sufficient condition that allow us to decide whether any call to the method satisfies the confidentiality requirement based on a set of program facts available \emph{whatever the calling context}.
For @f@, the latter set of facts includes, for instance, the security level of the effective argument for \cv v, or whether the call happens in a high context (\ie if it is guarded by a condition on high-sensitive information).
\newcommand\Sf{\ensuremath{S_{\mathtt f}}\xspace}%
\begin{figure}[t]
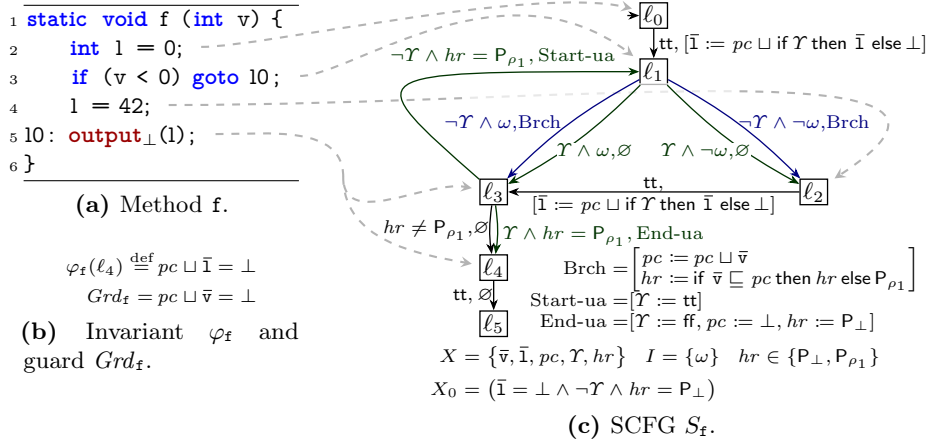

  \centering
  \begin{minipage}[t]{.3\linewidth}%
    \subfloat[Method \protect\inlinemeth{f}.]{%
      \input{fig/f}%
      \label{lst:f-example-code}%
    }%
    \\
    \subfloat[Invariant \(φ_{\cv f}\) and guard \SumGuard{\cv f}.]{%
      \begin{minipage}{\linewidth}
      \centering\smaller
      \(\begin{aligned}
	φ_{\cv f}(ℓ_4) ≝\,&\pc ⊔ \cplvl l = ⊥
	\\
	\SumGuard{\cv f} =\,&\pc ⊔ \cplvl v = ⊥
      \end{aligned}\)%
      \label{subfig:f-example-invariant-n-guard}%
    \end{minipage}%
  }%
  \end{minipage}%
  \hfill%
  \begin{minipage}[t]{.65\linewidth}%
    \raggedleft
  \subfloat[SCFG \Sf.]{%
    \begin{minipage}{\linewidth}%
      \centering
      \input{fig/f-example-setup}%
\input{fig/scfg-gen-setup}%
\begin{tikzpicture}[scale=.8, > = {Stealth[scale=.8]}, remember picture]
  \tikzstyle{label} = [font=\scriptsize, inner sep = 0pt, outer sep = 0pt]%
  \foreach \i/\x/\y in {1/2/0, 2/4.66/-2, 3/-.66/-2} {
    \node [location] at (\x, \y) (l\i) {\(ℓ_{\i}\)};
  }
  \node [location, above = 1.2em of l1] (l0) {\(ℓ_0\)};
  \node [location, below = 2em of l3] (l4) {\(ℓ_4\)};
  \node [location, below = 1.2em of l4] (l5) {\(ℓ_5\)};

  \begin{scope}[overlay, every path/.append style = {->, overlay-arrow, shorten > = 1mm}]

    \draw (f-l-init) .. controls +(1.2cm, 0) and ([xshift = -2cm, yshift = .5cm] l0) .. (l0.160);
    \draw (f-v-cond) .. controls +(1cm, 0) and ([xshift = -2.5cm, yshift = 1.9cm] l1) .. (l1);
    \coordinate [xshift = 3cm, yshift= .8cm] (x) at (f-l-sensitive-assign -| l2);
    \draw (f-l-sensitive-assign) .. controls +(7cm, 0) and (x) .. (l2);
    \fill [opacity = .7, white] ([xshift = 5cm, yshift = 2mm] f-l-sensitive-assign.north east) rectangle ([yshift = 1cm, xshift = -1cm] l2.north west);
    \coordinate [xshift = -2cm, yshift = 2mm] (x) at (l3);
    \draw (f-output-l)
    to [out = 0, in = 90] (x)
    to [out = -90, in = 170] (l3);
    \draw (x) to [out = -90, in = 170] (l4);
  \end{scope}

  \draw [basic-arrow] (l0.west)+(-2mm,0) to (l0);

  \draw (l0) edge [basic-arrow]
  node [right, label, xshift = 2pt, fill = white] (l01) {
    {\(\tt, [\cplvl l \assign \pc ⊔ \ite\uVar{\cplvl l}{⊥}]\)}%
  } (l1);

  \draw (l1) edge [nominal-arrow, bend left = 8]
  node [right, label] (l12-nominal) {
  \(¬\uVar∧¬ω\),\adjustedbg\TUAv } (l2);
  \draw (l1) edge [analysis-arrow, bend right = 8]
  node [left, label, pos = .63,xshift=2pt] (l12-analysis) {
  \adjustedbg{\(\uVar∧¬ω\),∅}} (l2);

  \draw (l1) edge [nominal-arrow, bend right = 8]
  node [left, label, xshift = -3pt] (l13-nominal) {
  \(¬\uVar∧ω\),\adjustedbg\TUAv} (l3);
  \draw (l1) edge [analysis-arrow, bend left = 8]
  node [right, label, pos = .63, xshift=-2pt] (l13-analysis) {
  \adjustedbg{\(\uVar∧ω\),∅}} (l3);

  \draw (l2) edge [basic-arrow]
  node [above, label] (l23) {\(\tt,\)}
  node [below, label, yshift=-1pt] (l23') {
    \([\cplvl l \assign \pc ⊔ \ite\uVar{\cplvl l}{⊥}]\)} (l3);

  \draw [analysis-arrow] (l3) .. controls ([xshift = -7em]l1 -| l3) ..
  node [above, label, pos = .8, yshift=1pt] (l3-startan) {
    \adjustedbg[1pt]{\(¬\uVar∧\rVar=\rMode{ρ_1},\Startua {}
    \)}%
  }
  (l1);

  \draw (l3) edge [analysis-arrow, bend left = 8]
  node [right, label, pos = .64, xshift = 2pt] (l3-endua) {%
    \(\uVar∧\rVar=\rMode{ρ_1},
    \Endua{} 
    \)%
  } (l4);
  \draw (l3) edge [basic-arrow, bend right = 8]
  node [left, label] (l34-nominal) {%
    \(\rVar≠\rMode{ρ_1}\),∅%
  } (l4);
  \draw (l4) edge [basic-arrow]
  node [left, label] (l45) {%
    \(\tt, ∅\)%
  } (l5);


  \node [font = \scriptsize, yshift=1em, right = .6em of l4, anchor = north west, align = left] (tuav) {%
    \(
    \begin{array}{r@{}l@{}}
    \TUAv =& ［
    \begin{array}{@{}r@{\,}l@{}}
      \pc \assign & \pc ⊔ \cplvl v\\
      \rVar \assign & \ite{\cplvl v ⊑ \pc}\rVar{\rMode{ρ_1}}
    \end{array}
    ］\\
    \Startua{} =& [\uVar\assign\tt]  \\
    \Endua{} =&［\uVar \assign \ff, \pc \assign ⊥, \rVar \assign \rNom］\\
    \end{array}%
  \)
  };

  \node [fit={(l0) (l5) (l01) (l34-nominal)}, inner sep = 0pt] (scfg-box) {};
  \node [below = 3pt of scfg-box.south, anchor = north
  , inner sep = 0pt, outer sep = 0pt,
  ] (init-states) {%
    \smaller
    \(
    \begin{aligned}
      X &= ｛\cplvl v, \cplvl l, \pc, \uVar, \rVar｝\ \ I = ｛ω｝\ \ \rVar \in \{\rNom, \rMode{ρ_1}\} \\
      X_0 &= （\cplvl l = ⊥ ∧ ¬\uVar ∧ \rVar = \rNom）
    \end{aligned}\)%
  };

\end{tikzpicture}%
\hrule width 0pt              
 %
      \label{fig:f-example-scfg}%
    \end{minipage}%
  }%
  \end{minipage}%
  \vspace*{-1ex}%
  \caption{Example method with SCFG, invariant, and resulting guard.}%
  \label{fig:f-example}%
\end{figure}%
\nberth{To achieve this, we build the SCFG 
that specifies the security semantics of any method in such a way that the set of all its potential initial states encodes all the possible calling contexts for the method.}
For instance, when encoding @f@ the 
variable \cplvl v assigned to the argument \cv v is left uninitialized (contrary to the level \cplvl l of local variable \cv l)
.
Another state variable 
that is 
left uninitialized is \pc, as it denotes the security~level~of~the~calling~context.
\nbrem[yep]{Next, the safety property φ expresses constraints on security levels at 
states of the transition system that typically correspond to sink statements in the method.
The security constraints require that high-sensitive data should not flow to sink statements.
Eventually, a co-reachability analysis finds the set of \emph{all initial states} 
from which no run ever leads to a violation of 
φ; a symbolic description of this set gives us the guard for the method in terms of the 
variables pertaining to the calling context, like \cplvl v and \pc for @f@.}%
%
\nberth{We
then associate a safety property φ that expresses constraints on security levels at states of the SCFG,
and use a co-reachability analysis to find the set of \emph{all initial states} 
from which no run ever leads to a violation of 
φ.
}\rmrk{Though it's still useful to recall what the invariant is; could be removed.}
We graphically represent in \figurename~\ref{fig:f-example-scfg} the SCFG \Sf 
obtained for @f@.
The associated invariant is given in \figurename~\ref{subfig:f-example-invariant-n-guard}, along with the inferred guard .

\subsection{Security Semantics}

\begin{leaveout}
  We use the method @m@ and declarations of classes @A@ and @B@ given in \figurename~\ref{lst:m-example-code} to illustrate our approach.
  \cplvl a, \cplvl b, \cplvl i and \cplvl r are state variables that respectively model the security levels of \emph{primitive} or \emph{reference variables} \cv a, \cv b, \cv i, and \cv r, and take their values in the security domain \LL.
  These level variables \emph{over-approximate} the true security levels: \ie that \(\cplvl i = ⊤\) indicates that the information held in \cv i \emph{may} be of highest confidentiality level; conversely, \(\cplvl i = ⊥\) indicates that \cv i \emph{must} hold low-sensitive data.
  Note that \cplvl a, \cplvl b, and \cplvl r associate levels with the variables themselves, not with the objects they might reach in the heap.
  In turn, \pc models the security level of the calling context: \ie \(\pc = ⊥\) indicates that the call to \cv m happens in a \emph{low context}, which means that the call does not depend on a high-sensitive condition.
  If on the other hand, the call happens in a high context (\ie \(\pc = ⊤\)), then any mutation of object or execution of a sink statement may leak high-sensitive information.
  On the transition from the location \(ℓ_0\) to \(ℓ_1\), \cplvl r is assigned to the context level: this reflects the implicit flow of information from the execution context to \cv r.

  The remaining elements of \Sm are dedicated to a heap abstraction
  predicate
  \sHeap that belongs to a \emph{symbolic abstract heap domain}  dedicated to the representation of abstract heaps pertaining to a given set of reference variables \(R\); for @m@ we have \(R = ｛\cv a, \cv b, \cv r｝\).
  The heap abstraction \sHeap is a predicate defined on two sets of state variables
  \GenericHLvlVars and \GenericRelVars. 
  \GenericHLvlVars associates a \emph{security level variable} \hlvl[\sHeap] r with each reference \(r ∈ R\), that represents an \emph{upper bound} on the security levels of any object that is reachable via \(r\).
  These variables model the security typing environments for the symbolic heaps modeled by \sHeap.
  In \Sm, we thus have the security level variables \chlvl[\sHeap] a, \chlvl[\sHeap] b, and \chlvl[\sHeap] r in addition to 
  \cplvl a, \cplvl b, and \cplvl r that hold security levels for the reference variables themselves (instead of for the sets of objects they may reach).  Variables from this set are also involved in expressing invariants on sink statements, as shown in \figurename~\ref{subfig:m-example-invariant}.
  In turn, the set \GenericRelVars consists of Boolean variables that describe \emph{over-approximations of heap-related relations} between the references in \(R\).
  A typical example of a heap-related relation is the aliasing relation, that we denote with the symbol \AliasRelSymbol, and which is defined as an equivalence relation where \(\AliasRel r s\) holds iff \(r\) and \(s\) point to the same object.
  For @m@, a symbolic abstract heap domain that over-approximates the aliasing relation shall make use of a Boolean variable \cAliasRel[\sHeap] b r to model whether \cv b may be an alias of \cv r
  .
  (We give more examples of heap-related relations below.)

  The notations ``\NoHeapEffect{⋅}\sHeap'' on transition labels are predicate transformers for 
  \sHeap
  ; they denote sets of assignments to variables in \(\GenericHLvlVars ∪ \GenericRelVars\).
  Similarly, ``\HeapInit{⋅}'' in the definition of \(X_0\) is a predicate on these variables.
  \green{%
    Observe that \Sm involves a single symbolic abstract heap \sHeap.
    Yet, such heaps are represented by means of predicates in a propositional logic.
    Therefore, any number of abstract heaps may be used in the definition of an SCFG without compromising the tractability of a (co-)reachability analysis---this, however, has an obvious impact on scalability.
    In particular, our approach for analyzing implicit flows via the heap relies on the ability of the symbolic model to simultaneously represent heap abstractions for taken and non-taken program branches; we detail this aspect in Section~\ref{sec:secur-guard-synth}.
  }
\end{leaveout}

\begin{forappendix}
  We note \(\junc m {ρ}\) the junction of a CDR ρ, and this junction always exists \red{as we assume that return statements lead to the empty sequence of statements \(surd\)}.
  \(\inducing m {ρ}\) is the branching statement that \emph{induces} ρ, and \region m \stm is the CDR that is induced by \stm.
  We further note \Regions m the set of all CDRs of \(m\); \red{we shall omit the index method \(m\) in CDR-related notations to ease readability}.
\end{forappendix}
Our encoding of security semantics captures implicit flows (\ie flows induced via the program control-flow structure) by constructing SCFGs that feature two \emph{execution modes}, encoded with the help of a state variable \uVar: (i) in \emph{nominal} mode (\(\uVar = \ff\)), updates to security levels reflect explicit information flows, and (ii) in \emph{upgrade analysis} mode (\(\uVar = \tt\)), the information flow from the \emph{high} execution context \pc to every variable updated in every possible execution path within the \emph{Control Dependence Regions} (CDRs) of the current context are captured.
A CDR ρ is a non-empty set of CFG (Control-Flow Graph) nodes that gathers every instruction that is control-dependent on a given branching statement.
This use of CDRs is inspired by previous works~\citep{DenningDenning1977CertificationofProgs4SecureIF,Barthe:2007:CLN:1762174.1762189,LiuMilanova2010StaticIFC4JavaWithImplicitFlows,Lortz2014Cassandra}\footnote{The classical algorithm of \citet{Ball:1993:CDRs} for computing CDRs works by identifying as a \emph{junction} each dominating node in the post-dominator tree of the CFG. 
Such a junction $j$ is reached by every execution path that starts from any node in the set ρ of nodes that $j$ post-dominates.
Further, one can always find a \emph{unique} branching node that precedes nodes in ρ and belongs to every path from the source of the CFG to any node in ρ, and ρ is therefore a CDR.}.
  The code in \figurename~\ref{lst:f-example-code} features a single conditional branching statement on line~\ref{f-line:v-cond}, which induces the CDR \(ρ_1\).
  The junction of \(ρ_1\) is the statement @output$_⊥$(l)@
  .
  Two execution branches are possible within \(ρ_1\): one branch executes no statement, whereas the other performs the assignment @l = 42@ on line~\ref{f-line:l-assign}.
  This means that the execution of the latter is dependent on the~condition~on~line~\ref{f-line:v-cond}.
  In the SCFG \Sf, the upgrade analysis of \(ρ_1\) 
  starts whenever the model reaches location \(ℓ_3\), which represents the \emph{junction} of \(ρ_1\), if the branching statement that induces \(ρ_1\) (encoded by \(ℓ_1\)) was subject to a high condition.
We use a state variable \rVar to record the CDR currently subject to a high-condition.
We make use of the input variable ω to
abstract away the actual branch condition in nominal mode (since our security semantics abstracts away the values of program variables).
This is for instance the case on location \(ℓ_1\) in \Sf when \(\uVar = \ff\).
The variable ω is also used to model upgrade analyses for \red{multiple possible program paths which can be taken non-deterministically}.
In \Sf, this is the case on location \(ℓ_1\) as well, when \(\uVar = \tt\).

\begin{figure}[!t]
  \centering
  \smaller
  \begin{freeruleset}
    \myfreerule{\GotoRule}{%
      \begin{array}{@{}c@{}}
	\NonJuncL \\
	ℓ = \semloc{\s{\cgoto~l};\stms, \_}%
      \end{array}
    }{%
      ℓ 
      \trans{\tt, ∅}%
      \newloc{\Target{} l}%
    }
    \qquad
    \myfreerule{\OutputRule}{%
      \begin{array}{@{}c@{}}
	\NonJuncL
	\quad
	ℓ = \semloc{\s{\coutput l (x)};\stms, \_}
	\\
      φ(ℓ) =
      \left(\plvl v\text{~if~}x = v\text, \hlvl[\hvar] r\text{~if~}x = r\right) ⊑ l
      ∧ \pc ⊑ l
      \end{array}
    }{%
      ℓ 
      \trans{\tt, ∅}\newloc{\stms}
    }%
  \end{freeruleset}
  \\[\myrulespace]
  \begin{ruleset}
    \myrule{\AssignRule}{%
      \begin{array}{@{}l@{}}
	T_\stm =
	\left\{%
	\begin{array}{@{}l@{\,}c@{\,}l@{\ \text{if~}\stm = \,}l@{}}
	  ［\plvl v \assignv \plvl e］ &&& \s{v = e}\\
	  ［\plvl v \assignv \plvl r ⊔ \flvl[\hvar]r{f_p}］ &&& \s{v = r.f_p} \\
	  ［\plvl r \assignv \plvl s ⊔ \flvl[\hvar]s{f_r}］ &\SQMergei u& \LoadRef r s {f_r}{\hvar}& \s{r = s.f_r} \\
	  ［\plvl r \assignv \plvl s］ &\SQMergei u& \CopyRef r s {\hvar} & \s{r = s}\\
	  ［\plvl r \assignv ⊥］& \SQMergei u& \NewRef r c {\pc}{\hvar} & \s{r = \cnew~c}\\
	  ［\plvl r \assignv ⊥］& \SQMergei u& \NullRef r {\hvar} & \s{r = \Null}\\
	  \multicolumn 3 {r@{\ \text{if~}\stm = \,}} {%
	  \mathllap{\NonJuncL\hspace{9em}}\!%
	  \StorePrim r {f_p} e {\nomlvl{\plvl{e}}}{\hvar}} & \s{r.f_p = e} \\
	  \multicolumn 3 {r@{\ \text{if~}\stm =  \,}} {%
	  \StoreRef r {f_r} s {\nomlvl{\plvl s ⊔ \hlvl[\hvar] s }}{\hvar}} & \s{r.f_r = s}
	\end{array}\right.
      \end{array}%
    }{%
      ℓ = \semloc{\stm;\stms, \_}%
      \trans{\tt, T_\stm}%
      \newloc{\stms}%
    }\\[\myrulespace]%
    \myrule{\BranchRule}{%
      \NonJuncL
      \quad
      ℓ = \semloc{\stm;\stms, \_}
      \quad
      \stm = \s{\cif~(e)~\cgoto~l}
    }{%
      \begin{array}{@{}l@{~}l@{}@{~}l@{}}%
        ℓ \trans{\phantom{¬}ω ∧ ¬\uVar ∧ \plvl e \,\not⊑\, \pc, \TBrch{\region{}\stm}} \newloc{\Target{} l} \quad\quad &
	ℓ\trans{\phantom{¬}ω ∧ ¬\uVar ∧ \plvl e \,⊑\, \pc ∨ \phantom{¬}ω ∧ \uVar, ∅} \newloc{\Target{} l} \\
        ℓ\trans{        ¬ ω ∧ ¬\uVar ∧ \plvl e \,\not⊑\, \pc, \TBrch{\region{}\stm}} \newloc{\stms} \quad\quad  &
        ℓ\trans{        ¬ ω ∧ ¬\uVar ∧ \plvl e \,⊑\, \pc ∨           ¬ ω ∧ \uVar, ∅} \newloc{\stms}
      \end{array}%
    }\\[\myrulespace]%
    \myrule{\JunctionRule}{%
      \begin{array}{@{}c@{}}
        \JuncL{ℓ}
        \quad
	ℓ = \semloc{\stms, \jsone}
	\quad
	J = \juncR{}{\stms}
	\quad
	P_J = ｛\rMode{ρ}｝_{ρ∈J}
      \end{array}
    }{%
        ℓ \trans{\rVar \,∉\, P_J, ∅}\semloc{\stms, \jstwo}
        \qquad
        ℓ \trans{¬\uVar ∧ \rVar \,=\, \rMode{ρ}, \Startua}\newloc{\inducing{}{ρ}}_{ρ∈J}
        \qquad
        ℓ \trans{\uVar ∧ \rVar \,∈\, P_J, \Endua{\pc'}}\semloc{\stms, \jstwo}
    }\\[\myrulespace]%
		\myrule{Call}{%
			\begin{array}{@{}c@{}}
				\NonJuncL\qquad
				T = 
				\subst{\SumEffect{m}^{r,w}}{\pc}{\pc ⊔ \hlvl[\hvar] r}
			\end{array}
		}{%
			\begin{array}{@{}r@{~}l@{}}%
				ℓ = \semloc{\s{
				r.m(\lits)};\stms, σ}%
				&\trans{\tt, T%
				}\newloc{\stms}\\
			\end{array}
		}%
		\quad%
		\begin{minipage}[c]{5cm}
		\setlength{\abovedisplayskip}{0pt}%
		\setlength{\belowdisplayskip}{0pt}%
        \[
	        \red{φ_{\textsc{Call}}} ≝~
            \subst{\SumGuard{m}^{r,w}}{\pc}{\pc ⊔ \hlvl[\hvar] r}
            \tag{\rname{φ-Call}}\label{eq:invariant4call}
        \]
		\end{minipage}
  \end{ruleset}
  \par
  \raggedright%
  \vspace{\myrulespace}
  where:\vspace*{-1.4em}
  \begin{align*}
    \plvl{p} ≝ ⊥ \qquad
    \plvl{⊖ e} ≝ \plvl e \qquad
    \plvl{e ⊕ x} ≝ \plvl e ⊔ \plvl x \qquad
    \plvl{r == s} ≝ \plvl r ⊔ \plvl s \span \\
    \nomlvl l ≝ （\ite{\uVar}{⊥}{l}） ⊔ \pc \qquad
    \plvl x \assignv l ≝ \plvl x \assign （\ite{\uVar}{\plvl x}{l}） ⊔ \pc\span
    \\
    \Junc{\stms}{ψ} ≝ \juncR{}{\stms} ≠ ∅ ∧ ψ = \jsone
		       &\qquad
      \TBrch{ρ} ≝ ［\rVar \assign \rMode{ρ}, \pc \assign
      ⊤
      , \SaveHeap{\hvar}］\\
    \Startua ≝ ［ \uVar \assign \tt, \SwapHeap{\hvar}］
    &\qquad
      \Endua{l} ≝ ［\rVar \assign \rNom, \uVar \assign \ff, \pc \assign 
      ⊥］
      \SQMergei u \HeapBulkUpgradeFrom[\hvar]{\hvar'}
  \end{align*}%
  \hrule width 0pt              
  \caption{Translation rules and safety properties for encoding the security semantics.}
  \label{fig:security-semantics-for-statements}
\end{figure}

We give in \figurename~\ref{fig:security-semantics-for-statements} the set of
translation rules that 
specify the security semantics of a program in terms of an SCFG.
Each location of the resulting SCFG 
corresponds to a \emph{semantic location}, that is defined as a
pair \semloc{\stms, ψ} where \stms
corresponds to a node in its CFG,
and ψ is a \emph{behavior mode} that belongs to \(｛\jsone, \jstwo｝\) (for
\textsf{n}ominal-or-\textsf{j}unction and \textsf{n}ominal
\textsf{b}ehaviors, respectively).
The junction step ψ is used in our encoding 
to distinguish the nominal mode from the upgrade
analysis stage of junctions.
Essentially, a semantic location that corresponds to a statement
\stm that is the junction of a CDR behaves as a junction when
\(ψ = \jsone\), and according to \stm when \(ψ = \jstwo\).
Thus, statements that are not junctions never give rise to
semantic locations where \(ψ = \jstwo\).
To clarify the translation rules, we define the helper predicate
\Junc{\stms}{ψ} in
\figurename~\ref{fig:security-semantics-for-statements} (where
\juncRName{} is the retraction of \juncName{}: \juncR{}\stms
gives the set of CDRs of which \stms is the junction), that
holds iff a semantic location \(\semloc{\stms, ψ}\) represents
an actual junction.
We use \Target{}l to denote the statement identified by a label \lbl l.

The \AssignRule rule encodes the security semantics of
assignments.
We use \plvl e to denote the security level of an expression \(e\). 
from upgrade analyses in the rules.
In nominal mode, \nomlvl l encodes the least upper-bound between \(l\) and the context level \pc, and \(\plvl x \assignv l\) models a \emph{strong update} of the security level assigned to \(x\) with \nomlvl l.
In upgrade analysis mode, however, \nomlvl l is equal to the context level (\ie \high), and \(\plvl x \assignv l\) encodes a \emph{weak update} of \plvl x with \pc.
Then, a statement \(v = r.f_p\) that loads a primitive field translates into a transition that updates \plvl v with: the upper-bound between \pc, \plvl r, and the level of any object potentially pointed to by \(r\) as maintained by the heap abstraction \hvar (\ie \hlvl[\hvar]r) when in nominal mode; the upper-bound between \pc and \plvl v otherwise.%
\BranchRule and \JunctionRule encode the alternation of nominal and upgrade analyses, and do so with the help of a placeholder abstract heap \(\hvar'\) that belongs to the same abstract heap domain as \hvar, and is also represented with state variables.
According to \BranchRule, when a high branch is reached, the transformer \TBrch{ρ}:
\begin{enumerate*}[(i)]
\item sets the state variable \rVar used to record the CDR currently subject to a high-condition to \rMode{ρ};
\item updates \pc; and
\item stores the current heap abstraction to \(\hvar'\) by copying the values of all variables \GenericHLvlVars[\hvar] (resp. \GenericRelVars[\hvar]) to \GenericHLvlVars[\hvar'] (resp. \GenericRelVars[\hvar']).
\end{enumerate*}
The join of abstract heaps that ends upgrade analyses is performed using a bulk upgrade.
In effect, \(\HeapBulkUpgradeFrom[\hvar]{\hvar'}\):
\begin{enumerate*}[(i)]
\item upgrades the security typing environment for referenced portions of the heap according to the result of the upgrade analysis in \hvar, by joining every security level from \hvar with the corresponding level in \(\hvar'\); and
\item restores every heap-related relation as saved in \(\hvar'\) when entering~the~upgrade~analysis~mode.
\end{enumerate*}
\rname{Call} encodes the security
semantics of invocation of a method $m$ based on its \emph{polymorphic information-flow summary}, which is a \emph{contract} that consists of:
\begin{itemize}[nosep]
 	\item an information-flow \emph{guard} \SumGuard m that specifies the invocation conditions under which the method call is secure, \ie there is no illegal information flow in the method.  This guard is described as constraints on the security types and heap structure of the method's formal arguments;
 	\item an \emph{effect} \SumEffect {m} about its \emph{worst potential} side-effects on security levels and heap structure, that is in principle a \emph{transformer} describing how the heap structure and security labels \emph{may be} updated by the method.
\end{itemize}
We 
use the 
guard to enforce the desired security properties upon an invocation of $m$: this boils down to ensure that Inv.~(\ref{eq:invariant4call}) holds for the location ℓ in which $m$ is called.
%
%
The effect is used in \rname{Call} to update the typing environment and the heap model.
In detail, \(\SumGuard {m}^{r,w}\) and \(\SumEffect {m}^{r,w}\) correspond to the aforementioned guard and effect—or a combination of several summaries in case of virtual method dispatch, where guards are combined using a conjunction, and transformers are merged—, and after substitutions \wrt \(m\)'s formal arguments.
Furthermore,
\subst e v l denotes the substitution of security level expression \(l\) for variable \(v\) in \(e\): this is required to upgrade the context \pc~\wrt the receiver object 
(the substitution in effects is performed in every expression on the right-hand side of assignments).
\nberth{For the sake of concision, we leave the computation of polymorphic effects out of the scope of this paper.  In that respect, we want to mention that this computation is achievable, even for the cases of recursion, via an extension of our security semantics, accompanied by a dedicated processing of the co-reachability analysis results.}
\nberth{Also note that a sound application of effects requires abstract heaps that capture object sharing relations, not just aliasing relations.}

\subsection{Guard Inference Procedure}
\label{sec:guard-synthesis}

\begin{figure}[t]
  \removelatexerror%
  \input{fig/synthesize-guard}
\end{figure}
We summarize the overall analysis procedure 
in \algorithmcfname~\ref{alg:guard-synthesis}, where \SecuritySemantics[]{m} denotes the specification of the security semantics of a method \(m\) as an SCFG \(S_m\) and invariant \(φ_m\) as described above.
We represent sets of states as mappings from locations to predicates on state variables, \eg \(\B_0\) is the set of all states that violate the invariant \(φ_m\).
\(\B_0\) associates every location that corresponds to a sink statement with a predicate on security levels for program variables that violate \red{security requirements \nberth{encoded in \(φ_m\)}}.
Then, the set of insecure states is back-propagated via a standard co-reachability analysis embodied by \Coreach{}, \ie  finding all states from which a given set of states may be reached, and is typically solved using a fixed-point~\citep{Ramadge89,PnueliRosner}.
On a symbolic finite-state system like \Sf or \Sm, this computation \nberth{always terminates}, and is traditionally performed using the least fixed-point (\(\mathrm{lfp}\))
{%
  \setlength{\abovedisplayskip}{1pt}%
  \setlength{\belowdisplayskip}{1pt}%
  \begin{equation}
    \B_∞ ≝ \mathrm{lfp}\ λ\B_i.\B_0 ∪ \mathrm{pre}(\B_i)\mbox,
  \end{equation}%
}%
where \(\mathrm{pre}(\B)\) gives all predecessor states of \B.
\(\B_∞\) associates each location with a predicate that must \emph{not} hold for every subsequent path in \(S_m\) to \red{represent secure executions}.
Therefore, the guard for \(m\) can be obtained by complementing \(\B_∞(ℓ_0)\) and eliminating every state variable that does not represent a proposition from \(m\)'s calling context.
This is done with the help of
\(\mathrm{cofactor}(f, g)\), which 
amounts to a partial evaluation of \(f\) \wrt all variables bound in \(g\),
\ie this gives a predicate \(f'\) that does not involve any variable fully determined by \(g\) and \st \((g = \tt) ⇒ (f = f')\).

\begin{leaveout}
\begin{example}[Guard inference for \cv f]
  \label{expl:sf-guard-synthesis}
  \setlength{\abovedisplayskip}{3pt}%
  \setlength{\belowdisplayskip}{3pt}%
  The application of \algorithmcfname~\ref{alg:guard-synthesis} on method @f@ first constructs the SCFG \Sf and invariant \(φ_{\cv f}\) of \figurename~\ref{fig:f-example}.
  Then, the co-reachability and partial evaluation give
  \begin{align*}
    \B_0 &= ｛ℓ_4 ⟼ \pc ⊔ \cplvl l ≠ ⊥｝ \red{∪ ｛ℓ_i ⟼ \ff｝_{i∈\{0,1,2,3,5\}}} \\
    \B_∞ &= ｛ℓ_0 ⟼ （\ite{¬\uVar}{\pc ⊔ \cplvl v}{\pc}）≠ ⊥｝ ∪ \B_∞' \tag{\(= \Coreach{\Sf,\B_0}\)} \\
    \SumGuard{\cv f} &= \pc ⊔ \cplvl v = ⊥ \tag{\(= \mathrm{cofactor}（¬\B_∞(ℓ_0), X_0）\)}\mbox.
  \end{align*}
  The guard \SumGuard{\cv f} holds whenever \(\pc = ⊥\) and \(\cplvl v = ⊥\): to satisfy the confidentiality requirement, @f@ must be called in a low context (this typically means that @f@ may execute a sink statement as is the case on line~\ref{f-line:output-l}).
  This also states that @f@ may leak information about the effective argument for \cv v.
\end{example}
\end{leaveout}


\begin{figure}[b]
  \begin{minipage}{0.67\textwidth}
    \smaller
    \centering
    \captionof{table}{Polymorphic guard inference for method \inlinemeth{m}, for different heap domains.\vspace*{-1em}}
    \label{tbl:m-example-coreach}
    \begin{tabular}{@{}r@{\,}|@{\,}Q@{}}
      \hline
      \multicolumn 2 c {with \(\B_0 = \big\{ℓ_3 ⟼ \pc ⊔ \cplvl b ⊔ \chlvl b ≠ ⊥\big\} ∪ \big\{ℓ_i ⟼ \ff\big\}_{i∈\{0,1,2,4\}}\)
      }
      \\ \hline
      \rotatebox[origin=c]{90}{\hdeep}
      & \begin{array}{@{}r@{\,=\,}l@{}l@{\,}l@{\,}l@{}l@{}}
	  \B_1 & \B_0\,∪ & \big\{ℓ_2 ⟼ &\pc ⊔ \cplvl b ⊔ \chlvl b ⊔ & (\ite{\cAliasRel b r}{\cplvl a ⊔ \chlvl a}{⊥}) & ≠ ⊥\big\} \\
	  \B_2 & \B_1\,∪ & \big\{ℓ_1 ⟼ &\pc ⊔ \cplvl b ⊔ \chlvl b ⊔ &(\ite{\cAliasRel b r}{\cplvl a ⊔ \chlvl a ⊔ \cplvl i}{⊥})\,⊔ & \\
	  \multicolumn{4}{c}{}
	       & (\ite{\cFieldAliasXRel b a}{\cplvl i}{⊥}) & ≠ ⊥\big\} \\
	  \B_∞ 
	       & \B_2\,∪ & \big\{ℓ_0 ⟼ &\pc ⊔ \cplvl b ⊔ \chlvl b ⊔ &(\ite{\cFieldAliasXRel b a}{\cplvl i}{⊥}) & ≠ ⊥\big\} \\
	  \SumGuard{\cv m}
	       &\span& \pc ⊔ \cplvl b ⊔ \chlvl b ⊔ &(\ite{\cFieldAliasXRel b a}{\cplvl i}{⊥}) = ⊥ \span
	\end{array}
      \\ \hline
      \rotatebox[origin=c]{90}{\(\mathclap{\phantom p}\)\hshal}
      & \begin{array}{@{}r@{\,=\,}l@{}l@{\,}l@{}l@{}l@{}}
	  \B_1 & \B_0\,∪ & \big\{ℓ_2 ⟼ &\pc ⊔ \cplvl b ⊔ & \chlvl b ⊔ (\ite{\cAliasRel b r}{\cplvl a ⊔ \chlvl a}{⊥}) & ≠ ⊥\big\} \\
	  \B_2 & \B_1\,∪ & \big\{ℓ_1 ⟼ &\pc ⊔ \cplvl b ⊔ \cplvl i ⊔ & \chlvl b ⊔ (\ite{\cAliasRel b r}{\cplvl a ⊔ \chlvl a ⊔ \cplvl i}{⊥}) & ≠ ⊥\big\} \\
	  \B_∞ 
	       & \B_2\,∪ & \big\{ℓ_0 ⟼ &\pc ⊔ \cplvl b ⊔ \cplvl i ⊔ & \chlvl b & ≠ ⊥\big\} \\
	  \SumGuard{\cv m} &\span& \pc ⊔ \cplvl b ⊔ \cplvl i ⊔ & \chlvl b = ⊥ \span
	\end{array}
      \\ \hline
      \rotatebox[origin=c]{90}{\(\mathclap{\phantom p}\)\hdumb}
      & \begin{array}{@{}r@{\,=\,}l@{}l@{\,}l@{}l@{}l@{}}
	  \B_1 & \B_0\,∪ & \big\{ℓ_2 ⟼ &\pc ⊔ \cplvl b ⊔ \cplvl a ⊔ & \chlvl b ⊔ \chlvl a & ≠ ⊥\big\} \\
	  \B_2 & \B_1\,∪ & \big\{ℓ_1 ⟼ &\pc ⊔ \cplvl b ⊔ \cplvl a ⊔ \cplvl i ⊔ & \chlvl b ⊔ \chlvl a & ≠ ⊥\big\} \\
	  \B_∞ 
	       & \B_2\,∪ & \big\{ℓ_0 ⟼ &\pc ⊔ \cplvl b ⊔ \cplvl a ⊔ \cplvl i ⊔ & \chlvl b ⊔ \chlvl a & ≠ ⊥\big\} \\
	  \SumGuard{\cv m} &\span& \pc ⊔ \cplvl b ⊔ \cplvl a ⊔ \cplvl i ⊔ & \chlvl b ⊔ \chlvl a = ⊥\span
	\end{array}
      \\ \hline
    \end{tabular}%
    \hrule width 0pt              
  \end{minipage}
  \hfill
  \begin{minipage}{0.3\textwidth}
    \centering
\input{fig/scfg-gen-setup}%
\begin{tikzpicture}[scale=.8, node distance = 2em, > = {Stealth[scale=.8]},
  ]
  \tikzstyle{label} = [font=\smaller, inner sep = 0pt, outer sep = 0pt]%

  \node [location] (l0) {\(ℓ_0\)};
  \foreach \i in {1,...,4} {
    \pgfmathtruncatemacro{\j}{\i-1}
    \node [location, below = of l\j] (l\i) {\(ℓ_{\i}\)};
  }

  \draw [basic-arrow] (l0.west)+(-2mm,0) to (l0);
  \draw (l0) edge [basic-arrow] node [left, label, xshift = -2pt] (l01) {%
    \(\tt, [\plvl\bl \assign \pc] \SQMergeiu\)\\
    \(\NewRef\bl{\mathtt{B}}\pc\sHeap\)%
  } (l1);
  \draw (l1) edge [basic-arrow] node [left, label, xshift = -2pt] (l12) {%
    \tt,
    \(\StorePrim{\cv a}{\fint}{\cv i}{\cplvl i ⊔ \pc}\sHeap\)%
  } (l2);
  \draw (l2) edge [basic-arrow] node [left, label, xshift = -2pt] (l23) {%
    \tt,\\
    \(\StoreRef\bl\fa{\cv a}{\cplvl a ⊔ \chlvl[\sHeap] a ⊔ \pc}\sHeap\)%
  } (l3);
  \draw (l3) edge [basic-arrow] node [left, label, xshift = -2pt] (l34) {%
    \(\tt, ∅\)%
  } (l4);
\end{tikzpicture}%
\hrule width 0pt              
 
    \captionof{figure}{SCFG for 
      \protect\inlinemeth{m}.}
    \label{fig:m-example-scfg}
  \end{minipage}
\end{figure}

\begin{example}[Guard inference for \cv m]
\def\bl{\cv r}%
\def\fint{\mathtt{fi}}%
\def\fa{\mathtt{fa}}%
 
%
  The analysis
  first builds the SCFG given in \figurename~\ref{fig:m-example-scfg}, and
  associates the invariant \(\pc ⊔ \cplvl b ⊔ \chlvl b = ⊥\)
  with
  location \(ℓ_3\).
  This states that,
  for @m@ to be secure, this statement must be executed in a low context and given a low-sensitive reference \cv b (\ie \(\cplvl b = ⊥\)) that must only reach low-sensitive objects (\ie \(\chlvl b = ⊥\)).
  This gives the unsafe states shown in the first row of \tablename~\ref{tbl:m-example-coreach}, where we report a trace of the co-reachability analysis and the resulting guard for each domain. 
  The guard obtained with the \hdumb domain is the least precise of all three, as it basically describes @m@ as insecure if it is called in high context, or whenever any of its effective arguments or objects they may reach in the heap is high-sensitive.
  With this domain, the statement \(\cv r.\fa = \cv a\) (location \(ℓ_2\)) may raise the security level \chlvl b since \(
  \cAliasRel r b
  ∈\GenericRelTauto\) (as \(\TAbase {} {\cAliasRel r b} = \TAmaybe\)
  ).
  On the other hand, the inference with \hshal is able to distinguish whether \cv b and \cv r may alias on location \(ℓ_2\), and then rules this case out thanks to the statement \(\cv r = \New~\cv B\) (location \(ℓ_0\)).
  However, the guard does not hold whenever \(\cplvl i ≠ ⊥\), as the domain cannot distinguish whether \(\cv b.\fa\) aliases \(\cv a\) or not: therefore, the statement \(\cv a.\fint = \cv i\) (location \(ℓ_1\)) always raises the level \chlvl b to that of \(\pc ⊔ \cplvl i\).
  At last, the domain \hdeep distinguishes whether \(\cv b.\fa\) aliases \(\cv a\) or not, and the guard indicates that @m@ may not be secure if \cv i is high-sensitive and \cv a and \cv b relate to each other via \(\cv b.\fa\).

  \begin{leaveout}
    which describes every state of the model that violates the security requirement (\red{the invariant}).
    The analysis then proceeds with 
    the computation of all states \(\B_∞\) that are co-reachable to \(\B_0\), \ie the states form which there exists a path in \red{\FSM\Sm} that reach \(\B_0\).
    Like \(\B_0\), \(\B_∞\) associates every location in \Sm with a predicate that describes \emph{unsafe states}.
    We can eventually obtain the guard for @m@ that only relates to its calling context (\ie only involving state variables pertaining to the calling context) via a partial evaluation \red{generalized cofactor} as on line~\ref{alg-line:cofactor-init} in \algorithmcfname~\ref{alg:guard-synthesis}.
\end{leaveout}
\end{example}



\section{Soundness}\label{sec::soundness}

We prove that any program guarded with a security guard inferred by our method guarantees termination-insensitive noninterference~\citep{sabelfeld2003language}.
This notion states that, for any initial states $\qstate$ and $\qstate'$ whose secret parts may only differ, the observation sequences of the program running from the states $\qstate$ and $\qstate'$ will either be the same, or one is a prefix of the other.
The reason for the latter case is that this notion is a termination-insensitive property.

\begin{figure}[t]
  \begin{minipage}{\linewidth}
    \begin{ruleset}
      \myrule{\mUpgrade}{
        \begin{array}{@{}c@{}}%
          ℓ \trans{
          g, T}
          ℓ'
        \end{array}
      }{%
        \begin{array}{@{}c@{}}%
          \mframe[]{ℓ}{\V}\mHeap
          \ttrans{\mode ∧ g} T
          \mframe[]{ℓ'}{\V}\mHeap
        \end{array}
      } , 
      \myrule{\mJunctionRule}{%
	\JuncL{ℓ} \quad ℓ \trans{g, T}ℓ'
      }{%
	\mframe[]{ℓ}\V \mHeap%
	\ttrans{¬\mode ∧ g} T%
	\mframe[]{ℓ'}\V \mHeap%
      }
      \\[\myrulespace]
      \myrule{\mBranchRule}{%
        \begin{array}{@{}c@{}}
          \NonJuncL \quad
          ℓ = \semloc{\s{\cif~(e)~\cgoto~\_};\;\_,\_} \quad
          c = \V(e) \\
          ℓ \trans{τ ∧ ¬\mode ∧ g_\tt, T_\tt} ℓ_\tt \quad
          ℓ \trans{¬τ ∧ ¬\mode ∧ g_\ff, T_\ff} ℓ_\ff
        \end{array}
      }{%
        \mframe[]{ℓ}\V \mHeap%
        \ttrans{¬\mode ∧ g_c}{T_c}
        \mframe[]{ℓ_c}\V \mHeap
      } , 
      \myrule{\mStmRule}{%
        \begin{array}{@{}c@{}}
          \NonJuncL \quad
          ℓ = \semloc{\stm;\;\_,\_} %
          \trans{g, T}%
          ℓ' \quad
          \stm ∈ \mathit{Stm}
          \\
          \V', \mHeap' = \evalStm\stm(\V, \mHeap)
        \end{array}
      }{%
        \mframe[]{ℓ}\V \mHeap%
        \ttrans{\neg \mode ∧ g} T%
        \mframe[]{ℓ'}{\V'}{ \mHeap'}%
      }
    \end{ruleset}
    \par
    \smallskip%
    \raggedright%
    where
    \[
      \mathit{Stm} ≝ ｛%
      \begin{array}{@{}c@{}}
        \s{v = e}, \s{v = r.f_p}, \s{r = s}, \s{r = s.f_r}, \s{r.f_p = e}, \s{r.f_r = s},\\
        \s{r = \cnew\,\_}, \s{r = \Null},
\s{\cgoto\,\_}, \s{\coutput l (\_)}
      \end{array}%
      ｝
    \]
    and\vspace*{-.2\baselineskip}%
    \begin{align*}
      \evalStm{a}(\V, \mHeap) ≝ \,
      \begin{cases}
        \V[v⟼{\V}(e)], \mHeap &\mbox{if~} a =\s{v=e}\\
        \V[v⟼\RefContent {r}{f_p} ], \mHeap &\mbox{if~} a= \s{v = r.f_p} \\
        \V, \mHeap &\mbox{if~} a \in \{{\s{\cgoto~\_}},
        \s{\coutput\_(\_)}
		\} \\
        \V, \MemHeapSem{a}\mHeap & \textnormal{otherwise.}
      \end{cases}
    \end{align*}
  \end{minipage}
  \caption{Full semantics.}
  \label{fig::full.semantics}
\end{figure}

To prove noninterference, we first define the full semantics of a program by  an SCFG that extends the security semantics with its operational semantics.
The full semantics basically extends the security semantics of a program with its execution semantics.
In addition to the symbolic variables that belong to the security semantics, we consider a concrete model for the \emph{heap}, as well as
 local variables.
A \emph{configuration} is therefore defined as \( \mframe[]{ℓ}\V\mHeap\) where ℓ denotes the semantic location,
\red{  \(\V
\)} is a {\emph{valuation}} for all primitive variables,
\mHeap is the heap value of the concrete heap domain \HeapDom \concrete.
We give the full semantics in \figurename~\ref{fig::full.semantics}, where \(\trans{g,T}\) and \(\ttrans g T\) are transitions of the security semantics and of the full semantics, respectively.
Apart from \mUpgrade and \rname{m-Sink}, every rule in this figure applies only when the SCFG is in the nominal mode (\ie \(\mode = \ff\)) --- this is reflected in the guards for the full semantics.
According to rule \mUpgrade, only the updates by the security semantics are included when the program runs in the upgrade analysis mode (\ie $\uVar = \tt$), \ie the program statements run  in nominal mode only.
The semantics of a branch statement is defined based on its corresponding rule of the security semantics with the addition that the {input} variable $\tau$ is substituted by the concrete guard \(e\) of the statement.
The rule \mStmRule handles assignments,
and other non-branching control-flow related statements, by extending the updates by the security semantics with the updates to the primitives \V and the concrete heap \mHeap.

Let \(\S \) denote the (symbolic) full semantics of a program,
and $\FSM\S=\FSMdef$ be an automaton that describes its concrete
semantics,
where \smStates is the set of states, $\mathcal{I}$ is the set of inputs,
$\smTrans \subseteq \smStates \times \smStates$ is the set of transitions and $\smStates_0$ is the set of initial states.
A program state \(\qstate 
\) is defined as
\(\mframeX {{\VarsTypes}} {ℓ} \V {\X} \mHeapVal\)
where
ℓ
is the current location,
\V is the
valuation for every one of the
primitive program variables $\PrimVars$,
{{\(\VarsTypes: \PrimVars \cup R
→ \LL\)} is the \emph{security typing environment} for {the primitive variables $\PrimVars$ and references $R$}},
and
\(\X
\)
is the current valuation of the state variables {of the security semantics} except the location and the typing environment for the references.
We use the notation
$\X_{\sHeap} {}$ to show
\X's \nberth{heap abstraction \sHeap},
and
\sHeapValLevs \X \sHeap to denote its security typing environment.
Furthermore, \mHeapVal is the valuation of the concrete heap's variables.
Let $\qstate \xrightarrow{\eta}_* \qstate'$ be an execution of full semantics  with a non-zero length (\ie the reflexive and transitive closure of {the concrete transition relation $\smTrans$}) from the state $\qstate$ to the state $\qstate'$, where $\eta \in \{o,\bot\}$.
This execution either ends by executing a statement that outputs on a channel (\ie $\eta=o$)
or makes no observation (\ie $\eta = \bot$).
We denote an execution that never reaches an observation point by $\qstate \xrightarrow{\bot}_* $.
We define noninterference based on a \emph{low-equivalence relation}, that states that the public parts of the two states $\qstate_1$ and $\qstate_2$ are indistinguishable.
\begin{definition}[{Indistinguishable Stores}]\label{def::low-equivalence-relation}
	We say  two valuations $\V_1 ∈ \Val \PrimVars$ and $\V_2 ∈ \Val \PrimVars$ are low-equivalent \wrt the {typing environment \(\VarsTypes: \PrimVars \cup R→ \LL\)},
	denoted by \(\V_1 =_{\VarsTypes} \V_2\),
	iff $\V_1(v) = \V_2(v)$ for all $v ∈ \PrimVars$ where \(\VarsTypes(v)=\low\).
\end{definition}


\begin{definition}[Low-Bisimulation]\label{def::low-bisimulation-relation}
	We say two states \( \qstate_i=\mframeX{{\VarsTypes_i}}{ℓ}{\V_i}{\X_i}{\mHeapVal_i},  i\in\{1,2\}\)
	are {compatible},
	denoted by \(\qstate_1 \approx \qstate_2 \), iff  \xspace
	(i) \( {\VarsTypes_1}=\VarsTypes_2\),
    (ii) ${\sHeapValLevs {\X_1} \sHeap}={\sHeapValLevs {\X_2} \sHeap}$,
	(iii) \(\V_1 =_{\VarsTypes_1} \V_2\),
	and (iv) \({\mHeapVal_1 =_{{\X_1}_{\sHeap}} \mHeapVal_2}\).
	They are called
	low-bisimilar, denoted by $\qstate_1 \lowbisim \qstate_2$, iff \(\qstate_1 \approx \qstate_2\), and  if $\qstate_1\xrightarrow{o}_* \qstate'_1$,
	then either (a) there exists $\qstate'_2$ such that $\qstate_2 \xrightarrow{o}_* \qstate'_2$
	and $\qstate_1' \lowbisim \qstate'_2$,
	or (b) $\qstate_2 \xrightarrow{\bot}_* $,
	and vice versa.
\end{definition}


\begin{theorem}[Noninterference]\label{thm::Non.Interference}
	For any method $m$ guarded by a security guard \SumGuard m, and any initial states \( \qstate_1\) and  \( \qstate_2\) where	\(\qstate_1\approx \qstate_2\) and
	$\qstate_i\models \SumGuard m $,
	$i \in \{1,2\}$, it holds $\qstate_1 \lowbisim\qstate_2$.
\end{theorem}
\begin{proof}
To prove this theorem, we show that there exists a witnessing bisimulation relation  
for $\qstate_1 \lowbisim \qstate_2$. See Appendix~\ref{sec::noninterference-proof}.
\end{proof}

\section{Implementation and Evaluation}
\label{sec:experimental-results}

To empirically validate our approach, we assess the respective performances of our three heap domains on actual code, both in terms of precision and scalability.

\subsection{Implementation Details}

We have first implemented a tool that relies on \soot~\citep{vallee2010soot} to obtain the \Jimple code of a program, and translates it into our input language.
\Jimple is an intermediate language to represent \Java byte-code at a higher level.
The semantics of its instructions and reference manipulations correspond to that of the \JVM.
One \Jimple statement roughly translates into one statement of our input language.
We have then implemented the guard inference algorithm in a prototype tool called \Guardies\footnote{Available as a software artifact~\citep{Guardies-VMCAI23-artifact}, with user documentation and source code at \url{http://nberth.space/symmaries}.}, that features multiple instantiations of our heap domains.
\Guardies's pre-analysis relies on a naive analysis of the class hierarchy to construct a graph that allows us to compute facts about heap-related relations (\ie function \TypeAnalysis {}{}).
%
This tool relies on \ReaX~\citep{ReaX-WODES-2014} to solve the co-reachability problems.
\ReaX uses (Multi-terminal) Binary Decision Diagrams---(MT)BDDs---~\citep{BryantBDDs,billon1987perfect} to represent symbolic expressions and compute the underlying fixed-points.
\begin{leaveout}
Can be removed now?
\subsubsection*{Inter\-procedural Analysis}
Simple \Java code often involves several methods, and our main objective for the evaluations consists in assessing the relative performances of applying our approach on actual code, using each heap domain.
Furthermore, a proper handling of precision benchmarks additionally requires the analysis to capture information-flows across method calls.
Therefore, \Guardies implements an \emph{inter-procedural} variant of the analysis formalized above.
Informally, the security semantics of method calls is encoded as follows:
First, \Guardies reuses the guard inferred for each called method (after some renaming of arguments) to associate an invariant with semantic locations that correspond to their potential call sites.
Second, to account for the effects of callees on the heap and returned values, \Guardies automatically generates predicate transformers based on signatures of called methods (\ie the \Java types of formal arguments).
Such transformers encode some predefined behaviors, where every portion of heap reachable via a reference argument may be mutated during the execution of the method, a high-sensitive value may be returned, etc.
\end{leaveout}
\red{
To deal with
guards and transformers that encode the semantics of library methods, we rely on
\emph{stubs}, given to \Guardies, that describe the effects of these methods at a high level.
We manually defined the security semantics of methods from the standard \Java and Android libraries (about \num{1200} methods in total) in this way.}

  
\subsection{Precision \& Recall}
%
We have employed the \IFSPEC benchmark suite~\citep{HamannHMM0T18} to
assess the \emph{precision} of our different heap domains and compare our results to
\KeY~\citep{KeY},
\Cassandra~\citep{Lortz2014Cassandra}, and
\Joana~\citep{Hammer:2009:FCO:1667545.1667547}.
The precision refers to a proportion
of test cases that are correctly classified.
The \emph{recall} is the fraction of true positive and false negative test cases that are categorized correctly.
\IFSPEC provides 232 test cases that showcase various information-flow vulnerabilities in \Java programs, with various forms of explicit 
and implicit information leaks.
%
\nbrem[yep]{\IFSPEC extends SecuriBench Micro~\citep{SecuriBenchMicro}, that is a benchmark suite for security analyzes of web applications; it provides 152 samples derived from the original 122 samples of SecuriBench Micro.
  \IFSPEC also features all 119 samples from DroidBench~\citep{Artz2014FlowDroid,DroidBench}, that is a benchmark suite for evaluating taint-analysis tools that target the Android platform.}%
%
We report in \tablename~\ref{tab::precision.experiments} the precision results that we obtain for different abstract heap domains for the various categories of leaks and language features that \IFSPEC 
covers.
We have checked 164 out of 232 test cases supported by our sub-language: the excluded cases involve reflection, static class initializers, exceptions and \red{method calls (11, 10, 9, and 39 samples respectively—we have excluded all cases in the latter category as they check the ability of the analysis to handle information-flows across method calls, 
while we 
left the problem of computing method  effects aside)}.
Note that a test case may belong to multiple categories.

\begin{wraptable}{R}{8.5cm}
  \vspace*{-\intextsep}%
  \begin{threeparttable}
    \caption[]{\IFSPEC Precision Results\vspace*{-1em}}%
    \label{tab::precision.experiments}%
  \smaller\smaller%
    \setlength{\tabcolsep}{0pt}%
    \setlength\aboverulesep{.06em}%
    \setlength\belowrulesep{.08em}%
    \renewcommand\arraystretch{0.92}%
    \let\BF\bfseries%
  \begin{minipage}{8.5cm}
    \begin{tabular*}{\linewidth}{@{}l@{\extracolsep{\fill}}ccccccc@{\,}}%
      \toprule
    Category       & \#Smpls &   \hdeep &   \hshal &   \hdumb & \KeY &   \Joana & \Cassandra \\ \midrule
    explicit-flows &   143   & \BF 80.4 &     79.7 &     78.3 & 70.6 &     77.6 &       72.7 \\
    implicit-flows &    21   & \BF 71.4 & \BF 71.4 & \BF 71.4 & 57.1 &     57.1 &       61.7 \\
    simple         &    51   &     72.5 &     72.5 &     72.5 & 64.7 & \BF 76.4 &       68.6 \\
    high-cond.     &    10   & \BF 80~~ & \BF 80~~ & \BF 80~~ & 60~~ &     60~~ &       60~~ \\
    arrays         &    26   &     73~~ &     73~~ &     69.2 & 65.3 & \BF 76.9 &       69.2 \\
    library        &    69   & \BF 88.4 & \BF 88.4 &     86.9 & 76.8 &     76.7 &       79.7 \\
    aliasing       &    7    & \BF 71.4 & \BF 71.4 &     57.1 & 57.1 &     42.8 &       42.8 \\ \hline
    average        &         & \BF 79.2 &     78.6 &     77.4 & 68.9 &     75~~ &       71~~ \\
      \bottomrule
    \end{tabular*}
  \end{minipage}
  \justify%
  The \#Smpls column shows the number of included samples for each category
  ; other figures are percentages.
  \end{threeparttable}
  \vspace*{-\intextsep}%
\end{wraptable}

Since our approach is sound, we obtain 100\% \emph{recall}, \ie we correctly detect every insecure flow.
Regarding precision, our experiments show that all the domains have close precision: the \hdeep domain offers the highest precision of {79.2}\% and \red{\hdumb} offers the lowest precision of 77.4\%.
The false positives (\ie the secure test cases that were restrictively classified as insecure) mainly occur because our domains are field-insensitive, value-insensitive, do not distinguish elements in some collections of data, or due to the over-approximations in heap-related relations.
The results for the aliasing category are rather similar; 3 test cases in this category are insecure that are classified correctly by all three domains, as our analysis is sound. Two of the remaining 4 secure test cases are classified as insecure in all domains due to value- and field-insensitivity.

On average, the domain \hdeep offers the best precision in five categories. It
slightly underperforms the state-of-the-art for simple and arrays test cases only, notably due to value- and field-insensitivity.
In some categories, the improvement is noticeable, \ie it
improves the best precision of the aliasing category offered by
the existing tools by 14.3\%, improves the library category by
8.7\% and enhances the implicit-flows category by 9.5\%. We
attribute these substantial results in part to our precise handling of implicit flows across method calls (unlike \Cassandra
which forbids method calls in high-contexts for instance), and
in part to our heap abstract domain, that is able to precisely
track some intricate aliasing relations.
That most of our domains obtain similar precision results on the aliasing category may indicate that these test cases are rather uniform in the facts about aliasing that need to be discovered to detect secure cases.
Our findings show that while different domains had close precision results, they offer different computational complexity though.
Further, 
IFSpec only partially covers the set of IFC problems one can encounter in practice; we therefore refrain from generalizing our results. Yet, IFSpec is the most extensive benchmark available for IFC that we know of.


\newcommand\webshopapp{\textsf{WebShop}\xspace}%
\newcommand\webshopname{\texttt{27823030\_WEB\_SHOP-CMS}\xspace}%
\newcommand\crmapp{\textsf{crm}\xspace}%
\newcommand\crmname{\texttt{44409894\_crm}\xspace}%
\newcommand\CMStudioapp{\textsf{CMStudio}\xspace}%
\newcommand\CMStudioname{\texttt{9783982\_CMStudio}\xspace}%
\newcommand\RadioCRMapp{\textsf{RadioCRM}\xspace}%
\newcommand\RadioCRMname{\texttt{2114204\_RadioCRM}\xspace}%
\newcommand\taglibsapp{\textsf{taglibs}\xspace}%
\newcommand\taglibsname{\texttt{9592899\_taglibs}\xspace}%
\newcommand\exccrmapp{\textsf{exc-crm}\xspace}%
\newcommand\exccrmname{\texttt{32859973\_exc-crm}\xspace}%
\newcommand\JavaCMSapp{\textsf{Java-CMS}\xspace}%
\newcommand\JavaCMSname{\texttt{32396294\_Java-CMS}\xspace}%
\newcommand\cmsapp{\textsf{cms}\xspace}%
\newcommand\cmsname{\texttt{34873434\_cms}\xspace}%
\newcommand\TravelERPapp{\textsf{TravelERP}\xspace}%
\newcommand\TravelERPname{\texttt{47907539\_Travel\_ERP}\xspace}%
\newcommand\cmsXapp{\textsf{cms'}\xspace}%
\newcommand\cmsXname{\texttt{50269397\_cms}\xspace}%
\newcommand\plateapp{\textsf{plate}\xspace}%
\newcommand\platename{\texttt{46690913\_plate}\xspace}%
\newcommand\NumMeth{\(|\!M\!|\)\xspace}%
\newcommand\NumVars{\(|\!V\!|\)\xspace}%
\newcommand\FootSize{\(|\!F\!|\)\xspace}%
\newcommand\MethLen{\(|\!m\!|\)\xspace}%
\newcommand\TotTime{\(T\)\xspace}%
\newcommand\AToM{\(t\!/\!m\)\xspace}%
\newcommand\Processed{\(P\)\xspace}%
\newcommand\NumSkipped{\#Skipped\xspace}%
\newcommand\NumUnsat{\#Unsat\xspace}%
\newcommand\NumSecure{\#Tauto\xspace}%
\newcommand\TaintExpr{{\sffamily TA}\xspace}%
\newcommand\ExplicitConf{{\sffamily EFA}\xspace}%
\newcommand\ImplicitConf{{\sffamily IFA}\xspace}%

\begin{figure}[t]
  \graphicspath{{results/}}%
  \centering\input{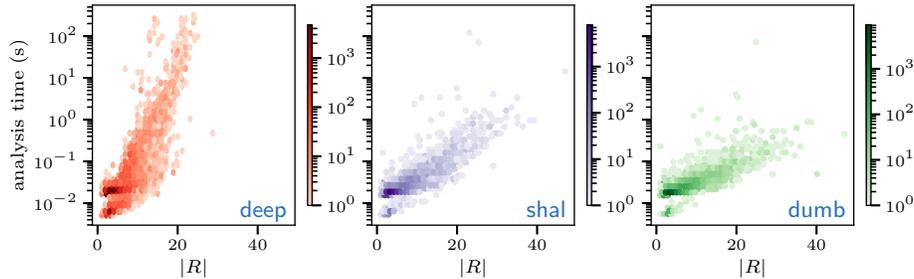}\vspace*{-1.5ex}%
  \caption{Density plots showing the distributions of analyzed ABM methods \wrt both the number of reference variables (horizontal axes) and the analysis time (vertical axes).  Note the \red{shared} log scales, including on the color-bars.}
  \label{fig:analysis-time-vs-nbrefs}
\end{figure}

\subsection{
  Scalability Evaluation%
  }
We have conducted experiments on real-life web applications to compare different heap abstract domains in terms of scalability.
To accommodate computationally intensive analyses, we 
interrupt any analysis after 5 minutes or if it uses more than 4GB of
memory.
We have used applications from the ABM benchmark~\citep{do2016toward},
a collection of 139 open-source projects that is dedicated to the
evaluation of static analyzers for \Java applications.
Its content is deemed representative of real-world software, and has
already been used for evaluating static taint analysis and dead code
elimination approaches~\citep{do2016toward}.
From this collection, we extracted the \Java code from {60}
applications with sizes ranging from 133 to 25K lines of \Java.
This provided us with a total of
\num{22512} analyzable methods.
Overall, the \hdeep domain led to 146 analyses being interrupted due
to the timeouts or memory limitations (3 for \hshal, 
0 for \hdumb).
We plot in \figurename~\ref{fig:analysis-time-vs-nbrefs}, for each domain, the distributions of successful analyses \wrt the number of reference variables and analysis time.
As expected, analysis times grow with the amount of references, and by factors that depend on the heap-related relations captured flow-sensitively by the domains,
\eg \hdeep 
is more expensive compared to \hdumb and \hshal.
Further, many methods have fewer than 10 reference variables, and as a result most analysis times do not exceed \num{0.1}s for every domain.
Those figures empirically support the applicability of our approach on real-life applications.

\begin{wrapfigure}{R}{5.29cm}
  \vspace*{-.8\intextsep}%
  \centering
  \graphicspath{{results/}}%
  \input{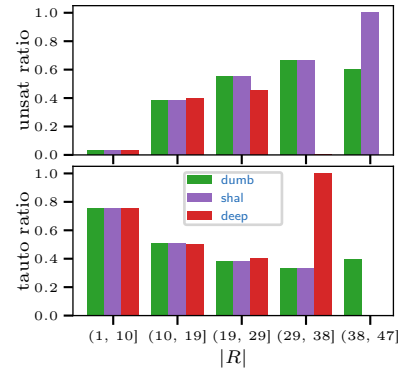}
  \caption{Plots showing for each domain, the proportions of un\-satisfiable (top) and tautological (bottom) guards \vs number of reference variables.}
  \label{fig:unsat_true_ratios}
  \vspace*{-1.1\intextsep}%
\end{wrapfigure}
Note that an ideal study on the scalability of different heap domains would compare the domains under different analysis techniques, provided by different tools.
This, however, requires the support of the existing tools for modeling different heap domains.
To the best of our knowledge, there was no such tool, as each implementation is typically tied with its own heap model, if any at all.
Otherwise, extending the tools to support different heap domains is virtually infeasible since most existing tools rely on store-based models.
To further compare the respective precision of each domain, we also report in \figurename~\ref{fig:unsat_true_ratios} the ratios of unsatisfiable and tautological guards obtained for each domain.
We observe that the precision of all domains seems similar when the number of reference variables is low, and diverges with growing numbers of references.
\nakh{
Moreover, \hdeep 
appears to be more permissive than \hdumb and \hshal for methods with many references.
However, that 
\hdeep 
 did not produce unsatisfiable guards for methods with more than 29 references indicates that many analyses of such methods were interrupted.
}

\section{Related Works}

Static analysis approaches to ensure noninterference have been studied
extensively in the community
.
The vast majority of suggested IFC solutions concentrates on
type-systems~\citep{pottier2003information,sabelfeld2003language,Barthe:2007:CLN:1762174.1762189
}, and
various tools that target realistic programming languages have been developed for verifying such properties.
Prominent examples include
JFlow JIF~\citep{sabelfeld2003language}, FlowCaml~\citep{pottier2003information}, \Cassandra~\citep{Lortz2014Cassandra}, and \KeY~\citep{
  KeY}.
Albeit 
sound
, the aforementioned approaches often lack
precision in practice (\eg~\cite{Lortz2014Cassandra}), or require 
user intervention, such
as the specification of loop invariants 
(\eg~\citep{sabelfeld2003language}).
Another line of research trades efficiency for soundness and/or precision, 
by exploiting more generic 
techniques like interprocedural dataflow analysis~\citep{Reps:1995:PID:199448.199462
} or program slicing~\citep{Kam:1977:MDF:2696874.2696926
}.
JOANA~\citep{Hammer:2009:FCO:1667545.1667547
}, DroidSafe~\citep{GordonKPGNR15DroidSafe} and \textsc{FlowDroid}~\citep{Artz2014FlowDroid} are prominent frameworks in this category.
%
{Other 
solutions are dedicated to web
  applications~\citep{
    hedin2014jsflow} or Android apps~\citep{li2017static}, although
  most of them do not handle implicit flows or lack soundness (\eg~\cite{Hammer:2009:FCO:1667545.1667547,GordonKPGNR15DroidSafe,Artz2014FlowDroid}).}
Tools that provide sound results via other forms of global program analyses include
\textsc{HornDroid}~\citep{Calzavara2016HornDroid,Calzavara2017HornDroidThreadsafeHeapAbstrNoImplicit%
}, which 
does not capture implicit flows.
In contrast to the above methods, our approach is proven sound, captures implicit flows (via heap), and our experiments show that it improves the state-of-the-art precision.
Further, the above approaches 
often rely on a simple store-based representation of the heap specified as a mapping from 
references (or abstract locations) to 
heap locations~\cite{DBLP:conf/popl/AmtoftBB06,Barthe:2007:CLN:1762174.1762189,Lortz2014Cassandra},
or do not rely on a flow-sensitive heap abstraction~\citep{Lortz2014Cassandra}.
By contrast, we use a store-less representation, where the structure of the heap {is} specified using a parameterizable family of (possibly over-approximated) relations.
\emph{This offers different levels of over-approximation and complexity, enabling the user to easily trade-off performance and scalability.} 
\begin{leaveout}
\nakh{	\textbf{Static Analysis} approaches for information flow control are often logic-based, or type system-based.
	A flow-sensitive information flow analysis for object-oriented programs via a Hoare-style logic has been presented in \cite{DBLP:conf/popl/AmtoftBB06}. The logic employs region assertions to describe possible aliasing, and independence assertions to describe indistinguishablity to formalize non-interference.
	A flow-insensitive type system for reasoning about a Java-like language is proposed in \cite{DBLP:conf/csfw/BanerjeeN02}, that requires manually annotating all fields, method parameters and method signatures with security types.
    Jif~\cite{jif,myers-popl99} is a well-known tool that extends Java with security types to support information flow control in Java applications. 
    While Jif is helpful to develop secure programs, it may require significant effort for annotating the programs.
	Joana~\cite{DBLP:journals/ijisec/HammerS09}, is a static analysis tool for information flow control in Java applications that uses program dependence graphs (PDGs) to represent the programs, and employs program slicing to check whether a source may influence a sink.
	 \Cassandra~\citep{Lortz2014Cassandra} is a sound solution that targets a realistic object-oriented low-level language.
	As shown by our experiments, our approach outperforms the existing work in terms of precision. Even the least precise instantiation of our work, \hdump offers a better precision. Further, we employ a synthesis-based approach to synthesize a guard rather than using a verification-based.
}
\end{leaveout}


Few works have addressed the problem of capturing implicit flows while exploiting flow-sensitive heap abstractions~\citep{DBLP:journals/compsec/Khakpour21,KhakpourC18SynthesisPermissiveSecurityMonitor,Zanioli2012SAILS}.
\citet{DBLP:journals/compsec/Khakpour21} synthesizes sound security monitors that enforce IFC by using a symbolic discrete control algorithm.
This work 
operates intra\-procedurally on high-level programs and 
uses an \adhoc field-sensitive heap abstraction that does not scale well.
\citet{Zanioli2012SAILS} advance an abstract-interpretation-based analysis, where the construction of the heap abstraction is delegated to a separate analysis.
Their analysis can operate on a flow-sensitive abstraction as produced by a TVLA-based shape analysis~\citep{10.1145/514188.514190}, yet it can only be applied to small, high-level programs.
Other forms of
symbolic heap abstractions have already been used in static program analysis.
Separation logic~\cite{Reynolds:2002:SLL:645683.664578} models a heap 
as a formula that comprises atomic predicates combined using the \emph{separation} operator.
While we use a store-less representation of the heap 
expressed using a proposition, symbolic heaps in separation logic are store-based, more expressive, and consequently are more complex for verification.
%
%
%
Store-less
heap abstractions are also polymorphic 
and enable us to operate on each method of the program in isolation.
This is to be contrasted with traditional data-flow analysis~\citep{Reps:1995:PID:199448.199462}, where flow functions must be distributive and expressed on finite domains (as typically provided by store-based abstractions).


\begin{leaveout}%
  Supervisory control has been increasingly applied in various domains, such as industrial automation~\citep{Frey2000FormalMeths4PLC}, or the control of computing systems, \eg for 
deadlock avoidance
~\citep{Wang2009DeadlockAvoidWithSupervControl
}.
Contrary to the works on IFC that follow a pragmatic, programming-language approach to define security properties , these notions in the DCS community are defined in terms of event traces (input/output sequences) and studied using Petri Nets and Timed or classical Automata formalisms.
\citet{Mantel:2003:UAS:959088.959094} give a very good overview of the various approaches for defining non-interference in this context.
%
%
In the DES community, however, recent works address \emph{opacity} properties, a notion that encompasses other security-related properties such as anonymity~\citep{Mazare2004Unification4Opacity,Bryans:2008:OGT:1452811.1452815} in addition to non-interference: a system is opaque if, for any secret behavior (execution trace), there exists at least one non-secret behavior (observable execution trace) that looks indistinguishable to the intruder.
This definition is more general as it is agnostic to the observational capabilities of attackers .
Several works address the problem of opacity enforcement, although often with varying definitions of this property~\citep{%
  Takai2008SupremalContrOpaqueSublanguage,%
  Tong:2018:COE:3220369.3220373}.
\end{leaveout}

\section{Conclusion}

We have introduced a generic abstract heap domain for modeling heaps and information flow via heap for low-level object-oriented programs, and instantiated it with different families of relations. Our experiments 
showed that our instantiated heap models
improve the state-of-the-art precision, and that the precision has an inverse relationship with scalability. 
We are currently investigating the computation of 
method summaries in order to obtain a fully modular inter\-procedural IFC analysis.
\begin{leaveout}
Indeed, our approach for symbolically abstracting the heap and encoding the security semantics effectively supports the design of \emph{sound} and \emph{modular} (bottom-up) inter\-procedural IFC analyses, where the analysis of a method provides a \emph{polymorphic} artifact.
Such an artifact is directly re-usable at \emph{any call site} of the method (modulo renaming of formal arguments), thereby limiting the needs for re-analyzing methods to the cases of (mutual) recursion.
Such a modular analysis is \emph{scalable} since it focuses on one method at a time.
%
\end{leaveout}
\Guardies 
can be improved by implementing a more advanced analysis to reduce the amount of symbolic variables involved 
to represent the heap, thereby improving scalability.
Further, the instantiated heap domains are field-insensitive, and a natural 
extension is introducing support for field-sensitive~analyses.




\begin{leaveout}
See \citet{10.1145/1057387.1057391} ``Polymorphic Predicate Abstraction'', for the use of the term ``polymorphic'' to denote the symbolic variables that represent the unknown context.
\citet{MizunoSchmidt1992SecurityFlowControlAlgo} also seem to be modular via symbolic variables.

\red{One family of static approaches for IFC mostly concentrates on
  type-systems~\citep{pottier2003information,sabelfeld2003language,Barthe:2007:CLN:1762174.1762189,Hedin:2012:ISC:2354412.2355236}.
  Another line of research focuses on
  program slicing~\citep{Kam:1977:MDF:2696874.2696926
  }
  , where a large graph (often called a Program Dependence Graph---PDG) is derived from the entire application, and then used to answer queries about information flows.
  \Joana~\citep{Hammer:2009:FCO:1667545.1667547
  }, DroidSafe~\citep{GordonKPGNR15DroidSafe}, \textsc{FlowDroid}~\citep{Artz2014FlowDroid}, and \PIDGIN~\citep{Johnson2015PIDGIN} are prominent frameworks in this category.
  \green{``PIDGIN PDGs are context sensitive, object sensitive, and field sensitive. They are flow sensitive for local variables \emph{and flow-insensitive for heap locations}.''}
  \red{XXX: \Cassandra~\citep{Lortz2014Cassandra} (individually types each field declared in the app; field-types are fixed for the whole app; object- and flow-insensitive \wrt heap abstraction. What about \KeY~\citep{KeY,Darvas2005TheoremProvingW4SecureIF-KeY} (needs manual inputs)?}
  \green{All the aforementioned approaches have one key ingredient in common: they ``consume points-to information'' provided by some other analysis.}


  About traditional models of the heap: Even points-to analyses ``rarely'' path- or flow-sensitive according to \citet{Gresh2018ShootingFromTheHeap} (yet they provide a flow-insensitive points-to analysis).}

\citet{NikolicSpoto2014ReachabilityAnalysisofProgramVariables} is a whole-program analysis, but is relevant in that they over-approximate flow-sensitive variable reachability information via abstract-interpretation; implemented in juliasoft, with many client analyses whose precision and efficiency is then improved ().  Abstract domain is simply a set of variable pairs, an extensional representation of the variable reachability relation.  Can also reuse pre-established facts about possible sharing and definite aliasing.  Not strictly related to security aspects.

\green{The only work we have found that does consider implicit flows with a flow-sensitive heap abstraction was developed by \citet{Zanioli2012SAILS}
; it however only operates intra-procedurally and on high-level structured programs.}
\citet{Zanioli2012SAILS,Cortesi2018CombiningSymbolicandNumericalDomainsforInformationLeakageAnalysis}: abstract interpretation approach with encoding of all variable dependencies via propositional formula and fixpoint over Pos domain for each statement --- specific to two-level lattice; re-uses TVLA-based heap abstract domain to dwal with \Java; implicit flows as well; structured programs only.
\end{leaveout}


\begin{leaveout}
What about optimization by limiting alias info to \emph{relevant context} (see \citet{Chatterjee1999RelevantContextInference})?
See \citet{ChengHwu2000PointsToSummariesWithAccessPaths} as well, similar to RCI but with summary conditions represented as access paths instead (Context insensitive modular points-to analysis for C with function pointers).

See \citet{CousotCousot2002ModularStaticProgramAnalysis}, Section 7, for a link to a seminal work on the Symbolic Relational Separate Analysis method (except that we use a backwards analysis to simultaneously compute the pre-condition and post-conditions).
\end{leaveout}



\subsubsection*{Acknowledgements} 
The first author was supported by the UK Engineering and Physical Sciences Research Council (EPSRC) through grant EP/M027287/1, and the second author was supported by the Swedish Knowledge Foundation (KKs) via the grant 20160186.

\ifieetranloaded\else           
\bibliography{IEEEabrv,ifc,sa,mc,des,rv,other}
\fi
\ifacmartloaded
\appendix
\else
\ifllncsloaded
\appendix
\else
\newpage
\ifieetranloaded
\printbibliography%
\appendices
\fi
\fi

\section{Proof of Secure Heap Abstraction}\label{sec::secure.heap.asbtraction-proof}
\subsection{Sound Security Typing}
We first define the soundness of security typing for an abstract heap 
\sHeap from a domain defined using aliasing and field-aliasing relations as follows:
\newcommand\objLevelAliasWF{\ensuremath{φ_{\mathit{\AliasRelSymbol=}}}\xspace}
\newcommand\objLevelFieldAliasWF{\ensuremath{φ_{\mathit{\FieldAliasXRelSymbol⊒}}}\xspace}
\begin{definition}[Sound Security Typing]
  \label{def:deepalias-sound-heap-typing-env}
  \def\hd{d}%
  An abstract heap 
  \(\sHeap \in \HeapDom \hd \) for \(\hdval \hd ∈ ｛\hdeep, \hshal, \hdumb｝\),
  is \emph{correctly typed}
  iff:
  \begin{itemize}[nosep,left=0pt]
  \item any pair of aliasing references
    have
    identical  security levels:\\
    \(\red{\objLevelAliasWF} ≝ ∀(r,s) ∈ R^2, \sHeap \models  \AliasRel[\sHeap] r s ⇒ \sHeap \models  \hlvl[\sHeap] r = \hlvl[\sHeap] s\mbox;\) and
  \item
    the security level associated with a reference
    may only be greater or equal than that of the references its
    fields may (transitively) alias:\\
    \(\objLevelFieldAliasWF ≝ ∀(r,s) ∈ R^2, \sHeap \models \FieldAliasXRel[\sHeap] r s
    ⇒ \sHeap \models \hlvl[\sHeap] r ⊒ \hlvl[\sHeap] s\mbox.\)
  \end{itemize}
\end{definition}
A predicate is an inductive invariant for a heap transformer of \SymbolicAbstractHeapDom R, if it is preserved by the transformer:
\begin{definition}[Inductive Invariant]
	A predicate \(φ\)
	 is an \emph{inductive invariant} for a transformer 
	\(T \in \Transformers \hd \) iff
		\(∀\sHeap∈\HeapDom \hd , (\sHeap ⊨ φ ⇒ \Eval{
		  {T}}{\sHeap} ⊨	φ)\mbox.\)

\end{definition}

We say a predicate is an inductive invariant for a heap domain \SymbolicAbstractHeapDom R, if it is an inductive invariant for all its transformers.
\begin{proposition}\label{prop:sound.object.levels}
	The predicates \objLevelAliasWF and
	\objLevelFieldAliasWF are inductive invariants for the abstract heap domain \SymbolicAbstractHeapDom [\hdeep] R.
\end{proposition}

\begin{leaveout}
\paragraph{Relations among heaps of the same domain}
\[
\sHeap ⊑_{\HH'} \hvar' ≝
\ ⋀_{\mathclap{r∈R'}}{\hlvl[\sHeap]r ⊑ \hlvl[\hvar']r}\ ∧
⋀_{\mathclap{\AliasRel r s ∈\AliasVars[\hvar']}}{\AliasRel[\sHeap] r s ⇒ \AliasRel[\hvar'] r s}\ ∧
\ ⋀\nolimits_{
	{\FieldAliasXRel r s ∈\FieldAliasXVars[\hvar']}}{\FieldAliasXRel[\sHeap] r s ⇒ \FieldAliasXRel[\hvar'] r s}
\]

\paragraph{Relations among heaps of different domain}
\todo[inline]{Consider one general relation by making the union of different relations and generalize the definition to the subsumption. Then one can define some relations between two domains/}

\end{leaveout}


\subsection{Proof of Theorem~\ref{thm::secure.heap.model}}

\begin{proof}
	Let \({\mHeap_1, \mHeap_2} \sqsubseteq { \HeapDom\concrete} \),  \(\sHeap ∈  { \HeapDom\hdeep}\), \({\mHeap_1 =_{\sHeap} \mHeap_2}\) where { \HeapDom\concrete} is the set of concrete heaps and { \HeapDom\hdeep} is the set of abstract heaps from the domain \hdeep.
	According to Def.~\ref{def::secure.heap.abstraction},  we should prove that
	for any operation \(( \mathit{as},\mathit{as}') \) of reference assignment statements and
	their corresponding operations on abstract heaps where \(\mathit{as} \in \{\s{r = s}, \s{r = s.f_r}, \s{r =	\Null} \} \), \({\mHeap'_1 =_{{\sHeap'}} \mHeap'_2}\)
	holds   where
	\(\mHeap'_i= \MemHeapSem {\mathit{as}}{\mHeap_i} \), $i \in \{1,2\}$, and \(\sHeap' = \NoHeapEffect{\mathit{as}'}{\sHeap}\).

	Let  \(\RefGraph{\mHeap_i}{R} ≝ \tuple{\Nodes {\mHeap_i}, \Edges{\mHeap_1}}, i \in \{1,2\}\). From \({\mHeap_1 =_{{\sHeap}} \mHeap_2}\), it follows that \(\RefGraph{\mHeap_1}{\low} \cong \RefGraph{\mHeap_2}{\low}\) from Definition~\ref{def::heap-low-equivalence-relation}. We  prove that \(\RefGraph{\mHeap_1'}{\low} \cong \RefGraph{\mHeap'_2}{\low}\) by a case analysis on $\mathit{as} $.
	Since, the proof of all cases are very similar and follows the same strategy, we limit ourselves to proving the case of reference assignment.
	For the proof of \(r = s \), two cases can happen:
	\begin{enumerate}[(i)]
		\item If $  \hlvl[\hvar] s = \low$, this means that $s \in \Nodes{{\mHeap_i}}$, $i \in \{1,2\}$. The truth value of the variables $\AliasRel {r'} s$ and $\FieldAliasRelConcrete {s} {r'} f$ respectively determine the presence of the edges \tuple{r',∼,s} and \(({s,\FieldAliasRelConcrete {}{}{f},r'})\) in the reference graph according to the definition of the reference graph. For any reference $r'$, $\tuple{s,\GenericRelSymb,r'}\in \Edges{{\mHeap_1}}$ iff
		$\tuple{s,\GenericRelSymb,r'} \in \Edges{{\mHeap_2}}$ according to \(\RefGraph{\mHeap_1}{{\low}} \cong \RefGraph{\mHeap_2}{\low}\)
		where $\GenericRelSymb \in \{\EquivSymbol, \FieldAliasRelConcrete {}{}{f} \}$.
		Hence, $\mHeap_1 \models {s \GenericRelSymb r'} \Leftrightarrow \mHeap_2 \models {s \GenericRelSymb r'}$, and
		$\mHeap_1 \models {r'\GenericRelSymb s} \Leftrightarrow \mHeap_2 \models {r' \GenericRelSymb s}$, for all $r' \in \R$, and the updates of \ICopyRef{r}{s} {\mHeap_1} and \ICopyRef{r}{s}{\mHeap_2} will be the same in both heaps (i).

		Form Definition~\ref{def::heap-low-equivalence-relation} (ii) and  \({\mHeap_1 =_{{\sHeap}} \mHeap_2}\), it follows that the primitive fields of $s$ hold the same valuations in both heaps, \ie $\forall x.~ \mHeap_1 \models	\RefContent{s}{f_p}= x \Leftrightarrow  \mHeap_2 \models	\RefContent{s} {f_p}=x $, for all $f_p \in \PrimFields s$ (ii).
		Furthermore, \hlvl[\hvar'] r will  be set to \hlvl s according to \figurename~\ref{fig::Symmaries.Abstract.Heap.Operations} (iii).
		From \({\mHeap_1 =_{{\sHeap}} \mHeap_2}\) and (i)-(iii), it follows that \({\mHeap'_1 =_{{\sHeap'}} \mHeap'_2}\).

		\item
		If  $ \hlvl[\hvar] s = \high$, this means that $s \notin \Nodes{{\mHeap_i}}$, $i \in \{1,2\}$ according to the definition of reference graphs, and Proposition~\ref{prop:sound.object.levels}. If $ \hlvl[\hvar] r = \high$, then $r \notin \Nodes{{\mHeap_i}}$  and subsequently \(\RefGraph{\mHeap_i}{\low}\) will not be updated by \ICopyRef{r}{s}{\mHeap_i}, \(i \in \{1,2\}\), \ie \(\RefGraph{\mHeap'_i}{\low} \cong \RefGraph{\mHeap_i}{\low}\) (i).
		If $\hlvl[\hvar] r = \low$, then $r$ will be removed from  \(\RefGraph{\mHeap'_i}{\low}\), \(i \in \{1,2\}\), because  all its connections will be updated to the corresponding ones of $s$ according to \ICopyRef{r}{s}{\mHeap_i}.
		Since, \(\RefGraph{\mHeap_1}{\low} \cong \RefGraph{\mHeap_2}{\low}\), it's obvious that the updated graphs will remain isomorphic (ii).
		Furthermore, \hlvl[\hvar'] r will  be \hlvl s according to \figurename~\ref{fig::Symmaries.Abstract.Heap.Operations} (iii).
		From \({\mHeap_1 =_{{\sHeap}} \mHeap_2}\) and (i)-(iii), it follows that \({\mHeap'_1 =_{{\sHeap'}} \mHeap'_2}\).
	\end{enumerate}
	For the proof of field load \(r=s.f_r\), we perform a case analysis on  \hlvl[\hvar] s, and the proof is similar to the one above.
	For the case \( r=\Null\), $\hlvl[\hvar] r$ will be updated to $\bot$ and $r$ will be included in  \(\RefGraph{\mHeap'_i}{\low}\), as an isolated node, according to 	\IResetRef{r \leftarrow .}{\mHeap} 	and \NullRef r \sHeap (i). Hence, from (i) and the fact that \(\RefGraph{\mHeap_1}{\low} \cong \RefGraph{\mHeap_2}{\low}\), it obviously follows that \(\RefGraph{\mHeap'_1}{\low} \cong \RefGraph{\mHeap'_2}{\low}\).

	We should also consider any mutation operation $\mathit{mu}$. To prove these, we do a case analysis on $\mathit{mu}$. We prove the case of a reference field store that is more complex, \ie \(r.f_r=s\) where ${\hlvl[\sHeap] s  } \sqsubseteq l$. 	For any reference $x$ \st \AliasRel x r ∨ \FieldAliasXRel x r holds, $\hlvl[ h ]x$ will  be  upgraded by $l$. Two cases can happen:
	\begin{itemize}[(i)]
		\item[\(l=\low\)] From ${\hlvl [\sHeap] s } \sqsubseteq l$, it follows that \hlvl[\sHeap] s = $\low$ and \( s \in \RefGraph{\mHeap_i}{\low}\).
		Since, $l=\low$, this means that no security level will be updated, \ie the nodes of \RefGraph{\mHeap'_i}{\low}  and \RefGraph{\mHeap'_i}{\low} will be the same (i).

		If \hlvl[\sHeap] r = \low, we can show that the updates of \IStoreRef r {f_r} s {\mHeap_1} and \IStoreRef r {f_r} s {\mHeap_2} will be the same in both concrete heaps (iii), similar to the case of reference assignment. Hence, from (i) and (iii), we can conclude that \(\RefGraph{\mHeap'_1}{\low} \cong \RefGraph{\mHeap'_2}{\low}\).

		If  \hlvl[\sHeap] r = \high, then $r \notin \RefGraph{\mHeap_i}{\low}$ according to the definition of reference graph.	From Proposition~\ref{prop:sound.object.levels}, it follows that \(\hlvl[\mHeap_i] x ⊒ \hlvl[\mHeap_i] r\), and hence, \(\hlvl[\mHeap_i] x =\high\) and \( x \notin \RefGraph{\mHeap_i}{\low}\) (ii).
		From (i) and (ii), it follows that $x \notin \RefGraph{\mHeap'_i}{\low}$ and $r \notin \RefGraph{\mHeap'_i}{\low}$, \ie \( x \in \RefGraph{\mHeap'_i}{\low}\) remains unmodified, and consequently \(\RefGraph{\mHeap'_1}{\low} \cong \RefGraph{\mHeap'_2}{\low}\).

		\item[\(l = \high\)] If $l=\high$, then \(\hlvl[\mHeap'_i] x =\high\) and consequently \( x \notin \RefGraph{\mHeap'_i}{\low}\). According to \HeapUpgradeObjLevel[\sHeap] r l, the level of any reference $x$ that aliases or transitively field-aliases $r$ will be upgraded to \high, and consequently will not belong to \RefGraph{\mHeap'_i}{\low} anymore.
		Informally, this means that $r$ and all references that can reach $r$ will be removed from  \RefGraph{\mHeap_i}{\low}.
		Since \(\RefGraph{\mHeap_1}{\low} \cong \RefGraph{\mHeap_2}{\low}\), we can conclude that this set of references and their incoming/outgoing edges will be the same in both concrete heaps.
		Hence, removing them from the reference graphs will still lead to isomorphic graphs, \ie \(\RefGraph{\mHeap'_1}{\low} \cong \RefGraph{\mHeap'_2}{\low}\).
	\end{itemize}
	Since, no primitive field is updated by a reference field store, the condition (ii) in Definition~\ref{def::heap-low-equivalence-relation} will still hold for the updated heaps ${\mHeap'_i}, i \in \{1,2\}$.

	The proof for  the case of the primitive field store \(r.f_p=v\) is similar to that of the reference field store. We prove \(\RefGraph{\mHeap'_1}{\low} \cong \RefGraph{\mHeap'_2}{\low}\) by performing case analysis on $l$.
	According to \IUpdatePrimFields {\mHeap} {op}, the primitive field $f_p$ of all references aliasing $r$ will be updated to $v$ in both heaps \({\mHeap'_i}\). This means that for any reference \(r'\) that belongs to \RefGraph{\mHeap_i'}{\low}, the updates to its primitive field $f_p$ will be the same in both heaps.
\end{proof}

\begin{leaveout}
\subsubsection{Symbolic Control Flow Graph}\label{sec::scfg}

\begin{definition}[Symbolic Control Flow Graph]
	A \emph{symbolic control flow graph} is a tuple \(\G = \SCFGtupleI
	I\) where:
	\begin{mathdesc}[nosep]
		\item[Λ] is a finite non-empty set of \emph{locations};
		\(X\) is a set of \emph{state variables}
		;
		\(I\) is a set of \emph{input variables};
		\item[Δ: Λ → 
		（\PDomX{X⊎I} × \UDom{X}{X⊎I} × Λ）^p, p ∈ ℕ^+,] is a total mapping
		from locations into 
		non-empty sets of \emph{symbolic transitions} \(δ = 〈g,u,ℓ'〉\).
		\(ℓ'∈Λ\) denotes the \emph{destination location} of δ, \(g ∈
		\PDomX{X⊎I}\) is a \emph{guard} for δ, and \(u ∈ \UDom{X}{X⊎I}\)
		is its associated \emph{update function} for state variables.
		The latter gives values to be assigned to a possibly empty subset
		of the state variables based on valuations for both \(X\) and
		\(I\);
		\ifSCFGshaveassertions{%
			\item[A: Λ → \PDom{X⊎I}] associates a predicate to each
			location, that encodes an assertion on the possible values of
			the inputs depending on the current memory variables;%
		}
		\item[ℓ_0 ∈ Λ] is \emph{the initial location}; and
		\item[x_0 ∈ \PDomX X] symbolically describes \emph{the set of
			initial valuations} for the state variables.
	\end{mathdesc}
\end{definition}
We formally define the semantics of an SCFG as a corresponding
automaton.
Given \(e ∈ \GEDomX\domain V\) and a valuation \(μ∈\Val{V⊎W}\), the
\emph{interpretation of \(e\) \wrt μ}, noted \Eval{e}{μ}, belongs to
\domain and can be computed according to the usual semantics of the
operators.
For convenience, we denote \(\TotalFunc u ∈ \UDom V {V⊎W}\) an
update function \(u\) lifted to a full domain \(V\) by adding
assignments \(v ≔ v\) for each \(v\) for which \(u\) is undefined.
\begin{definition}[SCFG Semantics as an Automaton]\label{def:SCFG.concrete.semantics}
	Given a reactive and deterministic SCFG \(\G = \SCFGtupleI{I}\), we
	define the corresponding \emph{automaton} \(\FSM{\G} = \FSMdef\)
	where:
	\begin{mathdesc}[nosep]
		\item[\Q = Λ × \Val X] is the \emph{state space} of \FSM{\G};
		\item[\I = \Val I] is the \emph{input space} of \FSM{\G};
		\item[\T: \Q × \I → \Q] is the \emph{transition function} defined as
		\((ℓ,x,ι) ⟼ \big(ℓ',\Eval{\TotalFunc u}{x⊎ι}\big)\) where
		\(〈g,u,ℓ'〉\) is the \emph{unique} transition \(Δ(ℓ)\) \st \(x⊎ι
		⊨ g\); \ifFSMshaveassertions{%
			\item[\A ⊆ \Q × \I] \(= ｛(ℓ,x,ι) ∈ \Q × \I｜x⊎ι ⊨ A(ℓ)｝\mbox{;}\)%
		}
		\item[\qstate_0 = ｛(ℓ_0, x)｜x⊨x_0｝] is the set of initial states
		of \FSM{\G}.
	\end{mathdesc}
\end{definition}
\FSM{\G} is initially in \(\qstate_0\).
Assuming that \FSM{\G} is in a state \(q = (ℓ,x) ∈ \Q\) (\ie in
location \(ℓ\), and with valuation of memory variables \(x\)),
\FSM{\G} evolves to state \(q' = \T(q,ι)\) upon reception of an
input \(ι ∈ \Val I\)%
\ifSCFGshaveassertions{ such that \((q,ι) ∈ \A\) (\ie ι is an
	admissible valuation for \(I\) in state \(q\))}.
\end{leaveout}

\renewcommand{\R}{\ensuremath{R}\xspace}
\renewcommand{\mHeap}{\mHeapVal}

\section{Proof of Noninterference}\label{sec::noninterference-proof}
 We use the notation $\qstate(a)$ to show the value of $a$ in the state $\qstate$, use $ \qstate(\hlvl \sHeap)$ to denote $ \qstate(\sHeapValLevs {\X}  \sHeap)$, and 
  $\qstate(\sHeap)$ to show $\qstate({\X_1}_{\sHeap})$.
Furthermore, we say   {$\qstate_1(\hlvl \sHeap) \sqsubseteq \qstate_{2}(\hlvl \sHeap)$}, if and only if for all $r \in R$, $\qstate_1(\hlvl[\sHeap] r) \sqsubseteq \qstate_2(\hlvl[\sHeap] r)$.
 To prove  Theorem~\ref{thm::Non.Interference}, we first need to define the notion of compatible states.

  \begin{definition}[Compatible States]\label{def::compatible.states}
  	We say two states \( \qstate_1\) and \( \qstate_2\) are {compatible},
  	denoted by \(\qstate_1 \approx \qstate_2 \), iff
  	\begin{enumerate}[(1)]
  		\item\label{compatible.states.item:same.locations} \(\qstate_1(ℓ)=\qstate_2(ℓ)\),
  		\item\label{compatible.states.item:same.memory.typing.environment} \(\qstate_1({\VarsTypes})=\qstate_2(\VarsTypes)\),
  		\item\label{compatible.states.item:low.equiv.memory} \(\qstate_1(V)=_{\qstate_1(\VarsTypes)}\qstate_2(V)\),
  		\item\label{compatible.states.item:same.heap.typing.environment} \(\qstate_1({\hlvl{\sHeap}})=\qstate_2(\hlvl{\sHeap})\),
  		\item\label{compatible.states.item:low.quiv.heap}  \({\qstate_1(\mHeap) =_{\qstate_1({\sHeap})} \qstate_2(\mHeap)}\),
  	\end{enumerate}
  \end{definition}


\nakh{Note that we assume that the information-flow summaries provided by the user are correct, \ie the summary update preserves the low-equivalence relation in the low context.}
  \begin{proof}
    To prove this theorem, we need to find a witnessing bisimulation relation $\InRelation$ that witnesses $\qstate_1 \lowbisim \qstate_2$. We define $\InRelation$ as the following and prove that it is a bisimulation relation:
    \[
      \InRelation=\left\{ \langle \qstate,\qstate' \rangle~|~
        \begin{array}{@{}ll@{}}
          \qstate \approx \qstate' \wedge \gamma(\qstate,\qstate')
        \end{array}%
      \right\}
    \]
    where \begin{equation}\label{compatible.states.item:same.pc.mode}
    \begin{array}{lll}
      \gamma(\qstate,\qstate') &=& \qstate(\mode)=\qstate'(\mode)=\ff \wedge\\
&&\qstate(\pc) =\qstate'(\pc) = \low \wedge \qstate(\rVar)=\qstate'(\rVar)=\rNom.
    \end{array}
    \end{equation}

    The initial states $\qstate_0$ and $\qstate'_0$ are obviously in the relation $\InRelation$ according to Def.~ \ref{def::low-bisimulation-relation}, $\qstate_0\models{ x_{0\mbox-\mathit{all}}}$ and $\qstate'_0\models{ x_{0\mbox-\mathit{all}}}$.
    Let $\langle \qstate, \qstate' \rangle \in \InRelation$. According to Def.~ \ref{def::low-bisimulation-relation}, we should prove
    %
    that,  if $\qstate\xrightarrow{o}_* t$,
    then either (a) there exists $t'$ such that $\qstate' \xrightarrow{o}_* t'$
    and $t \InRelation t'$,
    or (b) $\qstate' \xrightarrow{\low}_* $,
    and vice versa.
    If no observation is made from $\qstate'$, then the case (b) obviously holds.
    If an observation is made after $\qstate'$, then the conclusion is followed from Lemma~\ref{lemm::single.observation} which states that $t$ and  $t'$ are in the relation \(\InRelation\) too.
  \end{proof}

  We call a state \( \qstate \) a branching state, iff \(\qstate(\ell)=({\cif~(e)~\cgoto~l};\_,\_)\).
  A \emph{branch execution}  is an execution $\qstate_0{\overset{\rho}{\rightarrowtail}} \qstate_n = \qstate_0 \to \qstate_{1} \to \ldots \to \qstate_{n}$ in the CDR $\rho$ that starts in a branching state inducing a CDR $\rho$
  and exits from $\rho$, \ie $\qstate_{n-1}(\ell) =(\junc{}{\rho},\jsone)$ and $\qstate_{n}(\ell) =(\junc{}{\rho},\jstwo)$.

  \begin{lemma}[Single Observation]\label{lemm::single.observation}
    Let   \(\qstate_1 \InRelation \qstate_2 \).
    If $\qstate_1	\xrightarrow{o}_* \qstate'_1$
    and	$\qstate_2	\xrightarrow{o}_* \qstate'_2 $, then \(\qstate'_1 \InRelation \qstate'_2 \).
  \end{lemma}

  \begin{proof}
    We prove this by induction on the length of $\qstate_1	\xrightarrow{o}_* \qstate'_1$.
    Since, the execution does not terminate in $\qstate_1$, this means \(\qstate_1(\ell) \neq (\surd,\_) \) and \(\qstate_1(\ell) = (\stm;\_,\_) \).\\
    \textbf{Base Case} The base case (\ie the length of one) happens when $\stm$ is an output statement like \coutput l (\elts).
  \item	\red{In case of \coutput l (\elts), \plvl \elts can not be high-sensitive according to the invariant $φ(ℓ) $ in the rule \OutputRule.} 
    Hence, according to the conditions \ref{compatible.states.item:low.equiv.memory} and \ref{compatible.states.item:same.memory.typing.environment} in Def.~\ref{def::compatible.states}, the observations should be the same in both states.
    Since it does not update the state, apart from the symbolic location, the target states will remain compatible as well.

    \noindent
    \textbf{Inductive Step}  Let it hold for all executions $\qstate_1	\xrightarrow{o}_* \qstate'_1$ with the length of \(m\) or less than \(m\). We should prove the lemma for an execution with the length of  \(m+1\).

    Let \stm  be a junction point. From   \(\qstate_1 \InRelation \qstate_2 \) and the condition in Eq.~\ref{compatible.states.item:same.pc.mode}, it follows that \(\qstate_i(\mode) = \ff\) and \(\qstate_i(\rVar)=\rNom, i \in \{1,2\}\). According to \JunctionRule and \mJunctionRule, the first transition of \mJunctionRule is enabled in both states that leads to changing the symbolic location (Note that $P_J$ is obviously the same in both states).

    If  \stm  is not a junction point, we perform a case analysis on \stm .
    \begin{enumerate}[(i)]
    \item Let \stm  be an assignment where $\qstate_1	\to t_1$  and $\qstate_2	\to t_2$. From Lemma~\ref{lemm:assignments}, it follows that \(t_1 \InRelation t_2 \). The conclusion is derived by applying the inductive hypothesis on $t_1$ and $t_2$.


    \item
      \red{Let \stm  be an output statement  \coutput l (\elts)}. Since these states have not been avoided by the security guards,  this means that the invariant in the rule \OutputRule holds in $\qstate_i, i \in \{1,2\}$.
      However, observations should be produced in the last transition of this execution according to the definition of $	\xrightarrow{o}_* $. Since, this transition is not the last one, this means that \stm cannot produce any observation, \ie the invariant  in the rule \OutputRule  does not hold and \stm cannot be an output statement.

    \item\sloppy Let \stm  be \cgoto~\lbl{label}. Since, only the semantic location will change to \semloc{ \Target{}{label}, \jsone} according to \GotoRule and \mStmRule, it is obvious that \(t_1 \InRelation t_2 \), where $\qstate_1	\to t_1$  and $\qstate_2	\to t_2$. The conclusion is derived by applying the inductive hypothesis on $t_1$ and $t_2$.

      \item \red{ Let \stm  be a method call \(r.m'(\lits)\).  
      The update comprises executing a sequence of non-interfering assignments. We can prove that the two states after applying each assignment individually will still remain in relation, in a similar fashion to the proof of Lemma~\ref{lemm:assignments}.
    {Since the assignments by the summary update are non-interfering, it's trivial to show that the relation will be preserved after applying the update. The conclusion is derived by applying the inductive step hypothesis on the states after the update. }
    }
    
    \item Let \stm  be \cif~(e)~\cgoto~\lbl{label}. From $\qstate_1 \InRelation \qstate_2$ and the condition in Eq.~\ref{compatible.states.item:same.pc.mode}, it follows that \pc=\low and \mode = \ff in both states. Two cases can happen:
      \begin{itemize}
      \item    If \(\qstate_i (\plvl e) = \low\), from $\qstate_1 \InRelation \qstate_2$ and the condition~\ref{compatible.states.item:low.equiv.memory} in Def.~\ref{def::compatible.states}, it follows that $\qstate_1 \models e $ iff $\qstate_2 \models e$, \ie the same branch will be taken from both states. Let they respectively evolve into states $t_1$ and $t_2$. According to \mBranchRule and \BranchRule, the semantic location will change to the same value in both cases and no other update will be done, \ie it follows that $t_1 \InRelation t_2$.
        The conclusion is derived by applying the inductive hypothesis on the executions $t_1 \xrightarrow{o}_* \qstate'_1$ and $t_2\xrightarrow{o}_* \qstate'_2$.

      \item If $ \qstate_i(\plvl e) \,\not⊑ \qstate_i(\pc)$ holds, this means that \(\qstate_i(\plvl e) = \high\). Let $\qstate_i \to s_i, i \in \{1,2\}$ and $\rho$ be the CDR induced by \stm. According to \BranchRule and \mBranchRule, $s_i(\pc)= \qstate_i(\pc )\sqcup \qstate_i(\plvl e) = \high$. From Lemma~\ref{lemm::high.context.invariant1}-\ref{item.high.cond.pc.constraints}, \pc of all states of $\rho$'s execution is $\high$.
        This means that the execution in $\rho$ will lead to no observation, as the invariant in the rule \OutputRule  
        would have been violated, since $\pc \neq \low$.
        This execution will finally lead to an observation, according to the premises, \ie its \pc will become \low.
        {According to Lemma~\ref{lemm::high.context.invariant1}-\ref{item.high.cond.pc.constraints}, \pc will reset to \(\qstate_{i}(\pc) \) in the junction, \ie $\rho$'s junction will be visited and
          $\qstate_i	 {\overset{\rho}{\rightarrowtail}} t_i$  for some $t_i, i \in \{1,2\}$.}
        According to Lemma~\ref{lemm::branch}, it follows that \(t_1 \InRelation t_2 \). The conclusion is derived by applying the inductive hypothesis on $t_1$ and $t_2$.
      \end{itemize}


    \end{enumerate}
  \end{proof}

\begin{lemma}[Assignment Preserves Relation]\label{lemm:assignments}
  Let   \(\qstate_1 \InRelation \qstate_2 \) where \(\qstate_i(ℓ) = (\stm;\_,\_), i \in \{1,2\} \) and \stm is an assignment. If $\qstate_1	\to \qstate'_1$ and $\qstate_2	\to \qstate'_2 $, then \(\qstate'_1 \InRelation \qstate'_2 \).
\end{lemma}

\begin{proof}
  Case analysis on \stm and $i \in \{1,2\}$.

  \begin{enumerate}[(i)]
  \item\label{compatible.states.item:lem.assignments.primitives} \(v=e\): From $\qstate_1 \InRelation \qstate_2$, the conditions \ref{compatible.states.item:same.memory.typing.environment} and \ref{compatible.states.item:same.heap.typing.environment} in Def.~\ref{def::compatible.states} and the rule \mAssignRule in Fig.~\ref{fig::full.semantics}, \(\qstate_i(\VarsTypes)\) will be updated to the same value of \(\qstate_i(\plvl{e}) \sqcup \qstate_i(\pc)\), \ie the conditions \ref{compatible.states.item:same.memory.typing.environment} and \ref{compatible.states.item:same.heap.typing.environment} in Def.~\ref{def::compatible.states} will hold.
    Note that the heap will not be affected.

    If \(\qstate_i(\plvl{e})=\low\), this means that \(e\) contains no high-sensitive variables. Since, the low-sensitive variables have the same values in both states based on Def.~\ref{def::compatible.states}-\ref{compatible.states.item:low.equiv.memory} and Def.~\ref{def::compatible.states}-\ref{compatible.states.item:low.quiv.heap}, it follows that \(v\) will be updated to the same value in both cases, and the conditions \ref{compatible.states.item:low.equiv.memory} and \ref{compatible.states.item:low.quiv.heap} of Def.~\ref{def::compatible.states} will still hold for \(\qstate'_1\) and  \(\qstate'_1\).
    If \(\qstate_i(\plvl{e})=\high\), \(\qstate'_i(v)\) might differ. However, the conditions \ref{compatible.states.item:low.equiv.memory} and \ref{compatible.states.item:low.quiv.heap} of Def.~\ref{def::compatible.states} will still hold for \(\qstate'_1\) and  \(\qstate'_2\) according to Def.~\ref{def::low-equivalence-relation}, as  \(\qstate'_i(\VarsTypes)=\high\).

    The variables \pc, \rVar, \mode will remain unchanged, \ie  Eq.~\ref{compatible.states.item:same.pc.mode}
     will still hold. According to the semantics rule \mAssignRule, $ℓ$ will be updated to the same value in both states, \ie the condition~\ref{compatible.states.item:same.locations} will hold. Hence, all the conditions in Def.~\ref{def::compatible.states} will hold, and consequently, \(\qstate'_1 \InRelation \qstate'_2\).

  \item \(v = r.f_p\): Similar to the case (i), we can prove that all the conditions in Def.~\ref{def::compatible.states}  will hold  for the pair of $\qstate'_1$ and $\qstate'_2$.

  \item For the cases of \(r=s\),  $r = \cnew~c$,   \( s = r.f_r\),  \(r.f_r = s\) ,  \(r.f_p = e\) and $r = \Null$, we can show that the condition \ref{compatible.states.item:same.locations}  and  Eq. \ref{compatible.states.item:same.pc.mode} will hold in the new states similar to the first case.
    The conditions \ref{compatible.states.item:same.memory.typing.environment} and \ref{compatible.states.item:low.equiv.memory} {are followed from the fact that local primitive variables are not modified.}

    If \(r=r'\), \( s = r.f_r\), \(r.f_p = e\) or $r = \Null$, the conditions~\ref{compatible.states.item:same.memory.typing.environment}-\ref{compatible.states.item:low.quiv.heap}  are followed from Theorem~\ref{thm::secure.heap.model}.
    If $r = \cnew~c$, then $l$ in Def.~\ref{def::secure.heap.abstraction} will be set to $\pc$ according to the rule \AssignRule.
    If $s = r.f_r$, then $l=\nomlvl{\plvl s ⊔ \hlvl[\hvar] s } $ according to $T_\stm$ in \AssignRule. 		 Since, $\qstate_i(\pc)=\bot$ and $\mode=\bot$, then $l={\qstate_i(\plvl s) ⊔ \qstate_i(\hlvl s) } ⊔ \pc$ based on the definition of   \nomlvl ., \ie $l \sqsubseteq {\qstate_i(\plvl s) ⊔ \qstate_i(\hlvl s) }$ (i).
    Similarly, if $r.f_p = e$, then $l=\nomlvl{\plvl e} $, and subsequently, $l\sqsubseteq \qstate_i({\plvl e}) $ (ii). From (i), (ii) and Theorem~\ref{thm::secure.heap.model},  the conditions~\ref{compatible.states.item:same.memory.typing.environment}-\ref{compatible.states.item:low.quiv.heap}  are followed.
    Hence, all conditions hold,  we  conclude that   \(\qstate'_1 \InRelation \qstate'_2\).
  \end{enumerate}
\end{proof}


\newcommand{\context}{\ensuremath{\mathsf{context}}}

\newcommand{\regions}{\ensuremath{\Lambda}}

%

\begin{lemma}[High-Context Branch Invariants]\label{lemm::high.context.invariant1}
  Let $\qstate_0 \overset{\rho}{\rightarrowtail} (\qstate_n)$ be a (possibly finite) branch execution where \(\qstate_1(\pc) \sqcup \qstate_1(\plvl e)=\high\). For $1 \leq i( < n)$, the following properties hold:
  \begin{enumerate}
  \item\label{item.high.cond.pc.constraints} ${\qstate_i(\pc) =\high}$ and ${\qstate_i(\pc') =\qstate_0(\pc)}$ (and ${\qstate_n(\pc) =\qstate_0(\pc)}$).
  \item\label{item.high.cond.hr.mode.constraints1} ${\qstate_i(\rVar) =P_\rho}$.
  \item \label{item.high.cond.hr.stack}  If $\qstate_0 \overset{\rho}{\rightarrowtail} \qstate_n$, then
    \begin{enumerate}
    \item \label{item.high.cond.hr.stack.junctio.exists} 	there exists $j <n$ such that (i)
      for all $0 < k\leq j$,  $\qstate_k(\sHeap')=\qstate_0(\sHeap)$ and $\qstate_k(\mode) =\bot$,
      (ii) $\qstate_{j+1}(\sHeap)=\qstate_0(\sHeap)$,
      (iii) for all $j < k <n$,  $\qstate_k(\sHeap')=\qstate_j(\sHeap)$ and $\qstate_k(\mode) \neq\bot$, and
    \item
     \(	\qstate_n(\rVar)=\rNom\), and $\qstate_n(\mode) =\bot$.
    \end{enumerate}
  \end{enumerate}
\end{lemma}
\begin{proof}
  Trivial by induction on $i$ and then a case analysis on the statements.
  In the proof of item \ref{item.high.cond.hr.stack.junctio.exists}, we use contradiction to show that $j$ exists.

\end{proof}

\begin{lemma}[Updates in High-Context ]\label{lemm::high.context.invariants.general.vars}
  Let $\qstate_0 \overset{\rho}{\rightarrowtail} \qstate_n$ where \(\qstate_1(\pc) \sqcup \qstate_1(\plvl e)=\high\). For all $0 \leq i < n$, $i \neq j$ where $0 < j < n$, $\qstate_j(\mode) =\bot$, \(\qstate_j(\rVar)=\rMode{ρ}\) and \(\qstate_j (ℓ )=\semloc{\_, \jsone}\), the following properties hold:
  \begin{enumerate}
  \item\label{item.high.cond.upgrading.vars} For all $a\in\PrimVars$, $\qstate_i(\plvl a)  \sqsubseteq \qstate_{i+1}(\plvl a)$, and for all  $r \in R$, $\qstate_i(\plvl r)  \sqsubseteq \qstate_{i+1}(\plvl r)$ and $\qstate_i(\hlvl r)  \sqsubseteq \qstate_{i+1}(\hlvl r)$.
  \item\label{item.high.cond.unmodified.low.vars}  For all $a \in \PrimVars \cup \R$, $\qstate_{i+1}( a)=\qstate_{i}( a)$  if $\qstate_{i+1}(\plvl a)= \low$.

  \item\label{item.high.cond.unmodified.low.vars.in.whole.execution}  For all $a \in \PrimVars \cup \R$, $\qstate_{n}( a)=\qstate_{0}( a)$  if $\qstate_{n}(\plvl a)= \low$.

  \item\label{item.high.cond.refs.preserves.aliasing.relations} { For all $r,s \in \R$ where $\qstate_{i}(\hlvl r)=\low$, (i) $\qstate_{i}(\AliasRel r s) =\qstate_{0}(\AliasRel r s) $, and (ii)  $\qstate_{i}(\FieldAliasXRel r s)=\qstate_{0}(\FieldAliasXRel r s)$.}

  \end{enumerate}
\end{lemma}
\begin{proof}
  By induction on $i$ and then a case analysis on the statements. Informally, the security type of any variable updated in the execution will be at least upgraded by \pc. Since, \(\pc = \high\) according to Lemma~\ref{lemm::high.context.invariant1}-\ref{item.high.cond.pc.constraints}, then the label of no variable will be downgraded.
  Note that downgrading is not done in the junction state $\qstate_j$ where the junction is visited in the nominal mode and leads to applying the updates by \Startua {} that updates $\sHeap$ to $\sHeap'$.
\end{proof}

\begin{lemma}[Equal Typing Environment in High-Context]\label{lemm::hig.context.equal.typing.enrivonment}
  Let $\qstate_0 \overset{\rho}{\rightarrowtail} \qstate_n$ and $t_0 \overset{\rho}{\rightarrowtail}t_m$ be two executions where
  \(\qstate_0 \InRelation t_0\),
  and \(\qstate_0(\pc) \sqcup \qstate_0(\plvl e)=\high\).
  It holds that (i) for all $x \in \PrimVars \cup \R$, $\qstate_n(\plvl x)= \low$ iff $t_m(\plvl x)= \low$, and
  (ii) for all $r \in \R$, $\qstate_n(\hlvl r)= \low$ iff $t_m( \hlvl r)= \low$.
\end{lemma}
\begin{proof}
  We prove (i) by contradiction. Let $\qstate_n(\plvl x)= \low$ and $t_m(\plvl x)= \high$ for some $x$. From {\(\qstate_0 \InRelation t_0\) and the condition~\ref{compatible.states.item:same.heap.typing.environment} in Def.~\ref{def::compatible.states}}, it follows that $\qstate_0(\plvl x)= t_0(\plvl x)$.
  If $\qstate_0(\plvl x)=t_0(\plvl x)= \high$, then we can conclude that $\qstate_0(\plvl x) \sqsubseteq \qstate_n(\plvl x)$ using Lemma~\ref{lemm::high.context.invariants.general.vars}-\ref{item.high.cond.upgrading.vars} and the transitivity property of $\sqsubseteq$, \ie $\qstate_n(\plvl x) = \high$ that contradicts our assumptions and proves the conclusion.

  If $\qstate_0(\plvl x)=t_0(\plvl x)= \low$, from $t_m(\plvl x)= \high$, it follows that a statement \stm has upgraded the level of $x$.
  Based on the definition of valid executions that requires execution of all branches in the upgrade analysis mode, \stm should have been executed in $\qstate_0 \overset{\rho}{\rightarrowtail} \qstate_n$ as well.
  We show that $\qstate_{n}(\plvl x) =\high$ by performing a case analysis on \stm. The statement \stm cannot be a (conditional) \cgoto, \coutput{}. junction or $\surd$, as none can update \plvl x.
  \begin{itemize}
  \item Let \stm be an assignment. If \stm is not a field load, since \stm updates the security type of $x$ in $t_0 \overset{\rho}{\rightarrowtail} t_m$, then \stm = \s{x=x'} for some $x'$ according to \AssignRule.
  If \mode = \tt, then $\plvl x$ is upgraded by $\pc$ by the statement \stm in $\qstate_0 \overset{\rho}{\rightarrowtail} \qstate_n$ too according to \AssignRule, \ie $\qstate_{i}(\plvl x)= \qstate_{i-1}(\plvl x) \sqcup  \qstate_{i-1}(\pc)$.  From Lemma~\ref{lemm::high.context.invariant1}-\ref{item.high.cond.pc.constraints}, $\qstate_{i-1}(\pc) = \high$, and consequently $\qstate_{i}(\plvl x) = \high$. Using Lemma~\ref{lemm::high.context.invariants.general.vars}-\ref{item.high.cond.upgrading.vars}, it follows that $\qstate_{i}(\plvl x) \sqsubseteq \qstate_{n}(\plvl x)$ \ie  $\qstate_{n}(\plvl x) =\high$ that contradicts our assumption.





  \end{itemize}
  To prove (ii), we induct on the number of CDRs of $\rho$. \\
  \textbf{Base Case}
  According to the rules \mBranchRule and \BranchRule and the fact that the CDRs have unique junctions, the function \Endua{.} applies the final updates \HeapBulkUpgradeFrom[\sHeap']{\sHeap} to the heap in the states $\qstate_{n-1}$ and $t_{m-1}$, which is equal to \HeapCopyAliases[\sHeap]{\sHeap'} \SQMergei u
  \HeapRestoreObjLevels[\sHeap]{\sHeap'} in Fig.~\ref{fig::Symmaries.Abstract.Heap.Operations}.
  From Lemma~\ref{lemm::high.context.invariant1}-\ref{item.high.cond.hr.stack.junctio.exists}, it follows that $\qstate_{n-1}(\sHeap')=\qstate_j(\sHeap)$  where $j$ is $\rho$'s junction visited in the nominal mode (Note that $\rho$ has no inner CDR in the base case).
  Therefore, the updates to the heap by \Endua{.} will be equivalent to  \HeapCopyAliases[\qstate_{n-1}(\sHeap)]{\qstate_j(\sHeap)} \SQMergei u
  \HeapRestoreObjLevels[{\qstate_{n-1}(\sHeap)}]{{\qstate_j(\sHeap)}}{}.
  Using Lemma~\ref{lemm::high.context.invariant1}-\ref{item.high.cond.hr.stack.junctio.exists}-(ii),  and Lemma~\ref{lemm::high.context.invariants.general.vars}-\ref{item.high.cond.upgrading.vars} and the transitivity property of $\sqsubseteq$, we can show that {$\qstate_0(\hlvl \sHeap) \sqsubseteq \qstate_{n-1}(\hlvl \sHeap)$} and
  {$\qstate_0(\hlvl \sHeap) \sqsubseteq \qstate_{j}(\hlvl \sHeap)$}. Consequently, according to the definition of  \HeapRestoreObjLevels[ {\qstate_{n-1}(\sHeap)}]{{\qstate_j(\sHeap)}}, we can conclude that no downgrading will be done, \ie { $\qstate_0(\hlvl \sHeap) \sqsubseteq \qstate_{n}(\hlvl \sHeap)$ (I)}.

  We prove the base case by contradiction. Let $\qstate_n(\hlvl r)= \low$ and $t_m(\hlvl r)= \high$ for some $r$. 	From {\(\qstate_0({\hlvl{\sHeap}})=t_0(\hlvl{\sHeap})\)}, it follows that $\qstate_0(\hlvl r)= t_0(\hlvl r)$.
  If $\qstate_0(\hlvl r)=t_0(\hlvl r)= \high$, then we can conclude that $\qstate_0(\hlvl r) \sqsubseteq \qstate_n(\hlvl r)$ using (I), \ie $\qstate_n(\hlvl r) = \high$ that contradicts our assumptions and proves the conclusion.
  If $\qstate_0(\hlvl r)=t_0(\hlvl r)= \low$, from $t_m(\hlvl r)= \high$, it's followed that a statement \stm has upgraded the level of $r$ in the execution $t_0 \overset{\rho}{\rightarrowtail}t_m$. According to the definition of valid executions, this statement should have been executed by $\qstate_0 \overset{\rho}{\rightarrowtail}\qstate_n$ too.
  According to the definition of \HeapRestoreObjLevels[{\qstate_{n-1}(\sHeap)}]{{\qstate_j(\sHeap)}} and the reflexivity property of \AliasRel{}{} (\ie  $\AliasRel[ \qstate_{n-1}(\sHeap)] s s$), we can conclude that $\qstate_n({\hlvl x{}^{\qstate_{n-1}(\sHeap)})} = \qstate_{n-1}(\hlvl x{}^{\qstate_{n-1}(\sHeap)}) \sqcup \qstate_{n-1}(\hlvl x{}^{\qstate_{n-1}(\sHeap')}) $. This means that the label of $x$ should be \low in both \(\qstate_{n-1}(\hlvl x{}^{\qstate_{n-1}(\sHeap)}) \)  and \( \qstate_{n-1}(\hlvl x{}^{\qstate_{n-1}(\sHeap')})\).

  Let $a$ be a statement that modifies the level of $x$ in $t_0 \overset{\rho}{\rightarrowtail}t_m$. The only operation that can manipulate  \hlvl x should be a field store operation according to $T_\stm$ in the rule \AssignRule 
  (or other operations that might include field restore operations).
  Let $a$ be a field store $r.f_r = s$ executed in a state $t_j, 0 \leq j <m$ and led to upgrading \hlvl x.
  According to the semantics, either $t_j (\AliasRel x r)$ or $t_j(\FieldAliasXRel x r)$ should hold (II).
  Based on Lemma~\ref{lemm::high.context.invariants.general.vars}-\ref{item.high.cond.refs.preserves.aliasing.relations}, $t_j (\AliasRel x r)=t_0(\AliasRel x r)$ if $\AliasRel x r \in \GenericRelVars $ and $t_j(\FieldAliasXRel x r)=t_0(\FieldAliasXRel x r)$ if $\FieldAliasXRel x r $ (III). As mentioned earlier, the statement $a$ is executed in the execution $\qstate_0 \overset{\rho}{\rightarrowtail}\qstate_n$ too, say in a state $\qstate_k, 0 \leq k <n$. Similar to the case of first execution, we can show that  $\qstate_k (\AliasRel x r)=\qstate_0(\AliasRel x r)$ if $\AliasRel x r \in \GenericRelVars $ and $\qstate_k(\FieldAliasXRel x r)=\qstate_0(\FieldAliasXRel x r)$ if $\FieldAliasXRel x r $ (IV).
  From III, IV, {\(\qstate_0 \InRelation t_0\) and Def.~\ref{def::compatible.states}-\ref{compatible.states.item:low.quiv.heap}}, it follows that $\qstate_k (\AliasRel x r)=t_j(\AliasRel x r)$ if $\AliasRel x r \in \GenericRelVars $ and $\qstate_k(\FieldAliasXRel x r)=t_j(\FieldAliasXRel x r)$ if $\FieldAliasXRel x r $ (V).
  From (II) and (V), it follows that $\qstate_k (\AliasRel x r)$ or $\qstate_k(\FieldAliasXRel x r)$ holds.
  Since, $\qstate_k(\pc)=\high$, it follows that $l=\high$ in \AssignRule, 
   that leads to upgrading  \hlvl x = \high. Since no downgrading is done in non-junction states according to Lemma~\ref{lemm::high.context.invariants.general.vars}-\ref{item.high.cond.upgrading.vars}, this means that $\qstate_{n-1}(\hlvl x{}^{\qstate_{n-1}(\sHeap)}) \sqcup \qstate_{n-1}(\hlvl x{}^{\qstate_{n-1}(\sHeap')}) = \high$ which contradicts our assumption and proves the conclusion, \ie $\qstate_n({\hlvl x})=\high$.

  The case of the primitive field store operation is proven similarly.

  \textbf{Inductive Step} Straightforward.

\end{proof}
\begin{lemma}[High-Context Preserves Relation]\label{lemm::branch}
  Let \(\qstate_1 \InRelation \qstate_2 \) where
  (i) \(\qstate_i(ℓ) = (\cif~(e)~\cgoto~\lbl{label};\_,\_)\), \(i \in \{1,2\}\),
  and (ii) \(\qstate_1(\pc)=\low\) and \(\qstate_1(\plvl e)=\high\).
  If $\qstate_1 \overset{\rho}{\rightarrowtail} \qstate'_1$ and $\qstate_2 \overset{\rho}{\rightarrowtail}
  \qstate'_2 $, then \(\qstate'_1 \InRelation \qstate'_2 \).
\end{lemma}
\begin{proof}
  We show that all conditions in Def.~\ref{def::compatible.states} and Eq.~\ref{compatible.states.item:same.pc.mode} hold for the pair $( \qstate'_1,\qstate'_2)$.
  \begin{itemize}
  \item \(\qstate'_1(ℓ)=\qstate'_2(ℓ)\) follows from the definition of valid executions and the fact that each CDR has a unique junction.
  \item \(\qstate'_1(\mode)=\qstate'_2(\mode)=\ff\), \(\qstate'_1(\pc) =\qstate'_2(\pc) = \low\) are \(\qstate'_1(\rVar) =\qstate'_2(\rVar)=\rNom\) are  followed from Lemma~\ref{lemm::high.context.invariant1}.
  \item  \(\qstate'_1({\VarsTypes})=\qstate'_2(\VarsTypes)\) 
  is followed from Lemma~\ref{lemm::hig.context.equal.typing.enrivonment}.
  \item According to Lemma~\ref{lemm::high.context.invariants.general.vars}-\ref{item.high.cond.unmodified.low.vars.in.whole.execution}, it's followed that for all $a \in \PrimVars \cup \R$, $\qstate'_{i}( a)=\qstate_{i}( a)$,  if $\qstate'_{i}(\plvl a)= \low$ (i). From \(\qstate_1 \InRelation \qstate_2 \), Def.~\ref{def::compatible.states}-\ref{compatible.states.item:low.equiv.memory}, Def.~\ref{def::compatible.states}-\ref{compatible.states.item:low.quiv.heap}, it follows that $\qstate_{1}( a)=\qstate_{2}( a)$,  if $\qstate'_{i}(\plvl a)= \low$(ii). From (i), (ii) and Lemma~\ref{lemm::hig.context.equal.typing.enrivonment}, we can conclude that  for all $a \in \PrimVars \cup \R$, $\qstate'_{1}( a)=\qstate'_{2}( a)$,  if $\qstate'_{1}(\plvl a)= \low$.
    Consequently,  \(\qstate'_1(\V)=_{\qstate'_1(\VarsTypes)}\qstate'_2(\V)\) and { \({\qstate'_1(\mHeap) =_{\qstate'_1(\VarsTypes)} \qstate'_2(\mHeap)}\) }are followed.
  \end{itemize}

\end{proof}

\end{document}
